\documentclass[prd,nofootinbib,eqsecnum,superscriptaddress]{revtex4-2}
\usepackage{color}
\usepackage{xcolor}
\usepackage{amsmath}
\usepackage{amssymb}
\usepackage{graphicx}
\usepackage{setspace}

\usepackage{epsfig}

\usepackage{dcolumn}
\usepackage{bm}
\usepackage[breaklinks=true,colorlinks=true,linkcolor=blue,urlcolor=blue,citecolor=blue,pagebackref=true]{hyperref}

\newcommand{\w}{\omega}

\makeatother

\begin{document}
\title{Two-point function of a quantum scalar field in the interior
region of a Kerr black hole}
\author{Noa Zilberman}
\email{noazilber@campus.technion.ac.il}
\affiliation{Department of Physics, Technion, Haifa 32000, Israel}
\author{Marc Casals}
\email{marc.casals@ucd.ie}
\email{mcasals@cbpf.br}
\affiliation{Centro Brasileiro de Pesquisas F\'isicas (CBPF), Rio de Janeiro, CEP 22290-180, Brazil}
\affiliation{School of Mathematics and Statistics, University College Dublin, Belfield, Dublin 4, D04 V1W8, Ireland}
\affiliation{Laboratoire Univers et Th\'eories, Observatoire de Paris, CNRS, Universit\'e PSL,
Universit\'e de Paris, 92190 Meudon, France}
\author{Amos Ori}
\email{amos@physics.technion.ac.il}
\affiliation{Department of Physics, Technion, Haifa 32000, Israel}
\author{Adrian C. Ottewill}
\email{adrian.ottewill@ucd.ie}
\affiliation{School of Mathematics and Statistics, University College Dublin, Belfield, Dublin 4, D04 V1W8, Ireland}

\date{\today}
\begin{abstract}

 Quantum field effects on a classical background spacetime may be obtained from the semiclassical equations of General Relativity with
the expectation value of the  stress-energy tensor of the quantum field as a source. 
This expectation value can be calculated from Hadamard's elementary two-point function, which in practice is given in terms of sums of products of field modes evaluated at two spacetime points.
We derive expressions for the two-point function for a massless scalar field  in the Unruh state on a Kerr black hole spacetime.
Our main result in this paper is a novel expression valid when the two points lie {\it inside} the black hole; we also (re-)derive, using a new method, the known expression valid when the two points lie outside the black hole.
We achieve these expressions by finding relationships between Unruh modes, defined in terms of the retarded Kruskal coordinate, and Eddington modes, defined in terms of the Eddington coordinates.
While our starting expression for the two-point function is written in terms of the Unruh modes, we give our final expression in terms of the Eddington modes, which have the computational advantage that they decompose into factors that obey {\it ordinary} differential equations.  In an appendix we also derive expressions for the bare mode contributions to the flux components of the stress-energy tensor for a minimally-coupled massless scalar field inside the black hole.
Our results thus lay the groundwork for future calculations of quantum effects inside a Kerr black hole.
\end{abstract}

\maketitle

\section{Introduction}

In the semiclassical framework of quantum field theory on a curved
spacetime, a gravitational field is treated classically whereas matter
fields on the corresponding (background) spacetime are quantized.
In practice, the Einstein field equations of General Relativity
are sourced by the renormalized
expectation value of the stress-energy tensor (RSET) for the matter
fields in a certain quantum state. This framework is expected to provide
a good approximation to the Physics when the scales of the system are above
the Planck scales and it has yielded results as important as the emission
of quantum (Hawking) radiation by astrophysical black holes (BHs)~\cite{Hawking:1974,Hawking:1975}.

Within quantum field theory on a curved spacetime, one can define
Hadamard's elementary two-point function (HTPF) as the expectation
value of the anti-commutator of a (say, scalar) field $\hat{\Phi}$
in a certain quantum state $\left|\Psi\right\rangle $: $G_{\Psi}^{(1)}(x,x')=\left\langle \left\{ \hat{\Phi}\left(x\right),\hat{\Phi}\left(x'\right)\right\} \right\rangle _{\Psi}$, where $x$ and $x'$ are spacetime points, and curly brackets denote symmetrization with respect to $x$ and $x'$  \footnote{\label{ftn:antisymm}More explicitly,  $\left\{ \xi\left(x\right),\zeta\left(x'\right)\right\} \equiv\xi\left(x\right)\zeta\left(x'\right)+\xi\left(x'\right)\zeta\left(x\right)$,
where $\xi$ and $\zeta$ are two quantities that depend on the spacetime
point.}. The HTPF is a solution
of the homogeneous wave equation satisfied by the field $\hat{\Phi}$
 and is an important object for various reasons. First, it is physically
relevant in its own right, since it yields the quantum correlations
between different points on the background spacetime. Second, when
subtracting from it an appropriate, purely geometric, renormalization
term (also called the counterterm)~\cite{DeWittBook:1965} and taking the limit
$x'\rightarrow x$, the renormalized Wick product $\left\langle \hat{\Phi}^{2}\left(x\right)\right\rangle _\text{ren}^{\Psi}$
is obtained, which is a manifestation of the quantum vacuum fluctuations.
Last but not least, by applying a certain differential operator on
the HTPF minus the renormalization term  ~\cite{Christensen:1976,Christensen:1978}, and then taking the limit
$x'\rightarrow x$, the RSET is obtained, which is a source in the
semiclassical Einstein equations.

In principle, it is possible to define various states for a quantum
field on a BH background spacetime. In the case of a spherically-symmetric
(e.g., Schwarzschild) BH, the most commonly used states are: (i) the Boulware state~\cite{Boulware:1975a,Boulware:1975b},
which is meant to model the surrounding of a star-like object, since this state is empty at (past and
future) null infinity and it diverges on the (past and future) event
horizon (EH); (ii) the Unruh state~\cite{unruh1976notes}, which models an evaporating
BH via the emission of Hawking radiation;  and (iii) the Hartle-Hawking state~\cite{HH:1976}, which models a BH in equilibrium with its own radiation.
When the BH is rotating (Kerr), however, the corresponding Boulware
state~\cite{Unruh:1974bw,OttewillWinstanley:2000} is no longer empty
at future null infinity, as it contains the so-called Unruh-Starobinskii
radiation (essentially, the quantum version of classical superradiance~\cite{zel1971generation,starobinskii1973amplification}). Also, for
bosons, no state in thermal equilibrium with the rotating BH can be
constructed, i.e., no Hartle-Hawking-like state exists for bosons
in Kerr~\cite{KayWald:1991,OttewillWinstanley:2000} \footnote{For fermions, on the other hand, a state in thermal equilibrium can
be constructed sufficiently close (specifically, within the so-called
speed-of-light surface) to the rotating BH \cite{CDNOW}.}. Finally, the Unruh state can be constructed in Kerr (\cite{OttewillWinstanley:2000}
for a scalar field,~\cite{CDNOW} for a fermion field and~\cite{Casals:Ottewill:2005}
for the electromagnetic field) similarly to the spherically-symmetric
case. 

In this paper we shall focus on the Unruh state in Kerr spacetime, which is the state of relevance for astrophysical BHs.

Our treatment of the Unruh state 
in Kerr is also significant for the following reason. Often, in spherical symmetry,
one first calculates quantities in the Hartle-Hawking state since it allows for
the Euclideanization technique, whereby one merely needs to sum over a \textit{discrete} set of field modes thus facilitating the calculation; then, if one wishes
to calculate the quantity in another state -- such as Unruh -- one just
calculates the \textit{difference} between that quantity in that other
state and in the Hartle-Hawking state (such difference needs no renormalization
and so in principle it can be carried out relatively easily). 
In Kerr,
however, this method of Hartle-Hawking state subtraction is not applicable, because no Hartle-Hawking-like state exists in Kerr. 

Most calculations for quantum fields on a BH spacetime have focused
on quantities \textit{outside} the EH of the BH -- after
all, we are observers located far away from any BH! Expressions for
the HTPF with the two points outside the BH are known in both the
non-rotating~\cite{Candelas:1980,christensen1977trace} and the rotating
cases~\cite{FrolovThorne:1989,OttewillWinstanley:2000}  and they have been used to calculate $\Phi^2$ and
the RSET. While in the non-rotating case the RSET outside the BH has
been obtained in various physical settings, in the rotating case
the calculation is a lot more technically involved  and in fact it
has been achieved only recently and in one instance:~\cite{LeviEilonOriMeent:2017}. 

In its turn, the investigation of effects of quantum fields \textit{inside}
the EH of a BH may serve to address questions of fundamental
conceptual importance. Most notably, the question of whether the \textit{inner}
horizon (IH) of a rotating and/or electrically-charged BH is stable under
quantum perturbations. Beyond the IH, the Cauchy initial value
problem is not well-posed and so the Einstein field equations of General
Relativity cease to be deterministic. Quantum effects
have been seen to destroy the regularity of  the IH of a non-rotating
and electrically-charged (Reissner-Nordstr\"om, RN) BH~\cite{FluxesIH:2020}.
A similar behavior was also found in
 RN-de Sitter BH~\cite{hollands2020quantum} and, at least for
quantum perturbations approaching the IH from the inside,
of a 2+1-dimensional rotating BTZ BH~\cite{steif1994quantum,casals2017quantum}.
In all these cases the HTPF was known for the two points inside the
EH and served to calculate the RSET. However, in the most important case of a Kerr
BH, an expression for the HTPF with the two points inside the BH was
not known until the current work and, consequently, no  quantitative investigation of the quantum
effects on its IH has yet been carried out \footnote{There have been, however, {\it qualitative} investigations, see ~\cite{Hiscock:1980,OttewillWinstanley:2000}.}.

The main result in this paper is an expression for the HTPF $G_{U}^{(1)}(x,x')=\left\langle \left\{ \hat{\Phi}\left(x\right),\hat{\Phi}\left(x'\right)\right\} \right\rangle _{U}$
for a quantum massless scalar field in the Unruh state $\left|0\right\rangle _{U}$
with the two points $x$ and $x'$ located inside the Kerr BH between the EH
and the IH. One of the main values of this
expression is that it is given in terms of (Eddington) field modes which decompose
into factors that obey \textit{ordinary} differential equations and
so are relatively easy to calculate, at least numerically.
Thus, our expression for the HTPF is of practical use for potential
future calculations of the RSET inside the EH of a Kerr
BH in the Unruh state. (We perform a step in this direction in Appendix~\ref{app:fluxes}, where we derive expressions for the bare mode contribution to the flux components in the BH interior.)
Furthermore, once one achieves renormalization in the Unruh state via the HTPF provided in this paper, 
one can use that as the fiducial state with respect to which to calculate differences and thus easily achieve renormalization in another state.
Prior to obtaining this new expression for the HTPF inside a Kerr BH,
we 
derive an expression for 
the HTPF {\it outside} a Kerr BH;
although  this latter expression was already known, we 
(re-)derive it by employing
a new method
which 
is the one that we 
subsequently
apply
inside the BH. Moreover, 
in order to achieve these expressions for the HTPF,
 we 
 obtain
 relationships
between the Unruh family of modes (which are defined in terms of the retarded Kruskal coordinate and serve to define the Unruh state), and Eddington families of modes (which are defined in terms of the Eddington coordinates and, as mentioned, decompose by factors).
These relationships between families of field modes are useful in their own right in that they may be readily applicable to
the calculation of two-point functions
other than the HTPF, such as the Wightman
function (which is relevant, for example, for the calculation of the
transition probability rate of an Unruh-DeWitt quantum particle detector~\cite{BirrellDaviesBook:1982}).
For the reader who is just interested in the new expression for the
HTPF inside the BH, that expression is given in Eq.~\eqref{eq:G_U_int} or, equivalently, in Eq.~\eqref{eq:G_U_ftilde-1}.

The rest of this paper is organized as follows. Secs.~\ref{sec:The-Kerr-metric}-\ref{sec:Quantum states}
lay the foundations for the subject of the paper: the Unruh HTPF for
a scalar field on a Kerr BH interior. In Sec.~\ref{sec:The-Kerr-metric}
we review the Kerr metric and the associated wave equation satisfied
by a massless, uncharged scalar field. Sec.~\ref{sec:Families-of-modes}
introduces the various families of field modes which are relevant
to this paper. The modes allow us to define the Unruh quantum state
in Sec.~\ref{sec:Quantum states} and to derive the (already known)
expression for the HTPF outside a Kerr BH in Sec.~\ref{sec:TPF_exterior}
(specifically, Eq.~\eqref{eq:our_final_result_outside}). The paper culminates in Sec.~\ref{sec:TPF_interior}, where we obtain the
(new) expression for the HTPF inside a Kerr BH (specifically, Eq.~\eqref{eq:G_U_int}, or  ~\eqref{eq:G_U_ftilde-1}).
The paper also has two appendices: Appendix~\ref{App:small freq} addresses the issue of IR regularity (i.e. regularity at small frequencies) of our final expressions for the HTPF; and Appendix~\ref{app:fluxes} presents a derivation of the bare mode-sum expressions for the flux components of the RSET, based on the HTPF expression derived in this paper for the BH interior.

We use units where $c=G=1$ (while $\hbar$ is not taken to be equal
to 1) and metric signature $(-+++)$.

\section{The Kerr metric and the wave equation \label{sec:The-Kerr-metric}}

\subsection{The Kerr metric and coordinate systems\label{subsec:metric_and_coordinates}}

\begin{figure}
\centering\includegraphics{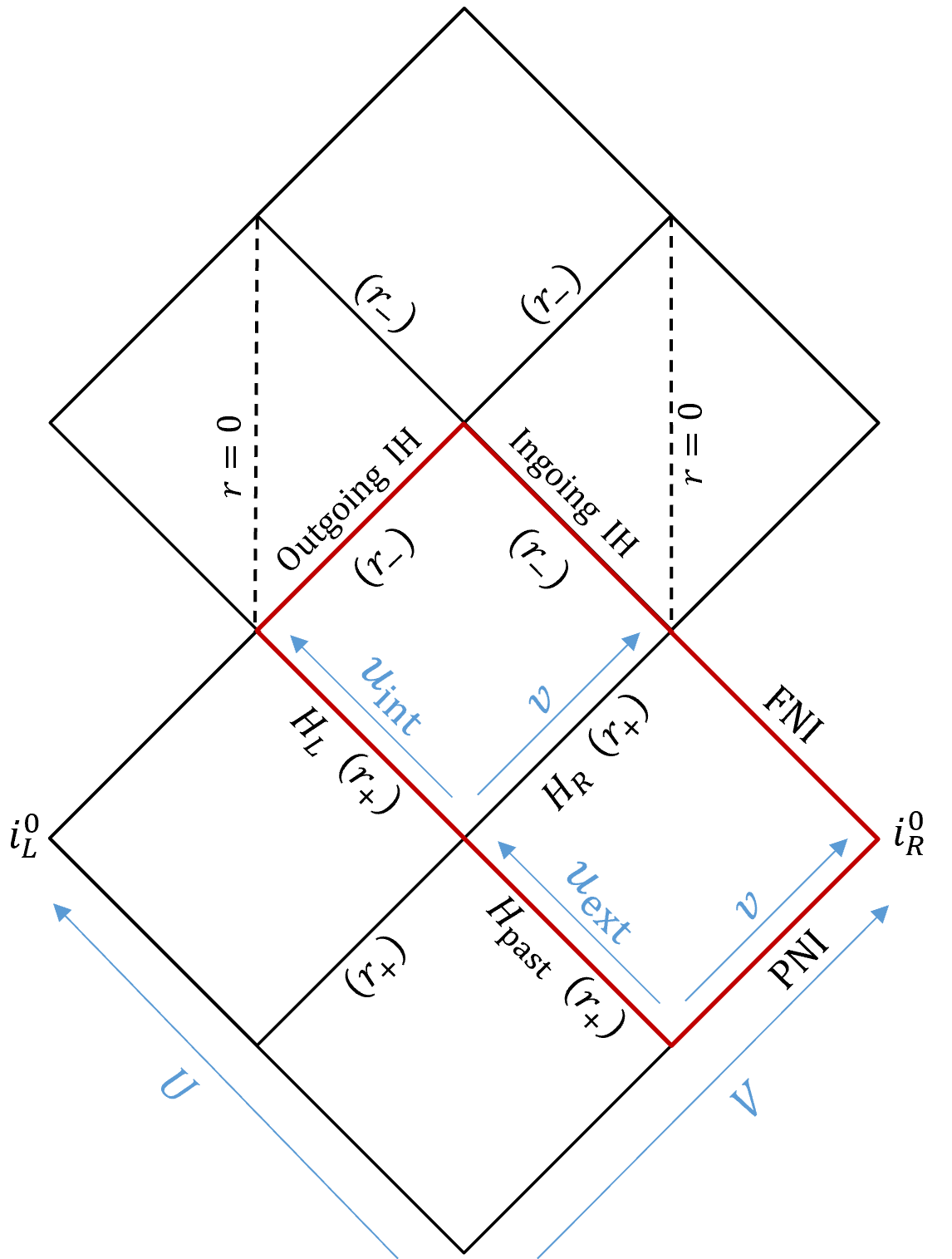}\caption{{\small{}A portion of the Penrose diagram of the analytically-extended
sub-extremal Kerr spacetime, with 3 systems of coordinates: the Kruskal
$U$ and $V$ (Eqs.~\eqref{eq:extKrusCoor} and \eqref{eq:intKrusCoor}),
the outer Eddington $u_{\text{ext}}$ and $v$ (Eq.~\eqref{eq:extEddCoor}),
and the inner Eddington $u_{\text{int}}$ and $v$ (Eq.~\eqref{eq:intEddCoor}). The spacetime regions relevant for this paper are within
the red frame, consisting of the BH exterior and interior. }}

\label{Fig:Kerr_Penrose}
\end{figure}

The Kerr metric is a vacuum solution to the classical Einstein field
equations, describing a BH of mass $M$ rotating with angular momentum
$J$. It is given by the line element in Boyer-Lindquist coordinates
$(t,r,\theta,\varphi)$:

\begin{equation}
\text{d}s^{2}=-\left(1-\frac{2Mr}{\rho^{2}}\right)\text{d}t^{2}+\frac{\rho^{2}}{\Delta}\text{d}r^{2}+\rho^{2}\text{d}\theta^{2}+\left(r^{2}+a^{2}+\frac{2Mra^{2}}{\rho^{2}}\sin^{2}\theta\right)\sin^{2}\theta\text{d}\varphi^{2}-\frac{4Mra}{\rho^{2}}\sin^{2}\theta\text{d}\varphi\text{d}t\,,\label{eq:Kerr_metric}
\end{equation}
where $a\equiv J/M$ and
\begin{align*}
\rho^{2} & \equiv r^{2}+a^{2}\cos^{2}\theta,\\
\Delta & \equiv r^{2}-2Mr+a^{2}\,.
\end{align*}
The horizon radii correspond to the roots of the equation $\Delta=0$,
yielding an EH at 
\[
r=r_{+}\equiv M+\sqrt{M^{2}-a^{2}}
\]
 and an IH  at 
\[
r=r_{-}\equiv M-\sqrt{M^{2}-a^{2}}.
\]
Note the resulting restriction on the BH parameters: $\left|a\right|/M\leq1$.
Throughout this paper we shall only treat the subextremal case, corresponding
to $\left|a\right|/M<1$, and restrict our attention to the region
bounded by $r\geq r_{-}$. We refer to the region $r>r_{+}$ (outside
the EH) as the external Universe or BH \emph{exterior}, whereas the
region bounded by the horizons $r_{-}<r<r_{+}$ is to be referred
to as the BH \emph{interior}. Note that we might occasionally use the term "exterior" for $r\geq r_+$ (namely, including $r=r_+$), and likewise the term "interior" for $r_{-}\leq r\leq r_{+}$, depending on the context.
See Fig.~\ref{Fig:Kerr_Penrose} for (a portion of) the Penrose diagram of the analytically-extended
sub-extremal Kerr spacetime.

We shall now briefly discuss the behavior of the standard Boyer-Lindquist
coordinates $\left(t,r,\theta,\varphi\right)$ for a free-falling
observer approaching the EH. As in the case of spherical symmetry (e.g.,
in the Schwarzschild and RN metrics), the $t$ coordinate diverges
at $r=r_{+}$ for an infalling observer, which motivates the definition
of Kruskal coordinates given below in \eqref{eq:extKrusCoor} and \eqref{eq:intKrusCoor}.
However, in Kerr, not only does $t$ diverge on approaching $r=r_{+}$ but, unlike in spherical symmetry,
also the azimuthal coordinate $\varphi$ diverges there. In other
words, a geodesic approaching the EH undergoes (fictitious) infinite spiraling when presented in the $\varphi$
coordinate (which is, however, a mere coordinate artifact). One may
compute the (constant) angular velocity with which the EH rotates
(or, more precisely, the geodesic's limiting value of $\text{d}\varphi/\text{d}t$
at $r\to r_{+}$) to be
\begin{equation}
\Omega_{+}\equiv\frac{a}{2Mr_{+}}=\frac{a}{r_{+}^{2}+a^{2}}\,.\label{eq:Omega+}
\end{equation}

This quantity can be used to construct a coordinate that remains regular
on approaching $r\to r_{+}$, defined by:

\begin{equation}
\varphi_{+}\equiv\varphi-\Omega_{+}t\,.\label{eq:phi+}
\end{equation}

Similar considerations apply at $r\to r_{-}$, where we analogously define
\begin{equation}
\Omega_{-}\equiv\frac{a}{2Mr_{-}},\,\,\,\,\,\,\varphi_{-}\equiv\varphi-\Omega_{-}t\,.\label{eq:Omega-phi-}
\end{equation}

For later use, we shall hereby define the tortoise coordinate $r_{*}$
in Kerr via $\text{d}r/\text{d}r_{*}=\Delta/\left(r^{2}+a^{2}\right)$.
We choose the constant of integration such that \footnote{We note that although this choice of constant of integration is common in the literature, it differs from other common choices such as that used in~\cite{ST}.}:
\begin{equation}
r_{*}=r+\frac{1}{2\kappa_{+}}\log\left(\frac{\left|r-r_{+}\right|}{r_{+}-r_{-}}\right)-\frac{1}{2\kappa_{-}}\log\left(\frac{\left|r-r_{-}\right|}{r_{+}-r_{-}}\right),\label{eq:rstarKerr}
\end{equation}
where $\kappa_{\pm}$ are the two corresponding surface gravity parameters,
given by
\begin{equation}
\kappa_{\pm}\equiv\frac{r_{+}-r_{-}}{2\left(r_{\pm}^{2}+a^{2}\right)}\,.\label{eq:kappaKerr}
\end{equation}

Note that $r=r_{+}$ corresponds to $r_{*}\to-\infty$, while $r=r_{-}$
(like $r\to\infty$ outside the BH) corresponds to $r_{*}\to\infty$.

From here we may define the Eddington coordinates \footnote{While these coordinates are usually known as ``Eddington-Finkelstein coordinates", we use ``Eddington" for abbreviation.}, given in the BH
exterior by
\begin{equation}
u_{\text{ext}}\equiv t-r_{*},\,\,\,v\equiv t+r_{*}\,,\label{eq:extEddCoor}
\end{equation}
and in the BH interior by 
\begin{equation}
u_{\text{int}}\equiv r_{*}-t,\,\,\,v\equiv r_{*}+t\,.\label{eq:intEddCoor}
\end{equation}

The coordinate $v$ is continuous across the EH (and parameterizes
it), whereas $u_{\text{ext}}$ and $u_{\text{int}}$ diverge there.
The regularity of the metric at the EH may be seen by transforming
to a set of Kruskal coordinates, which we shall denote by $U$ and
$V$, given in the BH exterior by:
\begin{equation}
U\left(u_{\text{ext}}\right)\equiv-\frac{1}{\kappa_{+}}\exp\left(-\kappa_{+}u_{\text{ext}}\right),\,\,\,V\left(v\right)\equiv\frac{1}{\kappa_{+}}\exp\left(\kappa_{+}v\right)\,,\label{eq:extKrusCoor}
\end{equation}
and in the BH interior by
\begin{equation}
U\left(u_{\text{int}}\right)\equiv\frac{1}{\kappa_{+}}\exp\left(\kappa_{+}u_{\text{int}}\right),\,\,\,V\left(v\right)\equiv\frac{1}{\kappa_{+}}\exp\left(\kappa_{+}v\right)\,.\label{eq:intKrusCoor}
\end{equation}

Note that both Kruskal coordinates $U$ and $V$ are continuous at
the EH: The former vanishes there (from both sides), whereas $V$, just like $v$,
regularly parametrizes the EH. Furthermore, the metric in the $(U,V,\theta,\varphi_{+})$
coordinates is regular and smooth across the EH.

The locus $r=r_{+}$ marks a four-arms cross in the Penrose diagram
in Fig.~\ref{Fig:Kerr_Penrose}. Out of these four arms, in this paper
we only concern with the three included in the red frame, being the
EH (or \emph{right horizon}, denoted $H_{R}$), the \emph{past horizon}
$H_{\text{past}}$ (the {\it white} hole horizon) and the \emph{left
horizon $H_{L}$}. The other arm, the one at the bottom-left, will
not concern us here as it is located outside the domain of dependence
relevant to the Unruh state (namely, the red frame in Fig.~\ref{Fig:Kerr_Penrose}).

We note that the IH is also a {\it Cauchy} horizon in the sense that it is the boundary of validity of the Cauchy initial value problem formulated on a spacelike hypersurface extending from $i^0_R$ to $i^0_L$, where $i^0_R$ ($i^0_L$) is spacelike infinity of the external universe at the right (left) side of Fig.~\ref{Fig:Kerr_Penrose}. 
However, in the more physically realistic case of a BH formed by gravitational collapse, which lacks a past horizon as well as the entire “left-side” external Universe, it is only the  {\it ingoing} section of the IH (see Fig.~\ref{Fig:Kerr_Penrose}) which retains the causal nature of a Cauchy horizon.

In Kruskal coordinates, the past horizon $H_{\text{past}}$ is found
at $V=0$ and $U<0$, the right horizon $H_{R}$ corresponds to $U=0$
and $V>0$, and the left horizon $H_{L}$ corresponds to $V=0$ and
$U>0$.

In the BH exterior there are, in addition, two null asymptotic boundaries
located at infinity: \emph{past null infinity} (PNI) is found at $U=-\infty$
and $V>0$, and \emph{future null infinity} (FNI) is at $V=\infty$
and $U<0$. See Fig.~\ref{Fig:Kerr_Penrose} for locating all the
mentioned null surfaces.

We now make a couple of related observations, relevant to constructing the
families of field modes further on in the paper. 

Each of the above mentioned asymptotic null surfaces (the three at
$r=r_{+}$ and the two at spacial infinity) can be regularly parameterized
by three coordinates -- which may be chosen to be two angular coordinates
($\theta$ and either $\varphi$ or $\varphi_{+}$) and one Eddington
coordinate -- as follows: $H_{\text{past}}$ by $\left(u_{\text{ext}},\theta,\varphi_{+}\right)$,
$H_{R}$ by $\left(v,\theta,\varphi_{+}\right)$, $H_{L}$ by $\left(u_{\text{int}},\theta,\varphi_{+}\right)$,
PNI by $\left(v,\theta,\varphi\right)$ and FNI by $\left(u_{\text{ext}},\theta,\varphi\right)$. We note that the Eddington coordinate in each of these null surfaces may also
be replaced by the corresponding Kruskal coordinate.

We also specify here the {\it affine} parameters along null geodesics generating
each of these asymptotic null surfaces: $U$ along $H_{\text{past}}$
and $H_{L}$ (both with fixed $\theta$ and $\varphi_{+}$) and $V$ along
$H_{R}$ (again with fixed $\theta$ and $\varphi_{+}$). At PNI and FNI,
asymptotic flatness implies that the affine parameters along these
surfaces are simply the Eddington coordinates $v$ and $u_{\text{ext}}$
respectively (both with fixed $\theta$ and $\varphi$).

\subsection{Separation of the wave equation\label{subsec:Separation-of-wave-equation}}

An uncharged scalar field $\Phi\left(x\right)$ of mass $m$ and coupling
$\xi$ to curvature obeys the Klein-Gordon (KG) equation,
\begin{equation}
\left(\square-m^{2}-\xi R\right)\Phi=0\,,\label{eq:generalKG}
\end{equation}
 where $\square$ is the covariant d'Alembertian associated with the background
metric with Ricci scalar  $R$. In the case of a massless field which is minimally-coupled
 $(\xi=0)$ and/or vanishing Ricci scalar (in particular, a vacuum spacetime), this equation becomes
\begin{equation}
\square\Phi=0\,.\label{eq:KG}
\end{equation}

Considering Eq.~\eqref{eq:KG} on a Kerr background, one readily obtains
the following explicit form:
\begin{align}
\square\Phi & =\left(\frac{\left(r^{2}+a^{2}\right)^{2}}{\Delta}-a^{2}\sin^{2}\theta\right)\frac{\partial^{2}\Phi}{\partial t^{2}}+\frac{4aMr}{\Delta}\frac{\partial^{2}\Phi}{\partial t\partial\varphi}+\nonumber \\
 & +\left(\frac{a^{2}}{\Delta}-\frac{1}{\sin^{2}\theta}\right)\frac{\partial^{2}\Phi}{\partial\varphi^{2}}-\frac{\partial}{\partial r}\left(\frac{\partial\Phi}{\partial r}\right)-\frac{1}{\sin\theta}\frac{\partial}{\partial\theta}\left(\sin\theta\frac{\partial\Phi}{\partial\theta}\right)=0\,.\label{eq:Teukolsky}
\end{align}

We shall refer to this equation as the scalar Teukolsky equation,
after the general-spin field case in~\cite{Teukolsky:1973}. Utilizing
the axial symmetry and time-translation invariance of the metric and
of the master equation, we may decompose the field into modes
\begin{equation}
\Phi_{\omega lm}\left(x\right)=\text{const}\cdot\frac{\psi_{\omega lm}\left(r\right)}{\sqrt{r^{2}+a^{2}}}e^{-i\omega t}Z_{lm}^{\omega}\left(\theta,\varphi\right)\,,\label{eq:decomKerr}
\end{equation}
indexed by the frequency $\omega\in\mathbb{R}$, the azimuthal number $m\in\mathbb{\mathbb{Z}}$ and the multipolar number
$l\in\mathbb{\mathbb{N}}_{\geq|m|}$, where $x$ is a spacetime point and
$\psi_{\omega lm}\left(r\right)$ is the so-called \emph{radial function};
the $\left(r^{2}+a^{2}\right)^{-1/2}$ factor has been introduced
to yield a convenient one-dimensional scattering-like\emph{ }equation
for $\psi_{\omega lm}\left(r\right)$ (see Eq.~\eqref{eq:radKerr}
to follow). The angular functions $Z_{lm}^{\omega}\left(\theta,\varphi\right)$
are the \emph{spheroidal} \emph{harmonics}, given by
\begin{equation}
Z_{lm}^{\omega}\left(\theta,\varphi\right)=\left(2\pi\right)^{-1/2}S_{lm}^{\omega}\left(\theta\right)e^{im\varphi}\,,\label{eq:spheroidalHar}
\end{equation}
where $S_{lm}^{\omega}\left(\theta\right)$ is the \emph{spheroidal
wave function} \cite{Berti:2005gp} solving the eigenvalue problem:

\begin{equation}
\frac{1}{\sin\theta}\frac{\text{d}}{\text{d}\theta}\left(\sin\theta\frac{\text{d}S_{lm}^{\omega}\left(\theta\right)}{\text{d}\theta}\right)+\left(a^{2}\omega^{2}\cos^{2}\theta-\frac{m^{2}}{\sin^{2}\theta}+E_{lm}\left(a\omega\right)\right)S_{lm}^{\omega}\left(\theta\right)=0,\label{eq:angKerr}
\end{equation}
with $E_{lm}\left(a\omega\right)$ the corresponding eigenvalue, obtained
by imposing regularity at $\theta=0,\pi$. Note that the angular equation
is real, and we shall only be concerned here with real angular functions
$S_{lm}^{\omega}$.

For a given $\omega$, the functions $Z_{lm}^{\omega}$ form a complete
basis of orthonormal functions on the two-sphere, fulfilling
\begin{align}
\int_{0}^{2\pi}\text{d}\varphi\int_{0}^{\pi}\text{d}\theta\sin\theta\,Z_{lm}^{\omega*}\left(\theta,\varphi\right)Z_{l'm'}^{\omega}\left(\theta,\varphi\right) & =\delta_{ll'}\delta_{mm'}\,.\label{eq:ang_orth}
\end{align}

There is no known closed form for the spheroidal harmonics \footnote{In fact, spheroidal harmonics may be expressed in terms of confluent
Heun functions but only with coefficients which are to be determinable numerically.}, but in the spherical
case (corresponding to $a\omega=0$) they reduce to the well-known
spherical harmonics $Y_{lm}\left(\theta,\varphi\right)$, whence the
spheroidal wave functions reduce, up to a normalization, to Legendre functions -- and the eigenvalue
$E_{lm}\left(a\omega\right)$ simplifies to $l\left(l+1\right)$.

The radial function $\psi_{\omega lm}\left(r\right)$ solves the \emph{radial
equation}

\begin{equation}
\frac{\text{d}^{2}\psi_{\omega lm}}{\text{d}r_{*}^{2}}+V_{\omega lm}\left(r\right)\psi_{\omega lm}=0\label{eq:radKerr}
\end{equation}
with the effective potential

\begin{equation}
V_{\omega lm}\left(r\right)\equiv \frac{K_{\omega m}^{2}\left(r\right)-\lambda_{lm}\left(a\omega\right)\Delta}{\left(r^{2}+a^{2}\right)^{2}}-G^{2}\left(r\right)-\frac{\text{d}G\left(r\right)}{\text{d}r_{*}}\,,\label{eq:KerrPot}
\end{equation}
where
\begin{equation}
K_{\omega m}\left(r\right)\equiv\left(r^{2}+a^{2}\right)\omega-am,\;\;\;\lambda_{lm}\left(a\omega\right)\equiv E_{lm}\left(a\omega\right)-2am\omega+a^{2}\omega^{2},\;\;\;G\left(r\right)\equiv\frac{r\Delta}{\left(r^{2}+a^{2}\right)^{2}}\,.\label{eq:KerrPotFun}
\end{equation}

From Eq.~\eqref{eq:angKerr} it is evident that  flipping
the signs of $\omega$ and/or $m$ leaves the angular equation invariant.
Since the imposed boundary conditions (regularity at the two poles)
has no explicit reference to either $\omega$ or $m$, it follows
that both the eigenvalue $E_{lm}\left(a\omega\right)$ and the angular
function $S_{lm}^{\omega}$ are invariant (modulo a sign)
under such 
sign flips. For our purposes, we focus on a simultaneous sign flip of $m$ and $\omega$:
\begin{equation}
S_{l\left(-m\right)}^{-\omega}=(-1)^mS_{lm}^{\omega}\quad,\qquad E_{l(-m)}\left(-a\omega\right)=E_{lm}\left(a\omega\right)\,\label{eq:  Ang_inv}
\end{equation}
where we have chosen the sign $(-1)^m$ for $S^\omega_{lm}$, so that, for $a\omega=0$, the spheroidal harmonics agree with the standard sign for the spherical harmonics: $\left(Z_{l(-m)}^{\omega=0}\right)^*=\left(Y_l^{-m}\right)^*=(-1)^mY_l^{m}=(-1)^mZ_{lm}^{\omega=0}$.

The situation with the radial equation is slightly more delicate.
This equation, too, is real, as can be seen in Eqs.~\eqref{eq:radKerr}-\eqref{eq:KerrPotFun}.
Therefore, if $\psi_{\omega lm}\left(r\right)$ is a solution, its
complex conjugate is a solution too. However, we are physically motivated
to choose complex boundary conditions to the radial solutions (which
would correspond to e.g. ``ingoing'' or ``upgoing'' waves) --
leading to \emph{complex} radial functions. The radial equation \eqref{eq:radKerr},
too, is invariant under a simultaneous change of signs of $\omega$
and $m$. This implies that if $\psi_{\omega lm}\left(r\right)$ solves
Eq.~\eqref{eq:radKerr}, it will also be a solution of the radial
equation with $\omega\mapsto-\omega$ and $m\mapsto-m$, and so will
be its complex conjugate. As will become evident in the next section
(see Eqs.~\eqref{eq:psi_in_inv}),
 from
the way the boundary conditions for the various modes are defined,
flipping the signs of both $\omega$ and $m$ will actually take us
from the original mode function $\psi_{\omega lm}\left(r\right)$
to its complex conjugate. 

Examining the effective potential (Eq.~\eqref{eq:KerrPot}) in the
asymptotic domains of the exterior region, $r_{*}\to\infty$ (corresponding
to $r\to\infty$) and $r_{*}\to-\infty$ (corresponding to $r\rightarrow r_{+}$),
we find: 
\begin{equation}
V_{\omega lm}^{\text{outside}}\rightarrow\begin{cases}
\omega^{2}, & r\to\infty\,\,\,\,\,\,\,\,\left(r_{*}\to\infty\right)\\
\omega_{+}^{2}, & r\to r_{+}\,\,\,\,\,\,\,\,\left(r_{*}\to-\infty\right)
\end{cases}\,,\label{eq:asymPot_ext}
\end{equation}
where we define 
\begin{equation}
\omega_{+}\equiv\omega-m\Omega_{+}\,.\label{eq:omega+}
\end{equation}
Thus \footnote{To be more precise, this is a consequence of the fact that the potential is short-range: $V_{\omega lm}=\omega^2+O(1/r^2)$ as $r\to\infty$ and $V_{\omega lm}=\omega_+^2+O\left(e^{2\kappa_+ r_*}\right)$ as $r\to r_+$ (corresponding to $r_{*}\to-\infty$).}, the asymptotic behavior of solutions to the radial equation
(Eq.~\eqref{eq:radKerr}) outside the BH is generally of the form
$e^{\pm i\omega r_{*}}$ at $r_{*}\to\infty$ and $e^{\pm i\omega_{+}r_{*}}$
at $r_{*}\to-\infty$, corresponding to free waves in both these domains. 

Similarly, when considering the effective potential in the BH interior,
we obtain
\begin{equation}
V_{\omega lm}^{\text{inside}}\rightarrow\begin{cases}
\omega_{-}^{2}, & r\to r_{-}\,\,\,\,\,\,\,\,\left(r_{*}\to\infty\right)\\
\omega_{+}^{2}, & r\to r_{+}\,\,\,\,\,\,\,\,\left(r_{*}\to-\infty\right)
\end{cases}\,,\label{eq:asymPot_int}
\end{equation}
where we define
\begin{equation}
\omega_{-}\equiv\omega-m\Omega_{-}\,.\label{eq:omega-}
\end{equation}

This is a crucial point for the definition of our modes in Kerr (see
Sec.~\ref{sec:Families-of-modes}), and it differs from the spherically
symmetric case (similar to the $m=0$ case here), where the asymptotic
behavior of the effective potential leaves $\omega^{2}$ in \textit{all}
asymptotic domains of the BH exterior and interior.

\section{Families of modes\label{sec:Families-of-modes}}

It is convenient to decompose the field into sets of modes, each providing
a complete set of solutions to Eq.~\eqref{eq:KG} on some spacetime
region, which are orthonormal with respect to the standard KG scalar product,
defined by
\begin{equation}
\left\langle \psi,\phi\right\rangle \equiv i\int_{\Sigma}\text{d}\sigma^{\mu}\left(\psi^{*}\phi_{;\mu}-\phi\, \psi_{;\mu}^{*}\right)\,,\label{eq:KGproduct}
\end{equation}
where 
$\Sigma$ is the spacelike hypersurface under consideration
and $\text{d}\sigma^{\mu}$ is a future directed normal to $\Sigma$.
Note that the prefactors of the various modes (e.g., the analog of the $\text{const.}$
appearing in Eq.~\eqref{eq:decomKerr}) are chosen such that orthonormality with respect to the KG inner product is satisfied.

\begin{figure}
\includegraphics[width=0.9\textwidth]{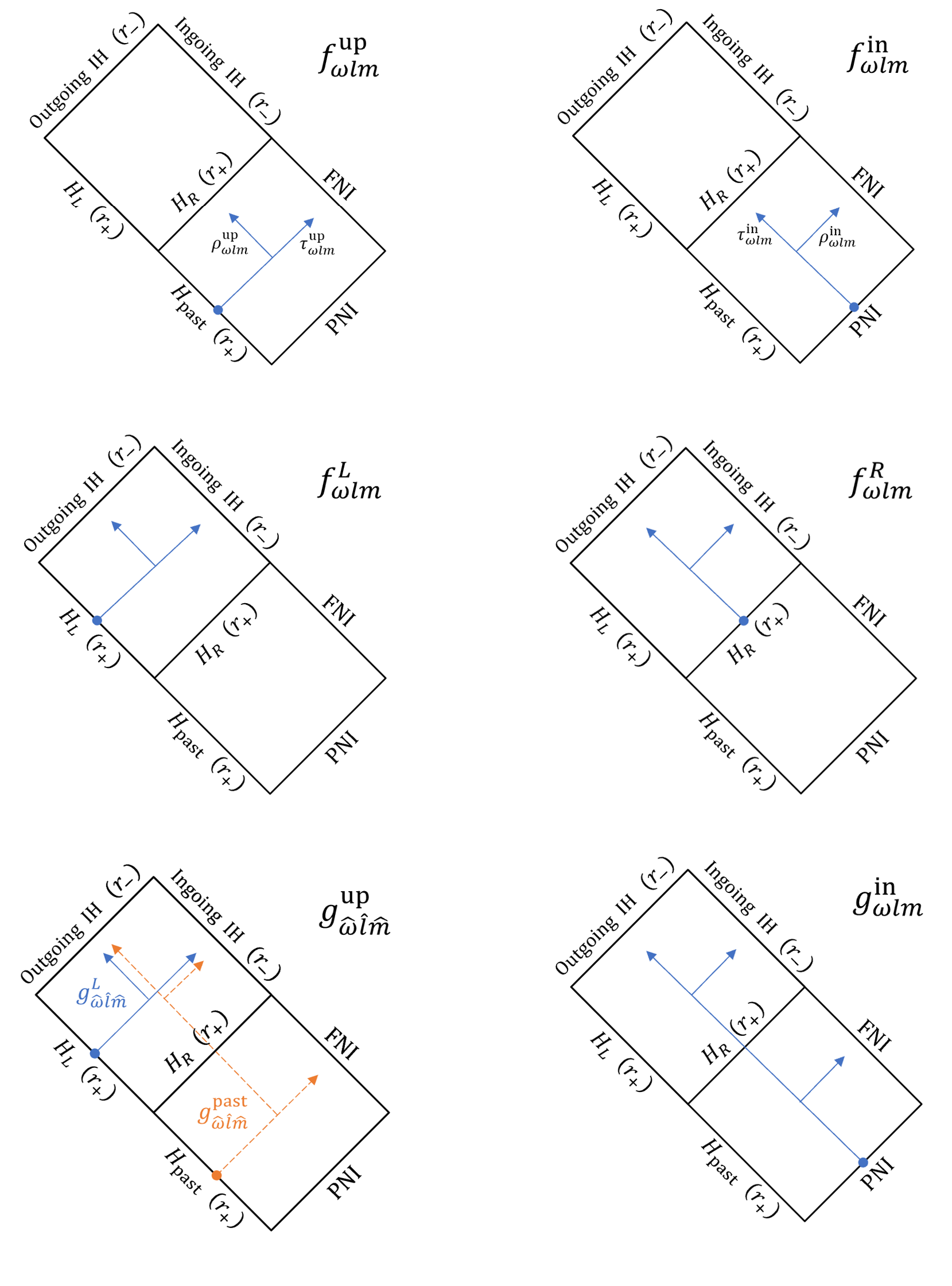}

\caption{Penrose diagrams 
for 
various field modes
on Kerr spacetime:
the outer Eddington
modes $f_{\w lm}^{\text{up}}$ \eqref{eq:fUP_past_asym} and $f_{\w lm}^{\text{in}}$ \eqref{eq:fIN_past_asym};
the
inner Eddington modes 
$f_{\w lm}^{L}$ \eqref{eq:fL_asym} and $f_{\w lm}^{R}$  \eqref{eq:fR_asym};
the Unruh modes
$g_{\hat{\omega}\hat{l}\hat{m}}^{\text{up}}$ \eqref{eq:gUP_asym} and $g_{\w lm}^{\text{in}}$ \eqref{eq:gIN_asym}.
All modes
start 
with unit  amplitude  on 
whichever hypersurface (PNI, $H_{\text{past}}$
or $H_{\text{L}}$) the 
corresponding 
dot 
lies;
part of the wave is reflected and 
part
is transmitted,
with the reflection $\rho$ and transmission
$\tau$ coefficients indicated for 
$f_{\w lm}^{\text{in/up}}$.
The modes $g_{\hat{\omega}\hat{l}\hat{m}}^{\text{up}}$
are defined equally throughout $H_{\text{past}}\cup H_{L}$
 in terms of 
 $U$ but they may also
be constructed as the sum of $g_{\hat{\omega}\hat{l}\hat{m}}^{L}$
(Eq.~\eqref{eq:gL_asym}, with no support on $H_{\text{past}}$) and $g_{\hat{\omega}\hat{l}\hat{m}}^{\text{past}}$
(Eq.~\eqref{eq:gPast_asym}, with no support on $H_{L}$), respectively coloured in blue and dashed
orange on the 
diagram for $g_{\hat{\omega}\hat{l}\hat{m}}^{\text{up}}$.}
\label{fig:modes}
\end{figure}

Three families of modes are of particular importance for our purposes:
the outer and inner Eddington modes, and the Unruh modes.
Each of these three families consists of two distinct sets, a ``left-moving''
and a ``right-moving'' one, as specified below. We begin here with
a brief general description of the three families of modes, to be
followed by a more detailed presentation. The various families of
modes are illustrated in Fig.~\ref{fig:modes}.

The \emph{outer Eddington modes} (see Subsec.~\ref{subsec:Outer-Eddington-mode})
are defined with respect to the Eddington coordinates on the BH \emph{exterior}, and consist of two sets of modes: the
outgoing \emph{up }modes $f_{\omega lm}^{\text{up}}$ which emerge
as free waves from $H_{\text{past}}$, and the ingoing\emph{ in }modes
$f_{\omega lm}^{\text{in}}$ which emerge as free waves from PNI.

The \emph{inner Eddington modes} (see Sec.~\ref{subsec:Inner-Eddington-mode})
are similarly defined with respect to the Eddington coordinates on
the BH \emph{interior}, and consist of two sets of modes: the outgoing
\emph{left} modes $f_{\omega lm}^{L}$ which emerge as free waves
from $H_{L}$, and the ingoing \emph{right }modes $f_{\omega lm}^{R}$
which emerge as free waves from $H_{R}$.

The \emph{Unruh modes} (see Sec.~\ref{subsec:Unruh-modes}) are
defined on the \textit{entire} combined interior and exterior domain
(i.e., the entire red frame in Fig.~\ref{Fig:Kerr_Penrose}), and
consist of two sets of modes: the outgoing\emph{ up} modes $g_{\hat{\omega}\hat{l}\hat{m}}^{\text{up}}$
(the indices $\hat{\omega}\hat{l}\hat{m}$ are to be introduced later
on) which emerge from $H_{\text{past}}\cup H_{L}$ as free waves with
respect to the affine parameter there, the Kruskal coordinate $U$;
and the ingoing \emph{in} modes $g_{\omega lm}^{\text{in}}$ which
emerge from PNI as free waves with respect to the affine parameter
there, the Eddington coordinate $v$ (see the end of Subsec.~\ref{subsec:metric_and_coordinates}). 

The Unruh modes are of crucial physical importance for a quantum scalar
field evolving on a BH background. See, in particular, the definition
of the Unruh vacuum in Subsec.~\ref{subsec:Unruh state}. Our motivation
for considering the Eddington modes is of a different
kind: Remarkably, both outer and inner Eddington modes
are entirely decomposable in terms of angular, radial and temporal
variables, and have a general form analogous to Eq.~\eqref{eq:decomKerr}.
Owing to this decomposition, the Eddington modes are easily
handled numerically -- which makes them convenient as a mathematical
tool for the decomposition of the Unruh modes -- whereas the Unruh
modes do not generally admit such decomposition.

Throughout the rest of the paper, whenever we have both $\omega_{+}$
and $\omega$ appearing in the same equation, they should be understood
to be related via the standard relation, $\omega=\omega_{+}+m\Omega_{+}$,
or vice versa, via $\omega_{+}=\omega-m\Omega_{+}$. In all of these
cases, there will always be a well-defined $m$ value, enabling this
transformation between $\omega$ and $\omega_{+}$.

\subsection{Outer Eddington modes \label{subsec:Outer-Eddington-mode}}

For the sake of separation of the wave equation ~\eqref{eq:Teukolsky}
in Kerr, we considered a particular decomposition of the field into
modes (Eq.~\eqref{eq:decomKerr}), which was given in terms of the
spheroidal harmonics $Z_{lm}^{\omega}\left(\theta,\varphi\right)=\frac{1}{\sqrt{2\pi}}S_{lm}^{\omega}\left(\theta\right)e^{im\varphi}$,
the temporal dependence $e^{-i\omega t}$ and the radial function
$\psi_{\omega lm}\left(r\right)$. This decomposition provides the
basis for the definition of the \emph{Eddington} modes,
which we shall generally denote by $f_{\omega lm}$. In this subsection
we introduce the \emph{outer} Eddington modes, $f_{\omega lm}^{\text{in}}$
and $f_{\w lm}^{\text{up}}$ defined exclusively on the BH
exterior.

Based on the asymptotic behavior of the effective potential in the
BH exterior (as given in Eq.~\eqref{eq:asymPot_ext}), we define two
spanning sets of solutions to the radial equation ~\eqref{eq:radKerr},
$\psi_{\omega lm}^{\text{in}}\left(r\right)$ and $\psi_{\w lm}^{\text{up}}\left(r\right)$,
uniquely determined by their boundary conditions:
\begin{equation}
\psi_{\omega lm}^{\text{in}}\left(r\right)\simeq\begin{cases}
\tau_{\w lm}^{\text{in}}e^{-i\omega_{+}r_{*}}\,, & r_{*}\to-\infty\\
e^{-i\omega r_{*}}+\rho_{\omega lm}^{\text{in}}e^{i\omega r_{*}}\,, & r_{*}\to\infty
\end{cases}\,,\label{eq:psiIN_asym}
\end{equation}
\begin{equation}
\psi_{\w lm}^{\text{up}}\left(r\right)\simeq\begin{cases}
e^{i\omega_{+}r_{*}}+\rho_{\w lm}^{\text{up}}e^{-i\omega_{+}r_{*}}\,, & r_{*}\to-\infty\\
\tau_{\omega lm}^{\text{up}}e^{i\omega r_{*}}\,, & r_{*}\to\infty
\end{cases}\,,\label{eq:psiUP_asym}
\end{equation}
where $\tau_{\w lm}^{\Lambda}$ and $\rho_{\w lm}^{\Lambda}$, with $\Lambda$ denoting
either ``in" or ``up", are the transmission and reflection
coefficients, respectively. Determination of these coefficients thus
requires numerically solving the one-dimensional scattering of the
$\psi_{\w lm}^{\Lambda}$ mode off the effective potential ~\eqref{eq:KerrPot}. 
We use the symbol ``$\simeq$" to denote asymptotic equivalence.

When the signs of $\omega$ and $m$ are flipped simultaneously, so
does the sign of $\omega_{+}$. This means that the emerging free
waves that initiate the \emph{in} and \emph{up} modes ($e^{-i\omega r_{*}}$
or $e^{i\omega_{+}r_{*}}$ respectively) simply undergo complex conjugation
under this transformation. We also recall from the previous section
that the radial equation is real, and is invariant under such a simultaneous
sign flip of $\omega$ and $m$. Therefore, the following symmetry
relations are satisfied:
\begin{align}
\psi_{\left(-\omega\right)l\left(-m\right)}^{\Lambda} & =\psi_{\omega lm}^{\Lambda*}\,
\label{eq:psi_in_inv}
\end{align}
for the radial solutions  and
\begin{align}
\rho_{\left(-\omega\right)l\left(-m\right)}^{\Lambda} & =\rho_{\omega lm}^{\Lambda*}\,\,\,\,,\,\,\,\,\,\tau_{\left(-\w\right)l\left(-m\right)}^{\Lambda}=\tau_{\w lm}^{\Lambda*}\,\label{eq:rho_inv}
\end{align}
for the corresponding reflection and transmission coefficients,
where $\Lambda$ stands here for either ``in" or ``up".

Having introduced the radial functions $\psi_{\omega lm}^{\text{in}}$
and $\psi_{\w lm}^{\text{up}}$ and specified their boundary
conditions in Eqs.~\eqref{eq:psiIN_asym} and \eqref{eq:psiUP_asym},
the complete orthonormal family of outer Eddington modes
are defined in accordance with Eq.~\eqref{eq:decomKerr}:
\begin{align}
f_{\omega lm}^{\text{in}}\left(x\right) & \equiv\frac{1}{\sqrt{4\pi\left|\omega\right|\left(r^{2}+a^{2}\right)}}Z_{lm}^{\omega}\left(\theta,\varphi\right)e^{-i\omega t}\psi_{\omega lm}^{\text{in}}\left(r\right),\label{eq:EddModesIU}\\
f_{\w lm}^{\text{up}}\left(x\right) & \equiv\frac{1}{\sqrt{4\pi\left|\omega_{+}\right|\left(r^{2}+a^{2}\right)}}Z_{lm}^{\omega}\left(\theta,\varphi\right)e^{-i\omega t}\psi_{\w lm}^{\text{up}}\left(r\right),\nonumber 
\end{align}
satisfying the boundary conditions (as emerge from Eqs.~\eqref{eq:psiIN_asym}
and \eqref{eq:psiUP_asym}) \footnote{In the RHS of Eq.~\eqref{eq:fUP_past_asym}, being non-zero only at
$H_{\text{past}}$, we could have replaced $r$ in the prefactor by
$r_{+}$. However, in Eq.~\eqref{eq:fIN_past_asym} this is not the
case, as well as in similar equations that follow, being non-zero
both at $r=r_{+}$ and at infinity. We thus choose, for the sake of
uniformity, to keep $r$ (rather than $r_{+}$) in the prefactor in
the RHS of Eq.~\eqref{eq:fUP_past_asym}, as well as in all similar
instances that follow.}:
\begin{equation}
f_{\omega lm}^{\text{in}}\left(x\right)\simeq\frac{1}{\sqrt{8\pi^{2}\left|\omega\right|\left(r^{2}+a^{2}\right)}}S_{lm}^{\omega}\left(\theta\right)\begin{cases}
e^{-i\omega v}e^{im\varphi}, & \text{PNI}\\
0, & H_{\text{past}}
\end{cases}\,,\label{eq:fIN_past_asym}
\end{equation}
\begin{equation}
f_{\w lm}^{\text{up}}\left(x\right)\simeq\frac{1}{\sqrt{8\pi^{2}\left|\omega_{+}\right|\left(r^{2}+a^{2}\right)}}S_{lm}^{\omega}\left(\theta\right)\begin{cases}
0, & \text{PNI}\\
e^{-i\omega_{+}u_{\text{ext}}}e^{im\varphi_{+}}, & H_{\text{past}}
\end{cases}\,.\label{eq:fUP_past_asym}
\end{equation}
 Eq.~\eqref{eq:fUP_past_asym} makes use of the relation $e^{-i\omega t}e^{im\varphi}=e^{-i\omega_{+}t}e^{im\varphi_{+}}$,
also useful later in the paper. Note that, in accordance with the
discussion towards the end of Subsec.~\ref{subsec:metric_and_coordinates},
the past asymptotic forms given above (as well as the future asymptotic
forms given below in Eqs.~\eqref{eq:fIN_future_asym}-\eqref{eq:fUP_future_asym})
are always expressed in terms of the three regular coordinates on
each of the asymptotic null surfaces. 


For future use, we also find it beneficial to write Eqs.~\eqref{eq:fIN_past_asym}
and \eqref{eq:fUP_past_asym} in a slightly different manner by absorbing
the $\varphi$- and $\varphi_{+}$-dependent factors into the angular functions:
\begin{equation}
f_{\omega lm}^{\text{in}}\left(x\right)\simeq\frac{1}{\sqrt{4\pi\left|\omega\right|\left(r^{2}+a^{2}\right)}}Z_{lm}^{\omega}\left(\theta,\varphi\right)\begin{cases}
e^{-i\omega v}, & \text{PNI}\\
0, & H_{\text{past}}
\end{cases}\,,\label{eq:fIN_past_asymZ}
\end{equation}
\begin{equation}
f_{\w lm}^{\text{up}}\left(x\right)\simeq\frac{1}{\sqrt{4\pi\left|\omega_{+}\right|\left(r^{2}+a^{2}\right)}}Z_{lm}^{\omega}\left(\theta,\varphi_{+}\right)\begin{cases}
0, & \text{PNI}\\
e^{-i\omega_{+}u_{\text{ext}}}, & H_{\text{past}}
\end{cases}\,,\label{eq:fUP_past_asymZ}
\end{equation}
where 
\begin{equation}
Z_{lm}^{\omega}\left(\theta,\varphi_{+}\right)\equiv\frac{1}{\sqrt{2\pi}}S_{lm}^{\omega}\left(\theta\right)e^{im\varphi_{+}}\,.\label{eq:Z(theta,phi+)}
\end{equation}
 Clearly, for any fixed $\omega\in\mathbb{R}$, $Z_{lm}^{\omega}\left(\theta,\varphi_{+}\right)$
is a complete family of functions on the 2-sphere, orthonormal in
the sense that
\begin{align}
\int_{0}^{2\pi}\text{d}\varphi_{+}\int_{0}^{\pi}\text{d}\theta\sin\theta\,Z_{lm}^{\omega*}\left(\theta,\varphi_{+}\right)Z_{l'm'}^{\omega}\left(\theta,\varphi_{+}\right) & =\delta_{ll'}\delta_{mm'}\,.\label{eq:ang_orth_phi+}
\end{align}

By inspecting Eq.~\eqref{eq:EddModesIU} along with Eqs.~\eqref{eq:psi_in_inv}
and \eqref{eq:  Ang_inv} we conclude that
the exterior Eddington modes are invariant (modulo a sign introduced by the angular functions) under simultaneously
flipping the signs of $\omega$ and $m$ along with complex conjugation.
That is,
\begin{align}
f_{\left(-\omega\right)l\left(-m\right)}^{\Lambda} & =(-1)^mf_{\omega lm}^{\Lambda*},\label{eq:ExtEddModesInvariance}
\end{align}
where $\Lambda$ stands for either ``in" or ``up".

For future use, we  also provide the asymptotic behavior of the
\emph{in} and \emph{up} Eddington modes on the future
null hypersurfaces:
\begin{align}
f_{\omega lm}^{\text{in}}\left(x\right)\simeq\frac{1}{\sqrt{8\pi^{2}\left|\omega\right|\left(r^{2}+a^{2}\right)}}S_{lm}^{\omega}\left(\theta\right)\begin{cases}
\rho_{\omega lm}^{\text{in}}e^{-i\omega u_{\text{ext}}}e^{im\varphi}, & \text{FNI}\\
\tau_{\w lm}^{\text{in}}e^{-i\omega_{+}v}e^{im\varphi_{+}}, & H_{R}
\end{cases}\,,\label{eq:fIN_future_asym}
\end{align}
\begin{align}
f_{\w lm}^{\text{up}}\left(x\right)\simeq\frac{1}{\sqrt{8\pi^{2}\left|\omega_{+}\right|\left(r^{2}+a^{2}\right)}}S_{lm}^{\omega}\left(\theta\right)\begin{cases}
\tau_{\omega lm}^{\text{up}}e^{-i\omega u_{\text{ext}}}e^{im\varphi}, & \text{FNI}\\
\rho_{\w lm}^{\text{up}}e^{-i\omega_{+}v}e^{im\varphi_{+}}, & H_{R}
\end{cases} & \,.\label{eq:fUP_future_asym}
\end{align}
The \emph{in} and \emph{up} outer Eddington modes are illustrated on the top row of Fig.~\ref{fig:modes}.
The \emph{in} mode may be interpreted as a (properly normalized) monochromatic
spherical wave propagating \emph{inwards} from infinity (PNI) and
being partially reflected back to infinity (FNI) and partially transmitted
across the horizon ($H_{R}$), with the relative coefficients of transmission
and reflection being respectively $\tau_{\w lm}^{\text{in}}$ and $\rho_{\omega lm}^{\text{in}}$.
Similarly, the \emph{up} mode may be interpreted as a monochromatic
spherical wave propagating \emph{upwards} from the past horizon ($H_{\text{past}}$)
and being partially reflected back to the future horizon ($H_{R}$)
and partially transmitted to infinity (FNI), with the relative coefficients
of transmission and reflection being respectively $\tau_{\omega lm}^{\text{up}}$
and $\rho_{\w lm}^{\text{up}}$.

\subsection{Inner Eddington modes\label{subsec:Inner-Eddington-mode}}

In a similar manner to the definition of the \emph{in} and \emph{up}
modes outside the BH, we may additionally define two sets of Eddington
modes confined to the BH interior. Note that, in this spacetime region,
the $r_{*}$ coordinate serves as a \textit{temporal} coordinate whereas
$t$ has a \textit{spatial} role. This means that a \textit{single}
initial condition is required for the radial equation ~\eqref{eq:radKerr},
which we simply take as a free wave at $r_{*}\to-\infty$ (corresponding
to $r\to r_{+}$): 
\begin{equation}
\psi_{\w lm}^{\text{int}}\simeq e^{-i\omega_{+}r_{*}},\,\,\,\,\,r\to r_{+}\,.\label{eq:psi_nearEH_asym}
\end{equation}

With this radial function we now define the \emph{right} ($R$) and \emph{left} ($L$) sets of orthonormal modes: \footnote{\label{fn:omega_unlike_RN}Note that in the treatment of the analogous
RN case in Ref.~\cite{Group:2018}, the \emph{right} and \emph{left} modes are defined with an essentially different temporal dependence
(see Eq.~(2.16) therein): the \emph{right} mode is decomposed with
respect to $e^{-i\omega t}$ while the \emph{left} mode is decomposed
with respect to $e^{i\omega t}$, and both share the same radial function.
This was possible since, in RN, the wave equation is invariant under
$\omega\mapsto-\omega$. In Kerr, however, the latter symmetry does
not apply, hence we stick with the canonical decomposition of Eq.~\eqref{eq:decomKerr}. This difference also leads to some differences
between several equations below and their counterparts in Ref.~\cite{Group:2018}.}

{\small{}
\begin{align}
f_{\w lm}^{R}\left(x\right) & \equiv\frac{1}{\sqrt{4\pi\left|\omega_{+}\right|\left(r^{2}+a^{2}\right)}}Z_{lm}^{\omega}\left(\theta,\varphi\right)e^{-i\omega t}\psi_{\w lm}^{\text{int}}\left(r\right),\label{eq:EddModesRL}\\
f_{\w lm}^{L}\left(x\right) & \equiv\frac{1}{\sqrt{4\pi\left|\omega_{+}\right|\left(r^{2}+a^{2}\right)}}Z_{lm}^{\omega}\left(\theta,\varphi\right)e^{-i\omega t}\psi_{\w lm}^{\text{int}*}\left(r\right).\nonumber 
\end{align}
}These modes admit the following asymptotic forms at the right and
left horizons:

{\small{}
\begin{equation}
f_{\w lm}^{R}\left(x\right)\simeq\frac{1}{\sqrt{8\pi^{2}\left|\omega_{+}\right|\left(r^{2}+a^{2}\right)}}S_{lm}^{\omega}\left(\theta\right)e^{im\varphi_{+}}\begin{cases}
0, & H_{L}\\
e^{-i\omega_{+}v}, & H_{R}
\end{cases}\,,\label{eq:fR_asym}
\end{equation}
\begin{equation}
f_{\w lm}^{L}\left(x\right)\simeq\frac{1}{\sqrt{8\pi^{2}\left|\omega_{+}\right|\left(r^{2}+a^{2}\right)}}S_{lm}^{\omega}\left(\theta\right)e^{im\varphi_{+}}\begin{cases}
e^{i\omega_{+}u_{\text{int}}}, & H_{L}\\
0, & H_{R}
\end{cases}\,.\label{eq:fL_asym}
\end{equation}
}{\small\par}

As for the exterior Eddington modes, we shall find it
beneficial when constructing the HTPF to present a slightly different
form for Eqs.~\eqref{eq:fR_asym}-\eqref{eq:fL_asym} by absorbing
the $\varphi$-and $\varphi_{+}$-dependent factors into angular functions:
\begin{equation}
f_{\w lm}^{R}\left(x\right)\simeq\frac{1}{\sqrt{4\pi\left|\omega_{+}\right|\left(r^{2}+a^{2}\right)}}Z_{lm}^{\omega}\left(\theta,\varphi_{+}\right)\begin{cases}
0, & H_{L}\\
e^{-i\omega_{+}v}, & H_{R}
\end{cases}\,,\label{eq:fR_asymZ}
\end{equation}
\begin{equation}
f_{\w lm}^{L}\left(x\right)\simeq\frac{1}{\sqrt{4\pi\left|\omega_{+}\right|\left(r^{2}+a^{2}\right)}}Z_{lm}^{\omega}\left(\theta,\varphi_{+}\right)\begin{cases}
e^{i\omega_{+}u_{\text{int}}}, & H_{L}\\
0, & H_{R}
\end{cases}\,.\label{eq:fL_asymZ}
\end{equation}

Analogously to the symmetry in Eq.~\eqref{eq:ExtEddModesInvariance}
satisfied by the exterior Eddington modes, the interior
Eddington modes satisfy
\begin{equation}
f_{\left(-\w\right)l\left(-m\right)}^{\Lambda}=(-1)^mf_{\w lm}^{\Lambda*}\,,\label{eq:IntEddModesInvariance}
\end{equation}
where $\Lambda$ stands here for either ``\emph{R}" or ``\emph{L}".
The \emph{right} and \emph{left} inner Eddington modes are illustrated on the middle row of Fig.~\ref{fig:modes}.

We have readily settled the definition of the inner Eddington
family of modes. However, it is interesting to also inspect the form
of the radial function $\psi_{\w lm}^{\text{int}}\left(r\right)$
at $r_{*}\to\infty$ (i.e., on approaching the IH). In correspondence
with the asymptotic behavior of the effective potential \eqref{eq:asymPot_int},
the solution of Eq.~\eqref{eq:radKerr} admits the free asymptotic
form:
\begin{equation}
\psi_{\w lm}^{\text{int}}\simeq A_{\w lm}e^{i\omega_{-}r_{*}}+B_{\w lm}e^{-i\omega_{-}r_{*}},\,\,\,\,\,\,r\to r_{-},\label{eq:psi_nearIH_asym}
\end{equation}
where $A_{\w lm}$ and $B_{\w lm}$ are constant complex
coefficients.

\subsection{Unruh modes\label{subsec:Unruh-modes}}

The \emph{Unruh modes} are the basic modes for the field expansion
involved in the definition of the Unruh quantum state (see Subsec.~\ref{subsec:Unruh state}).
There are two distinct sets of Unruh modes and both inhabit the \textit{entire}
union of the BH interior and exterior (namely, the red
frame in Fig.~\ref{Fig:Kerr_Penrose}). We shall occasionally refer
to this domain as the ``united domain''. This domain has two null
boundaries in its past: The one on the right is PNI. The other boundary,
on the left, is located at $r=r_{+}$; it is the union of $H_{\text{past}}$
and $H_{L}$. Recall that, as discussed at the end of Subsec.~\ref{subsec:metric_and_coordinates},
the corresponding affine parameters are Eddington $v$ along the past-right
boundary (PNI), and Kruskal $U$ along the past-left boundary ($H_{\text{past}}\cup H_{L}$). We also note that both these affine parameters span the
entire range $\left(-\infty,\infty\right)$ along their respective
past null boundaries (namely $v$ along PNI and $U$ along $H_{\text{past}}\cup H_{L}$). 

The two sets of Unruh modes naturally emerge from these two past null
boundaries, with positive frequencies in each set defined with respect
to the affine parameter along the corresponding boundary. This is
unlike the Eddington modes (introduced above), which are
always defined asymptotically with respect to the corresponding Eddington
coordinates $v$ and $u$ \footnote{Since the Unruh modes are introduced here directly for the construction
of a quantum state, we only need to define them with positive frequencies.
The Eddington modes, however, were introduced to be utilized
as a mathematical tool for decomposition. Thus, the latter were defined
in Subsecs. \ref{subsec:Outer-Eddington-mode}-\ref{subsec:Inner-Eddington-mode}
for negative frequencies as well (hence the absolute value in their
normalization constant).\label{fn:positive_w_Unruh_modes}}. We now introduce the two sets of Unruh modes, \emph{in} and \emph{up},
as outlined above.

\subsubsection{The \emph{in} Unruh modes\label{subsec:The-in-Unruh}}

The \emph{in} Unruh mode $g_{\omega lm}^{\text{in}}$ originates at
PNI as a free wave with respect to the affine parameter there, the
Eddington $v$ coordinate (i.e., $\propto e^{-i\omega v}$) and vanishes
on both $H_{\text{past}}$ and $H_{L}$. That is, it is endowed with
the following past boundary conditions: 
\begin{equation}
g_{\omega lm}^{\text{in}}\left(x\right)\simeq\frac{1}{\sqrt{4\pi\omega\left(r^{2}+a^{2}\right)}}Z_{lm}^{\omega}\left(\theta,\varphi\right)\begin{cases}
e^{-i\omega v}, & \text{PNI}\\
0, & H_{\text{past}}\cup H_{L}
\end{cases}\,.\label{eq:gIN_asym}
\end{equation}
Recall that here $\omega$ attains only positive values (see footnote
\ref{fn:positive_w_Unruh_modes}).
The \emph{in} Unruh modes are illustrated at the bottom right diagram of Fig.~\ref{fig:modes}.

Evidently, these boundary conditions are regular. Since the d'Alembertian
operator in Eq.~\eqref{eq:KG} is regular as well, the regularity
of the $g_{\omega lm}^{\text{in}}$ modes is guaranteed throughout
the (interior of the) united domain. 

We now restrict our attention to the BH exterior. Notably, the \emph{in}
Unruh mode $g_{\omega lm}^{\text{in}}$, when constrained to the BH
exterior, has the same boundary conditions (on the past asymptotic
null surfaces PNI and $H_{\text{past}}$) as the \emph{in} Eddington
mode $f_{\omega lm}^{\text{in}}$: compare Eqs.~\eqref{eq:fIN_past_asymZ}
and \eqref{eq:gIN_asym}. That is,
\begin{align*}
\left.g_{\omega lm}^{\text{in}}\right|_{\text{PNI}} & =\left.f_{\omega lm}^{\text{in}}\right|_{\text{PNI}}\\
\left.g_{\omega lm}^{\text{in}}\right|_{H_{\text{past}}} & =0=\left.f_{\omega lm}^{\text{in}}\right|_{H_{\text{past}}}\,.
\end{align*}
Since $g_{\w lm}^{\text{in}}$ and $f_{\w lm}^{\text{in}}$
satisfy the same wave equation \eqref{eq:Teukolsky}, it follows that
these two quantities are identical not only on the initial null hypersurfaces
$H_{\text{past}}$ and PNI but at every
spacetime point in the BH exterior:

\begin{equation}
g_{\omega lm}^{\text{in}}\left(x\right)=f_{\omega lm}^{\text{in}}\left(x\right)\,,\,\,\,\,\,\,\,\,\,\,r\geq r_{+}\,.\label{eq:gIN-fIN}
\end{equation}

In order to find the behavior of $g_{\omega lm}^{\text{in}}$ in the
BH interior, we carry it using Eq.~\eqref{eq:gIN-fIN}
to $H_{R}$, where it fulfills
\begin{equation}
\left.g_{\omega lm}^{\text{in}}\right|_{H_{R}}=\left.f_{\omega lm}^{\text{in}}\right|_{H_{R}}=\sqrt{\frac{\left|\omega_{+}\right|}{\omega}}\tau_{\w lm}^{\text{in}}\left.f_{\w lm}^{R}\right|_{H_{R}}\label{eq:gIN-fR-HR}
\end{equation}
(for the last equality, compare Eq.~\eqref{eq:fIN_future_asym} with
Eq.~\eqref{eq:fR_asym}). Again, since $g_{\omega lm}^{\text{in}}$
and $\sqrt{\frac{\left|\omega_{+}\right|}{\omega}}\tau_{\w lm}^{\text{in}}f_{\omega lm}^{R}$
coincide on $H_{R}$ and on $H_{L}$ as well (since they both vanish
on the latter hypersurface, see Eq.~\eqref{eq:gIN_asym} and Eq.~\eqref{eq:fR_asym}),
these solutions are identical everywhere in the BH interior:
\begin{equation}
g_{\omega lm}^{\text{in}}\left(x\right)=\sqrt{\frac{\left|\omega_{+}\right|}{\omega}}\tau_{\w lm}^{\text{in}}f_{\w lm}^{R}\left(x\right)\,,\,\,\,\,\,\,\,\,\,r_{-}\leq r\leq r_{+}\,.\label{eq:gIN-fR}
\end{equation}

Equations \eqref{eq:gIN-fIN} and \eqref{eq:gIN-fR} demonstrate a
useful property of the \emph{in} Unruh modes, namely that we may match
each \emph{in} Unruh mode, at any given neighborhood, with a particular
Eddington mode: with an \emph{in} mode at $r\geq r_{+}$
and with a \emph{right} mode (up to a specified multiplicative constant)
at $r_{-}\leq r\leq r_{+}$. In particular, this means that, throughout
the united domain, $g_{\omega lm}^{\text{in}}$ decomposes into radial,
angular and temporal terms, as may also be anticipated from its initial
conditions given in Eq.~\eqref{eq:gIN_asym} (because $e^{-i\omega v}$
decomposes naturally into a temporal factor $e^{-i\omega t}$ times
a function of $r$, and this separable form of the $t$-dependence
is preserved as $g_{\omega lm}^{\text{in}}$ evolves according to
the $t$-independent wave equation). This situation changes when considering
the \emph{up} Unruh modes, as we shall do next. 

\subsubsection{The \emph{up} Unruh modes\label{subsec:The-up-Unruh}}

The \emph{up} Unruh modes are solutions to Eq.~\eqref{eq:Teukolsky}
emerging from $H_{\text{past}}\cup H_{L}$ with positive frequency,
which we denote by $\hat{\omega}$ (to distinguish it from the Killing
frequency, $\omega$). That is, the \emph{up} Unruh modes originate
from (the ingoing arms of) $r=r_{+}$ as $\propto e^{-i\hat{\omega}U}$,
with the Kruskal $U$ as defined in Eqs.~\eqref{eq:extKrusCoor} and
\eqref{eq:intKrusCoor}. 

The desired orthonormal set of \emph{up} modes is conveniently defined
by specifying the initial value of each of the modes at the initial
null hypersurface $H_{\text{past}}\cup H_{L}$ -- as a function
of the three regular coordinates $U,\theta,\varphi_{+}$ which span
it (see the end of Subsec.~\ref{subsec:metric_and_coordinates}).
This initial-value setup for the \emph{up} modes is complemented by
requiring these modes to vanish at PNI, see Eq. \ref{eq:UnruhUP} below. In order for the
\emph{up} modes to provide (when combined with the \emph{in} modes
defined in the previous subsection) a complete KG-orthonormal set
of solutions to the wave equation, the aforementioned 3-parameter
set of \emph{up}-modes initial functions has to be in itself a complete
KG-orthonormal set of functions of $U$, $\theta$ and $\varphi_{+}$ on $H_{\text{past}}\cup H_{L}$.
We already chose the modes' initial $U$-dependence at $H_{\text{past}}\cup H_{L}$
to be $\propto e^{-i\hat{\omega}U}$, so all that is left is to specify
a complete orthonormal set of functions of the remaining coordinates
$\theta$ and $\varphi_{+}$ on the 2-sphere. This set can be chosen quite
arbitrarily (and in principle it could also depend on $\hat{\omega}$).
It should depend on two discrete parameters (reflecting the dimensionality
of the 2-sphere), which we here schematically denote by $\hat{l},\hat{m}$;
hence we may generally denote such a set of ``initial'' angular
functions as $\hat{Z}_{\hat{l}\hat{m}}^{\hat{\omega}}(\theta,\varphi_{+})$ \footnote{\label{fn: angular_independence}As mentioned, the Unruh modes introduced
here are to be utilized in Subsec.~\ref{subsec:Unruh state} for construction
of the Unruh quantum state. Generally speaking, what determines a
quantum state is the \emph{frequency} (and thereby the implied choice
of \emph{positive-frequency modes}), which was here chosen to be the
parameter $\hat{\omega}$ appearing in $e^{-i\hat{\omega}U}$. Then,
the remaining choice of angular functions $\hat{Z}_{\hat{l}\hat{m}}^{\hat{\omega}}(\theta,\varphi_{+})$
for the \emph{up} Unruh modes' initial conditions does not affect
the resultant quantum state. In particular, the final mode-sum structure
of the Unruh-state HTPF, as appears in Eq.~\eqref{eq:TPFoutside}
or \eqref{eq:G_U_int}, does not depend at all on the choice of $\hat{Z}_{\hat{l}\hat{m}}^{\hat{\omega}}$ -- we show this explicitly in Subsecs.~\ref{subsec:Invariance-to-Zwlm-choice} and \ref{subsec:Invariance-to-Zwlm-choice-int}. }.
We choose it to be orthonormal in the usual sense, 
\begin{equation}
\int_{0}^{2\pi}\text{d}\varphi_{+}\int_{0}^{\pi}\text{d}\theta\sin\theta\,\,\hat{Z}_{\hat{l}\hat{m}}^{\hat{\omega}*}\left(\theta,\varphi_{+}\right)\hat{Z}_{\hat{l}'\hat{m}'}^{\hat{\omega}}\left(\theta,\varphi_{+}\right)=\delta_{\hat{l}\hat{l}'}\delta_{\hat{m}\hat{m}'}\,.\label{eq:Zhat_ort}
\end{equation}
 The set of \emph{up} modes is then generally defined via its initial conditions
at $H_{\text{past}}\cup H_{L}$ and PNI by

\begin{equation}
g_{\hat{\omega}\hat{l}\hat{m}}^{\text{up}}\left(x\right)\simeq\frac{1}{\sqrt{4\pi\hat{\omega}\left(r^{2}+a^{2}\right)}}\hat{Z}_{\hat{l}\hat{m}}^{\hat{\omega}}\left(\theta,\varphi_{+}\right)\begin{cases}
0, & \text{PNI}\\
e^{-i\hat{\omega}U}, & H_{\text{past}}\cup H_{L}
\end{cases}\,.\label{eq:UnruhUP}
\end{equation}
Recall that $\hat{\omega}$ attains only positive values (see footnote
\ref{fn:positive_w_Unruh_modes}).
The \emph{up} Unruh modes are illustrated at the bottom left diagram of Fig.~\ref{fig:modes}.

We shall choose our arbitrary angular functions $\hat{Z}_{\hat{l}\hat{m}}^{\hat{\omega}}(\theta,\varphi_{+})$
to be the simplest complete orthonormal set of angular functions on
the two-sphere -- namely the conventional \emph{spherical harmonics},
\[
Y_{\hat{l}\hat{m}}\left(\theta,\varphi_{+}\right)=
\sqrt{
\frac{(2\hat{l}+1)}{4\pi}
\frac{(\hat{l}-\hat{m})!}{(\hat{l}+\hat{m})!}
}
P_{\hat{l}}^{\hat{m}}\left(\cos\theta\right)e^{i\hat{m}\varphi_{+}}\,,
\]
where $P_{\hat{l}}^{\hat{m}}$ are the associated Legendre polynomials.

Thus, we define $g_{\hat{\omega}\hat{l}\hat{m}}^{\text{up}}$ as
a solution to Eq.~\eqref{eq:Teukolsky} with the initial conditions
\begin{equation}
g_{\hat{\omega}\hat{l}\hat{m}}^{\text{up}}\left(x\right)\simeq\frac{1}{\sqrt{4\pi\hat{\omega}\left(r^{2}+a^{2}\right)}}Y_{\hat{l}\hat{m}}\left(\theta,\varphi_{+}\right)\begin{cases}
0, & \text{PNI}\\
e^{-i\hat{\omega}U}, & H_{\text{past}}\cup H_{L}
\end{cases}\,.\label{eq:gUP_asym}
\end{equation}

Conveniently, any given \emph{up} mode may be written as a sum of
two other solutions to Eq.~\eqref{eq:Teukolsky}, denoted $g_{\hat{\omega}\hat{l}\hat{m}}^{\text{past}}$
and $g_{\hat{\omega}\hat{l}\hat{m}}^{L}$, which are endowed with
the following initial conditions:
\begin{equation}
g_{\hat{\omega}\hat{l}\hat{m}}^{L}\left(x\right)\simeq\frac{1}{\sqrt{4\pi\hat{\omega}\left(r^{2}+a^{2}\right)}}Y_{\hat{l}\hat{m}}\left(\theta,\varphi_{+}\right)\begin{cases}
0, & \text{PNI}\\
0, & H_{\text{past}}\\
e^{-i\hat{\omega}U}, & H_{L}
\end{cases}\label{eq:gL_asym}
\end{equation}
\begin{equation}
g_{\hat{\omega}\hat{l}\hat{m}}^{\text{past}}\left(x\right)\simeq\frac{1}{\sqrt{4\pi\hat{\omega}\left(r^{2}+a^{2}\right)}}Y_{\hat{l}\hat{m}}\left(\theta,\varphi_{+}\right)\begin{cases}
0, & \text{PNI}\\
e^{-i\hat{\omega}U}, & H_{\text{past}}\\
0, & H_{L}
\end{cases}\,.\label{eq:gPast_asym}
\end{equation}
That is, while the initial support of $g_{\hat{\omega}\hat{l}\hat{m}}^{\text{up}}\left(x\right)$
is on the entire null surface $H_{\text{past}}\cup H_{L}$, the function
$g_{\hat{\omega}\hat{l}\hat{m}}^{L}\left(x\right)$ has its initial
support on $H_{L}$ alone whereas $g_{\hat{\omega}\hat{l}\hat{m}}^{\text{past}}\left(x\right)$
is initially supported on $H_{\text{past}}$ only.

One might be slightly confused about our choice of spherical harmonics
as the angular functions here, because it is customary to use the
\emph{spheroidal} harmonics for a Kerr BH. But the only reason for
this common use of spheroidal harmonics in Kerr is to achieve angular
separability of the wave equation, as in Eq.~\eqref{eq:decomKerr}.
Recall, however, that
since the spheroidal harmonics explicitly depend on $\omega$,
this angular separability can only be achieved
when the temporal dependence is
precisely of the form $e^{-i\omega t}$ (multiplying some function of $r$, $\theta$ and $\varphi$). Quite unluckily, an \emph{up} Unruh
mode admits a more intricate temporal dependence, which does not fit
any single Killing frequency $\omega$. Instead, each such mode is
a superposition of Eddington modes with potentially all
possible $\omega$ values; this may be seen already at the initial
hypersurface $H_{\text{past}}\cup H_{L}$, where an \emph{up} Unruh
mode is $\propto e^{-i\hat{\omega}U}$, recalling that $U$ cannot
be expressed as a sum of $t$-dependent and $r$-dependent pieces,
unlike the Eddington coordinates. Since angular separability cannot
be achieved anyway in this case, there is no advantage in using the
spheroidal harmonics for the \emph{up} Unruh modes -- and we choose
the much simpler (and frequency-independent) spherical harmonics instead.

We point out, however, that despite our choice of spherical harmonics
for the definition of the \emph{up} Unruh modes, our final mode-sum
expressions (in Eddington modes) for the HTPF  are actually given in terms of \emph{spheroidal}
harmonics (as one would naturally expect for Kerr) -- as may explicitly
be seen in Eqs.~\eqref{eq:TPFoutside} and \eqref{eq:G_U_int} (along
with Eqs.~\eqref{eq:EddModesIU} and \eqref{eq:EddModesRL}). In fact,
these final mode-sum expressions for the Unruh-state HTPF are entirely
independent of the choice of angular functions $\hat{Z}_{\hat{l}\hat{m}}^{\hat{\omega}}\left(\theta,\varphi_{+}\right)$
at this stage of constructing the \emph{up} Unruh modes, see footnote
\ref{fn: angular_independence} above. 

It is clear from the discussion above that, unlike what we did for
the \emph{in} Unruh modes (see Eqs.~\eqref{eq:gIN-fIN} and \eqref{eq:gIN-fR}),
it is not possible to match a specific \emph{up} Unruh mode with a
single Eddington mode (neither in the exterior nor in
the interior of the BH). This reflects our inability to reduce the
PDE \eqref{eq:Teukolsky} to an ODE (such as \eqref{eq:radKerr}) for the \emph{up} Unruh modes. Therefore, in order
to allow a convenient numerical implementation involving the solution
of ODEs rather than PDEs, we shall later Fourier-decompose the $g_{\hat{\omega}\hat{l}\hat{m}}^{\text{up}}\left(x\right)$
modes in terms of the (separable) Eddington modes.

\subsection{Wronskian relations\label{subsec:Wronskian-relations}}

The absence of a first derivative in the radial equation \eqref{eq:radKerr}
leads to $r_{*}$-independence of the Wronskian of any pair of solutions.
This Wronskian conservation yields well-known relations involving
the exterior reflection and transmission coefficients $\rho_{\w lm}$ and
$\tau_{\w lm}$ (defined via Eqs.~\eqref{eq:psiIN_asym} and \eqref{eq:psiUP_asym}),
as well as relations involving the interior near-IH coefficients $A_{\w lm}$
and $B_{\w lm}$ (defined via Eq.~\eqref{eq:psi_nearIH_asym}).

\paragraph*{Relations involving $\rho_{\w lm}$ and $\tau_{\w lm}.$}

Using the Wronskian conservation on pairs of solutions chosen from
$\psi_{\omega lm}^{\text{in}}$, $\psi_{\w lm}^{\text{up}}$,
$\psi_{\omega lm}^{\text{in}*}$ and $\psi_{\w lm}^{\text{up}*}$
(in particular, equating their Wronskian at $r_{*}\to-\infty$ with
their Wronskian at $r_{*}\to\infty$, using the asymptotic forms given
in Eqs.~\eqref{eq:psiIN_asym} and \eqref{eq:psiUP_asym}), yields
the following constraints on the reflection and transmission coefficients: 

\begin{align}
 & \left|\rho_{\omega lm}^{\text{in}}\right|^{2}+\frac{\omega_{+}}{\omega}\left|\tau_{\w lm}^{\text{in}}\right|^{2}=1\,,\label{eq:rho_tau_Wron}\\
 & \left|\rho_{\w lm}^{\text{up}}\right|^{2}+\frac{\omega}{\omega_{+}}\left|\tau_{\omega lm}^{\text{up}}\right|^{2}=1\,,\nonumber \\
 & \tau_{\omega lm}^{\text{up}}=\frac{\omega_{+}}{\omega}\tau_{\w lm}^{\text{in}}\,,\nonumber \\
 & \frac{\rho_{\w lm}^{\text{up}*}}{\rho_{\omega lm}^{\text{in}}}=-\frac{\tau_{\omega lm}^{\text{up}*}}{\tau_{\omega lm}^{\text{up}}}\,.\nonumber 
\end{align}
The last equation yields, in particular, $ \left|\rho_{\w lm}^{\text{up}}\right|= \left|\rho_{\w lm}^{\text{in}}\right|$.

Notably, from the first (second) constraint, modes with $\omega\, \omega_{+}<0$
have $\left|\rho_{\omega lm}^{\text{in}}\right|^{2}>1$ $\left(\left|\rho_{\w lm}^{\text{up}}\right|^{2}>1\right)$.
That is, the reflected \emph{in} (\emph{up}) wave has, at FNI ($H_{R}$),
an amplitude greater than it originally had at PNI ($H_{\text{past}}$).
This is the classical phenomenon of \emph{superradiance}~\cite{zel1971generation,starobinskii1973amplification}. \footnote{Note that in RN we have $\omega=\omega_{+}$, which leads to the simple
analogous relation $\left|\rho_{\omega l}\right|^{2}+\left|\tau_{\omega l}\right|^{2}=1$,
implying that there exist \emph{no} superradiant modes.} 

\paragraph*{Relations involving $A_{\w lm}$ and $B_{\w lm}$.}

Similarly, Wronskian conservation of the interior radial function
$\psi_{\w lm}^{\text{int}}$ (see Eq.~\eqref{eq:psi_nearEH_asym})
and its conjugate $\psi_{\w lm}^{\text{int}*}$ relates the
internal scattering coefficients $A_{\w lm}$ and $B_{\w lm}$
(defined through Eq.~\eqref{eq:psi_nearIH_asym}) as follows:
\begin{equation}
\left|B_{\w lm}\right|^{2}-\left|A_{\w lm}\right|^{2}=\frac{\omega_{+}}{\omega_{-}}\,.\label{eq:A=000026B_Wron}
\end{equation}

\section{Quantum states in a Kerr spacetime\label{sec:Quantum states}}

All topics outlined in the paper so far were basically purely classical. We shall
now promote our scalar field from a classical field $\Phi$ to a quantum
field operator $\hat{\Phi}$. We first provide a brief review of its
decomposition via annihilation and creation operators and then introduce
various quantum states in Kerr spacetime.

\subsection{Generic construction of quantum states}

Consider a space of generic positive-frequency mode solutions, $\Phi_{i}$,
with respect to some temporal coordinate (clearly, in curved spacetime,
this choice is not unique). These solutions fulfill the following
orthonormality relations with respect to the KG inner product:
\begin{equation}
\left\langle \Phi_{i},\Phi_{j}\right\rangle =\delta_{ij},\,\,\,\,\left\langle \Phi_{i}^{*},\Phi_{j}^{*}\right\rangle =-\delta_{ij},\,\,\,\,\left\langle \Phi_{i},\Phi_{j}^{*}\right\rangle =0\,,\label{eq:orthonRelations}
\end{equation}
so that the union of the set $\Phi_{i}$ (for all $i$) and the set $\Phi^*_j$ (for all $j$)   is a complete
family of orthonormal solutions to the KG equation \eqref{eq:KG}.
We may now expand the field in terms of this basis of solutions via
creation ($\hat{a}_{i}^{\dagger}$) and annihilation ($\hat{a}_{i}$)
operators as follows: 

\begin{align*}
\hat{\Phi}\left(x\right) & =\sum_{i}\left(\hat{a}_{i}\Phi_{i}\left(x\right)+\hat{a}_{i}^{\dagger}\Phi_{i}^{*}\left(x\right)\right)\,,
\end{align*}
where the following commutation relations are imposed:
\[
\left[
\hat{a}_{i},\hat{a}_{j}^{\dagger}
\right]
=\hbar\,\delta_{ij}\hat{1},\,\,
\Big[
\hat{a}_{i},\hat{a}_{j}
\Big]
=0,\,\,\left[\hat{a}_{i}^{\dagger},\hat{a}_{j}^{\dagger}\right]=0\,.
\]

The vacuum with respect to this family is a state $\left|0\right\rangle $
such that 
\[
\hat{a}_{i}\left|0\right\rangle =0,\ \text{for all}\ i\,.
\]

Acting on the vacuum state with the creation operators $\hat{a}_{i}^{\dagger}$
yields the one-particle states
\[
\left|1_{i}\right\rangle =\hbar^{-1/2}\hat{a}_{i}^{\dagger}\left|0\right\rangle \,,
\]
and from here one may construct the entire many-particle Fock space. 

\subsection{The Boulware and (lack of) Hartle-Hawking quantum states}

The outlined decomposition scheme is utilized in the construction
of quantum states for a scalar field. As a concrete example, consider
the field decomposition in the Kerr BH exterior in terms of the outer
Eddington modes (Eq.~\eqref{eq:EddModesIU})
\begin{align}
\hat{\Phi}\left(x\right) & =\sum_{l=0}^{\infty}\sum_{m=-l}^{l}\int_{0}^{\infty}\text{d}\omega\left(\hat{b}_{\omega lm}^{\text{in}}f_{\omega lm}^{\text{in}}\left(x\right)+\hat{b}_{\omega lm}^{\text{in}\dagger}f_{\omega lm}^{\text{in}*}\left(x\right)\right)\nonumber \\
 & +\sum_{l=0}^{\infty}\sum_{m=-l}^{l}\int_{0}^{\infty}\text{d}\omega_{+}\left(\hat{b}_{\w lm}^{\text{up}}f_{\w lm}^{\text{up}}\left(x\right)+\hat{b}_{\w lm}^{\text{up}\dagger}f_{\w lm}^{\text{up}*}\left(x\right)\right)\,,\label{eq:Boulware-sum}
\end{align}
for some operator coefficients $\hat{b}_{\omega lm}^{\text{in}}$
and $\hat{b}_{\w lm}^{\text{up}}$. Note that the \emph{up}
modes are defined with respect to the positive frequency $\omega_{+}$
rather than $\omega$ (this is a direct result of the asymptotic behavior of the effective
potential \eqref{eq:asymPot_ext}). Therefore,
the corresponding integration is over positive $\omega_{+}$.  The decomposition in Eq.
\eqref{eq:Boulware-sum} serves to define the so-called (past) Boulware
state $\left|0\right\rangle _{B}$ (see Refs. \cite{Boulware:1975a,Boulware:1975b}
for the Schwarzschild case and Refs. \cite{Unruh:1974bw,OttewillWinstanley:2000}
for Kerr) via
\begin{align*}
\hat{b}_{\omega lm}^{\text{in}}\left|0\right\rangle _{B}&=0,\quad
\text{for all $\omega>0$};
\\
\hat{b}_{\w lm}^{\text{up}}\left|0\right\rangle _{B}&=0,\quad \text{for all $\omega_+>0$} \,.
\end{align*}
In non-rotating BHs (i.e., Schwarzschild and RN), this state is irregular
on both $H_{\text{past}}$ and $H_{\text{R}}$ and is empty on both
PNI and FNI (and so it is said to model a cold star); in the rotating
case, it continues to be empty on PNI but it contains quantum superradiance
at FNI (the Unruh-Starobinskii effect). 

The focus of this paper is another state: the Unruh state, which we
define in the next subsection. Before turning to the Unruh state,
however, we wish to give the following remark on another, third state.
In non-rotating BHs, one may consider the \emph{Hartle-Hawking} (HH)
state \cite{HH:1976,Israel:1976}, which corresponds to a BH in thermal
equilibrium, coupled to an infinite bath of radiation. Although not
too realistic, the HH state provides (in the non-rotating case) relative
simplicity due to its time-reversal and time translational invariance,
and so historically it was used to make some progress in the study
of the RSET. However, a state analogous to HH is ill-defined in Kerr
(see Ref.~\cite{KayWald:1991}, as well as Refs. \cite{OttewillWinstanley:2000}
and \cite{FrolovThorne:1989}). This may be intuitively understood
from the existence of superradiant modes (see Subsec.~\ref{subsec:Wronskian-relations}),
for which waves are reflected back to infinity with increased amplitude,
conflicting with the feasibility of a state of thermal equilibrium.
We shall thus consider only the (highly physically relevant) Unruh
state from now on. 

\subsection{The Unruh quantum state\label{subsec:Unruh state}}

The \emph{Unruh} state \cite{Unruh:1976} is widely recognized as
a physically realistic vacuum quantum state, describing an evaporating
BH (and thus, by definition, is not time-reversal invariant). The
Unruh state is constructed to resemble the quantum state arising at
late times for a BH formed by gravitational collapse. It is defined
by taking positive frequencies with respect to the affine parameters
along both initial null hypersurfaces (see the formulation of the
Unruh modes in Subsec.~\ref{subsec:Unruh-modes}). I.e., positive
frequencies are defined with respect to $v$ (the affine coordinate
on PNI) for incoming modes, and with respect to $U$ (the affine coordinate
on $H_{\text{past}}$ and $H_{L}$) for outgoing modes.

For a straight-forward definition of the Unruh vacuum state, we decompose
the metric in terms of the Unruh modes (Eqs.~\eqref{eq:gIN_asym}
and \eqref{eq:gUP_asym}):
\begin{align}
\hat{\Phi}\left(x\right)= & \sum_{l=0}^{\infty}\sum_{m=-l}^{l}\int_{0}^{\infty}\text{d}\omega\left(\hat{a}_{\omega lm}^{\text{in}}g_{\omega lm}^{\text{in}}\left(x\right)+\hat{a}_{\omega lm}^{\text{in}\dagger}g_{\omega lm}^{\text{in}*}\left(x\right)\right)\label{eq:FieldUnruhDec}\\
 & +\sum_{\hat{l}=0}^{\infty}\sum_{\hat{m}=-\hat{l}}^{\hat{l}}\int_{0}^{\infty}\text{d}\hat{\omega}\left(\hat{a}_{\hat{\omega}\hat{l}\hat{m}}^{\text{up}}g_{\hat{\omega}\hat{l}\hat{m}}^{\text{up}}\left(x\right)+\hat{a}_{\hat{\omega}\hat{l}\hat{m}}^{\text{up}\dagger}g_{\hat{\omega}\hat{l}\hat{m}}^{\text{up}*}\left(x\right)\right)\,,\nonumber 
\end{align}
where $\hat{a}_{\omega lm}^{\text{in}}$ and $\hat{a}_{\omega lm}^{\text{in\ensuremath{\dagger}}}$
($\hat{a}_{\hat{\omega}\hat{l}\hat{m}}^{\text{up}}$ and $\hat{a}_{\hat{\omega}\hat{l}\hat{m}}^{\text{up\ensuremath{\dagger}}}$)
are the creation and annihilation operators corresponding to the \emph{in} (\emph{up}) Unruh modes, and the set of quantum numbers $\hat{\omega}\hat{l}\hat{m}$
classifying the \emph{up} Unruh modes are as discussed in Subsec.
\ref{subsec:The-up-Unruh}.

The Unruh state $\left|0\right\rangle _{U}$ is then defined as the
vacuum state with respect to the Unruh decomposition \eqref{eq:FieldUnruhDec},
namely, it is the state annihilated by all Unruh-modes annihilation
operators:
\begin{align*}
\hat{a}_{\omega lm}^{\text{in}}\left|0\right\rangle _{U}&=0,\quad \text{for all $\omega>0$};
\\
\hat{a}_{\hat{\omega}\hat{l}\hat{m}}^{\text{up}}\left|0\right\rangle _{U}&=0,\quad  \text{for all $\hat{\omega}>0$}\,.
\end{align*}

The Unruh state involves no incoming flux at PNI, and an outgoing
flux of thermal radiation at FNI, in correspondence with Hawking radiation
of an evaporating BH. The corresponding RSET is expected to be regular
across the interior of the united domain (since the united domain
is the future domain of dependence of the two Unruh-state initial
null hypersurfaces, PNI and $H_{\text{past}}\cup H_{L}$). In particular,
this expectation for regularity applies at $H_{R}$, but not at $H_{\text{past}}\cup H_{L}$.

\section{Constructing the Unruh state HTPF in the exterior of a Kerr BH\label{sec:TPF_exterior}}

From the decomposition of the field $\hat{\Phi}$ in terms of Unruh
modes (Eq.~\eqref{eq:FieldUnruhDec}), applying the commutation relations
$\left[\hat{a}_{I}^{\Lambda},\hat{a}_{I'}^{\Lambda'\dagger}\right]=\hbar\,\delta_{II'}\delta_{\Lambda\Lambda'}\hat{1}$
(where $I$ denotes the set of all three quantum numbers and $\Lambda$
is either ``up" or ``in"), we obtain the mode-sum expression
of the Unruh state HTPF in terms of the Unruh modes $g_{\omega lm}^{\text{in}}$
and $g_{\hat{\omega}\hat{l}\hat{m}}^{\text{up}}$ (defined in Eqs.
\eqref{eq:gIN_asym} and \eqref{eq:gUP_asym}):

\begin{equation}
G_{U}^{(1)}(x,x')\equiv\left\langle \left\{ \hat{\Phi}\left(x\right),\hat{\Phi}\left(x'\right)\right\} \right\rangle _{U}=\left\langle \hat{\Phi}\left(x\right)\hat{\Phi}\left(x'\right)+\hat{\Phi}\left(x'\right)\hat{\Phi}\left(x\right)\right\rangle _{U}=G_{U}^{\text{in}}\left(x,x'\right)+G_{U}^{\text{up}}\left(x,x'\right),\label{eq:G_U}
\end{equation}
where $x$ and $x'$ are spacetime points and we denote 
\begin{align}
G_{U}^{\text{in}}\left(x,x'\right) & \equiv\hbar\sum_{l,m}\int_{0}^{\infty}\text{d}\omega\left\{ g_{\omega lm}^{\text{in}}\left(x\right),g_{\omega lm}^{\text{in}*}\left(x'\right)\right\} ,\label{eq:G_U^in}\\
G_{U}^{\text{up}}\left(x,x'\right) & \equiv\hbar\sum_{\hat{l},\hat{m}}\int_{0}^{\infty}\text{d}\hat{\omega}\left\{ g_{\hat{\omega}\hat{l}\hat{m}}^{\text{up}}\left(x\right),g_{\hat{\omega}\hat{l}\hat{m}}^{\text{up}*}\left(x'\right)\right\} \,,\label{eq:G_U^up}
\end{align}
and, recall, $\left\{ \xi\left(x\right),\zeta\left(x'\right)\right\} \equiv\xi\left(x\right)\zeta\left(x'\right)+\xi\left(x'\right)\zeta\left(x\right)$.

Throughout the paper, we shall use the shorthand notation $\sum_{l,m}\equiv\sum_{l=0}^{\infty}\sum_{m=-l}^{l}$
(and likewise for $\sum_{\hat{l},\hat{m}}$). Remarkably, it has been
found (see Eq.~(3.22) in Ref.~\cite{FrolovThorne:1989}, as well
as Eq.~(3.18c) in Ref.~\cite{OttewillWinstanley:2000} \footnote{Note that Ref.~\cite{OttewillWinstanley:2000} considers a slightly different two-point function (TPF) from our $G_{U}^{(1)}\left(x,x'\right)$, namely the \emph{non-symmetrized} TPF,  usually called the Wightman function $G^+(x,x') = \langle \hat{\Phi}(x)\hat{\Phi}(x')\rangle$ rather than the Hadamard  two-point function $G^{(1)} (x,x')= \langle \bigl\{\hat{\Phi}(x),\hat{\Phi}(x')\bigr\}\rangle$.  In particular, it is trivial to derive the symmetrized TPF from the non-symmetrized
one provided in Ref.~\cite{OttewillWinstanley:2000}, thereby obtaining Eq.~\eqref{eq:TPFoutside}. }, with the analogous Schwarzschild case in Refs.~\cite{christensen1977trace,Candelas:1980})),
that the mode-sum expression in Eq.~(\ref{eq:G_U}) may be decomposed
in terms of the more manageable outer Eddington modes
(defined and discussed in Subsec. \ref{subsec:Outer-Eddington-mode}),
yielding the expression: 
\begin{align}
G_{U}^{(1)}\left(x,x'\right)= & \hbar\sum_{l,m}\left[\int_{0}^{\infty}\text{d}\omega\left\{ f_{\omega lm}^{\text{in}}\left(x\right),f_{\omega lm}^{\text{in}*}\left(x'\right)\right\} +\int_{0}^{\infty}\text{d}\omega_{+}\coth\left(\frac{\pi\omega_{+}}{\kappa_{+}}\right)\left\{ f_{\omega lm}^{\text{up}}\left(x\right),f_{\omega lm}^{\text{up}*}\left(x'\right)\right\} \right] .\label{eq:TPFoutside}
\end{align}

In what follows in this section, we shall present the derivation of
Eq.~\eqref{eq:TPFoutside} in the Kerr BH exterior via a procedure
different from the methods previously used in the mentioned references.
This procedure is the same as that which we use later on to derive
the HTPF in the Kerr \textit{interior} and is an extension to Kerr
of the procedure used in Ref.~\cite{Group:2018} for the interior
of a RN BH. In doing so, we recover the known result (Eq.~\eqref{eq:TPFoutside}).
Proceeding in this way will allow us to demonstrate our method on
a Kerr BH background in a simpler case (i.e. outside the BH) before
delving into the BH interior, and to handle various issues special
to Kerr that arise already in the BH exterior.

\subsection{Mode decomposition of the exterior Unruh HTPF}

Concentrating on the BH exterior, we wish to express Eqs.~\eqref{eq:G_U^in}
and \eqref{eq:G_U^up} in terms of the exterior Eddington
modes, $f_{\omega lm}^{\text{in}}$ and $f_{\omega lm}^{\text{up}}$,
defined in Subsec.~\ref{subsec:Outer-Eddington-mode}.

In various stages of the computation to be carried out, it turns out
to be very useful to define a new version of $f_{\omega lm}^{\text{up}}$,
which carries an index $\omega_{+}$ rather than $\omega$. We shall
use the notation $f_{\omega_{+}lm}^{\text{up}\left(+\right)}$ as
the $\omega_{+}$-indexed version of $f_{\omega lm}^{\text{up}}$.
That is, for a certain set $\omega lm$, we define
\begin{equation}
f_{\omega_{+}lm}^{\text{up}\left(+\right)}\equiv f_{\omega\left(\omega_{+},m\right)lm}^{\text{up}}\label{eq:f^up_wp}
\end{equation}
 where $\omega$ on the RHS is related to $\omega_{+}$ and $m$ on
the LHS by the standard relation, $\omega\left(\omega_{+},m\right)=\omega_{+}+m\Omega_{+}$.
All relations and equations from the previous sections that include
$f_{\omega lm}^{\text{up}}$ may now be carried to this section with
the simple replacement $f_{\omega lm}^{\text{up}}\mapsto f_{\omega_{+}lm}^{\text{up}\left(+\right)}$  \footnote{A clarification regarding our notation may be in order here (particularly
related to the interchangeability of $f_{\omega lm}^{\text{up}}$
and $f_{\omega_{+}lm}^{\text{up}\left(+\right)}$): Once the object
$f_{\omega_{+}lm}^{\text{up}\left(+\right)}$ has been defined here
(in Eq.~\eqref{eq:f^up_wp}), our notational rules allow us to write
equations of the form, e.g., $f_{\omega_{+}lm}^{\text{up}\left(+\right)}=f_{\omega lm}^{\text{up}}$
(just to give a simple illustrative example). The exact meaning of
an equality of this type has been clarified in Sec. \eqref{sec:Families-of-modes},
and we repeat it here for clarity: Whenever a part of a given equation
depends on $\omega$ and another part depends on $\omega_{+}$, the
latter is to be viewed as given by $\omega-m\Omega_{+}$. (Or, if
one prefers, the other way around: $\omega$ may be viewed to be given
by $\omega_{+}+m\Omega_{+}$.) \label{fn:wp_index_explained}}. In what follows, we shall use the object $f_{\omega_{+}lm}^{\text{up}\left(+\right)}$
(rather than $f_{\omega lm}^{\text{up}}$) as a tool until we reach
the final expression (Eq.~\eqref{eq:result_outside_wp}), which will
then be re-expressed in terms of the usual $f_{\omega lm}^{\text{up}}$.

We begin with the \emph{in} contribution. We may readily use the equality
presented in Eq.~\eqref{eq:gIN-fIN} between $g_{\omega lm}^{\text{in}}$
and $f_{\omega lm}^{\text{in}}$ in the BH exterior to express Eq.~\eqref{eq:G_U^in}
as the mode sum 
\begin{equation}
G_{U}^{\text{in}}\left(x,x'\right)=\hbar\sum_{l,m}\int_{0}^{\infty}\text{d}\omega\left\{ f_{\omega lm}^{\text{in}}\left(x\right),f_{\omega lm}^{\text{in}*}\left(x'\right)\right\} \,.\label{eq:G_in_ext}
\end{equation}

Likewise, the \emph{up} counterpart $G_{U}^{\text{up}}\left(x,x'\right)$
requires establishing a relation between the \emph{up} Unruh modes
$g_{\hat{\w} \hat{l}\hat{m}}^{\text{up}}$ and the exterior Eddington
modes. However, as discussed at the end of Subsec.~\ref{subsec:The-up-Unruh},
that task is a more complicated one -- compared to the \emph{in} Unruh
modes, but also compared to the spherical symmetry counterpart.

We shall now introduce a notation to be used occasionally throughout
the rest of the paper. In determining a quantum state, the frequency
has a special role over the other quantum numbers. Let us then denote
collectively all other quantum numbers by $J$ (or sometimes $\hat{J}$).
Here, more specifically, $J\equiv\left(l,m\right)$ and $\hat{J}\equiv\left(\hat{l},\hat{m}\right)$
(where the latter indices were introduced in Subsec.~\ref{subsec:The-up-Unruh}).
Then, we shall also use the shorthand notation $\sum_{J}\equiv\sum_{l,m}=\sum_{l=0}^{\infty}\sum_{m=-l}^{l}$, and likewise for $\sum_{\hat{J}}$.

To relate the \emph{up} Unruh modes $g_{\hat{\omega}\hat{J}}^{\text{up}}$
and the Eddington modes on the BH exterior, we turn our
attention to the relevant asymptotic surfaces. Recall that on the
past horizon, we have (see Eqs.~\eqref{eq:gUP_asym} and \eqref{eq:fUP_past_asymZ})

\begin{equation}
\left.g_{\hat{\omega}\hat{J}}^{\text{up}}\right|_{H_{\text{past}}}=\frac{1}{\sqrt{4\pi\hat{\omega}\left(r_{+}^{2}+a^{2}\right)}}Y_{\hat{J}}\left(\theta,\varphi_{+}\right)e^{-i\hat{\omega}U\left(u_{\text{ext}}\right)}\label{eq:gup_Hpast}
\end{equation}
and 
\begin{align}
\left.f_{\omega_{+}J}^{\text{up}\left(+\right)}\right|_{H_{\text{past}}} & =\frac{1}{\sqrt{4\pi\left|\omega_{+}\right|\left(r_{+}^{2}+a^{2}\right)}}Z_{J}^{\omega}\left(\theta,\varphi_{+}\right)e^{-i\omega_{+}u_{\text{ext}}}\,.\label{eq:fup_Hpast}
\end{align}

As was already spelled out in Eq.~\eqref{eq:ang_orth_phi+}, the
set of spheroidal harmonics $Z_{J}^{\omega}\left(\theta,\varphi_{+}\right)$
is a complete orthonormal family on the 2-sphere for any fixed $\omega$.
For the analysis below, it is important to note that this set also
forms a complete orthonormal family for any fixed $\omega_{+}$.
It is orthonormal in the sense

\begin{align}
\int_{0}^{2\pi}\text{d}\varphi_{+}\int_{0}^{\pi}\text{d}\theta\,\sin\theta\,\left[Z_{lm}^{\omega\left(\omega_{+},m\right)}\left(\theta,\varphi_{+}\right)\right]^{*}Z_{l'm'}^{\omega\left(\omega_{+},m'\right)}\left(\theta,\varphi_{+}\right) & =\delta_{ll'}\delta_{mm'}\,,\label{eq:ang_omega+}
\end{align}
where, as usual, $\omega\left(\omega_{+},m\right)\equiv\omega_{+}+m\Omega_{+}$.
This is so because $\int_{0}^{2\pi}e^{i\left(m'-m\right)\varphi_{+}}\text{d}\varphi_{+}=2\pi\delta_{mm'}$,
hence the $\theta$-integral in Eq.~\eqref{eq:ang_omega+} needs
only to be carried out for $m'=m$ -- in which case the equality
of $\omega_{+}$ (in the two $Z$ functions) is fully equivalent to
the equality of $\omega$. A fairly similar argument may be used to
show that the completeness of the family $Z_{J}^{\omega}\left(\theta,\varphi_{+}\right)$
with fixed $\omega$ also implies its completeness with fixed $\omega_{+}$.

Now, in order to decompose $g_{\hat{\omega}\hat{J}}^{\text{up}}$
in terms of $f_{\omega_{+}J}^{\text{up}\left(+\right)}$ on $H_{\text{past}}$,
we introduce two sets of coefficients, $\alpha_{\hat{\omega}\omega_{+}}$
and $C_{\hat{J}J}^{\omega_{+}}$, aimed at handling the frequential
and angular factors respectively. The Fourier coefficients $\alpha_{\hat{\omega}\omega_{+}}$
are given by the inverse Fourier transform 
\begin{equation}
\alpha_{\hat{\omega}\omega_{+}}\equiv\int_{-\infty}^{\infty}\text{d}u_{\text{ext}}\,e^{-i\hat{\omega}U\left(u_{\text{ext}}\right)}e^{i\omega_{+}u_{\text{ext}}}\,.\label{eq:alpha_def}
\end{equation}
This integral may be evaluated as described in Ref.~\cite{Group:2018}
 \footnote{\label{fn:alpha^past_notation}See Eq.~(3.3) therein, and apply the
notation change $\tilde{\omega}\mapsto\omega_{+},\omega\mapsto\hat{\omega}$.}, yielding

\begin{equation}
\alpha_{\hat{\omega}\omega_{+}}=\frac{1}{\kappa_{+}}\left(\frac{\hat{\omega}}{\kappa_{+}}\right)^{i\omega_{+}/\kappa_{+}}e^{\pi\omega_{+}/2\kappa_{+}}\Gamma\left(-i\frac{\omega_{+}}{\kappa_{+}}\right)\,.\label{eq:alpha^past_res}
\end{equation}

Similarly, $C_{\hat{J}J}^{\omega_{+}}$, the coefficients translating
between the two bases of orthonormal functions on the two-sphere,
$Z_{J}^{\omega}$ and $Y_{\hat{J}}$, are defined by 
\begin{equation}
C_{\hat{J}J}^{\omega_{+}}\equiv\int_{0}^{2\pi}\text{d}\varphi_{+}\int_{0}^{\pi}\text{d}\theta\sin\theta\,\,\left[Z_{J}^{\omega\left(\omega_{+},m\right)}\left(\theta,\varphi_{+}\right)\right]^{*}Y_{\hat{J}}\left(\theta,\varphi_{+}\right)\label{eq:C_def}
\end{equation}
(recall that in the square brackets the information about the value
of $m$ in $\omega\left(\omega_{+},m\right)$ is encoded in $J$).

The $\alpha_{\hat{\omega}\omega_{+}}$ and $C_{\hat{J}J}^{\omega_{+}}$
coefficients, as defined in Eqs.~\eqref{eq:alpha_def} and \eqref{eq:C_def},
allow a translation from the Unruh to the Eddington frequential
and angular factors via the relations 
\begin{equation}
e^{-i\hat{\omega}U\left(u_{\text{ext}}\right)}=\frac{1}{2\pi}\int_{-\infty}^{\infty}\text{d}\omega_{+}\,\alpha_{\hat{\omega}\omega_{+}}e^{-i\omega_{+}u_{\text{ext}}}\,\label{eq:alpha_int_w+}
\end{equation}
and 
\begin{equation}
Y_{\hat{J}}\left(\theta,\varphi_{+}\right)=\sum_{J}C_{\hat{J}J}^{\omega_{+}}Z_{J}^{\omega(\omega_{+},m)}\left(\theta,\varphi_{+}\right)\,,\label{eq:C_sum_J}
\end{equation}
where, recall, in the last equality the sum over $J$ is carried out
with fixed $\omega_{+}$. The first of these relations is just the
inversion of the (inverse) Fourier transfom in Eq.~\eqref{eq:alpha_def}.
To derive the second relation, we use the (fixed-$\omega_{+}$) completeness
of the spheroidal harmonics to decompose the spherical harmonics as
\begin{equation}
Y_{\hat{J}}\left(\theta,\varphi_{+}\right)=\sum_{J}P_{\hat{J}J}^{\omega_{+}}Z_{J}^{\omega(\omega_{+},m)}\left(\theta,\varphi_{+}\right)\,,\label{eq:Y_decom_Z}
\end{equation}
where $P_{\hat{J}J}^{\omega_{+}}$ denote the coefficients of the decomposition.
Substituting this decomposition into the RHS of Eq.~\eqref{eq:C_def},
and recalling the spheroidal harmonics orthonormality, one readily
sees that $C_{\hat{J}J}^{\omega_{+}}=P_{\hat{J}J}^{\omega_{+}}$,
hence Eq. \eqref{eq:Y_decom_Z} reduces to the desired relation \eqref{eq:C_sum_J}.
In a similar manner (this time employing the orthonormality and completeness
of the \emph{spherical} harmonics), one can decompose the spheroidal
harmonics and show that 
\begin{equation}
Z_{J}^{\omega(\omega_{+},m)}\left(\theta,\varphi_{+}\right)=\sum_{\hat{J}}C_{\hat{J}J}^{\omega_{+}*}Y_{\hat{J}}\left(\theta,\varphi_{+}\right)\,.\label{eq:Z_decomposition}
\end{equation}

For future use, we also note that the following relation holds: 
\begin{equation}
\sum_{\hat{J}}C_{\hat{J}J}^{\omega_{+}}C_{\hat{J}J'}^{\omega_{+}*}=\delta_{JJ'}\,.\label{eq:sum_C*C}
\end{equation}
To see this, we substitute Eq.~\eqref{eq:C_sum_J} into the RHS of
Eq.~\eqref{eq:Z_decomposition} (with index renaming $J\mapsto J'$
in the latter) and obtain 
\begin{equation}
Z_{J'}^{\omega(\omega_{+},m')}\left(\theta,\varphi_{+}\right)=\sum_{J}\left[\sum_{\hat{J}}C_{\hat{J}J}^{\omega_{+}}C_{\hat{J}J'}^{\omega_{+}*}\right]Z_{J}^{\omega(\omega_{+},m)}\left(\theta,\varphi_{+}\right)\,.\label{eq:Z'_Z_decomposition}
\end{equation}
Recalling the orthonormality of the spheroidal harmonics, the term
in square brackets must be the identity matrix with components $\delta_{JJ'}$, yielding Eq.~\eqref{eq:sum_C*C}.

Substituting Eqs.~\eqref{eq:alpha_int_w+} and \eqref{eq:C_sum_J}
in Eq.~\eqref{eq:gup_Hpast} and comparing to Eq.~\eqref{eq:fup_Hpast},
the decomposition of $g_{\hat{\omega}\hat{J}}^{\text{up}}$ in terms
of $f_{\omega_{+}J}^{\text{up}\left(+\right)}$ on the null hypersurface
$H_{\text{past}}$ may now be written as follows:
\begin{align}
\sqrt{\hat{\omega}}\left.g_{\hat{\omega}\hat{J}}^{\text{up}}\right|_{H_{\text{past}}} & =\frac{1}{2\pi}\sum_{J}\int_{-\infty}^{\infty}\text{d}\omega_{+}\sqrt{\left|\omega_{+}\right|}\, \alpha_{\hat{\omega}\omega_{+}}C_{\hat{J}J}^{\omega_{+}}\left.f_{\omega_{+}J}^{\text{up}\left(+\right)}\right|_{H_{\text{past}}}\,.\label{eq:gUP-fUP-Hpast}
\end{align}
In addition, recall that both $g_{\hat{\omega}\hat{J}}^{\text{up}}$
and $f_{\omega_{+}J}^{\text{up}\left(+\right)}$ vanish on PNI: see
Eqs.~\eqref{eq:fUP_past_asymZ} and \eqref{eq:gUP_asym}. Thus, 
\begin{equation}
\sqrt{\hat{\omega}}\left.g_{\hat{\omega}\hat{J}}^{\text{up}}\right|_{\text{PNI}}=0=\frac{1}{2\pi}\sum_{J}\int_{-\infty}^{\infty}\text{d}\omega_{+}\sqrt{\left|\omega_{+}\right|}\, \alpha_{\hat{\omega}\omega_{+}}C_{\hat{J}J}^{\omega_{+}}\left.f_{\omega_{+}J}^{\text{up}\left(+\right)}\right|_{\text{PNI}}\,.\label{eq:gUP-fUP-PNI}
\end{equation}
Since $g_{\hat{\omega}\hat{J}}^{\text{up}}$ and $f_{\omega_{+}J}^{\text{up}\left(+\right)}$
satisfy the same wave equation \eqref{eq:Teukolsky}, it follows that
these two quantities are related as prescribed in Eqs.~\eqref{eq:gUP-fUP-Hpast}
and \eqref{eq:gUP-fUP-PNI} not only on the initial null hypersurfaces
$H_{\text{past}}$ and PNI but throughout the BH exterior. That is,
\begin{equation}
\sqrt{\hat{\omega}}\,g_{\hat{\omega}\hat{J}}^{\text{up}}(x)=\frac{1}{2\pi}\sum_{J}\int_{-\infty}^{\infty}\text{d}\omega_{+}\sqrt{\left|\omega_{+}\right|}\, \alpha_{\hat{\omega}\omega_{+}}C_{\hat{J}J}^{\omega_{+}}f_{\omega_{+}J}^{\text{up}\left(+\right)}(x)\,\,\,,\,\,\,\,\,\,\,\,\,\,r\geq r_{+}\,.\label{eq:gUP-fUP}
\end{equation}

Next, the HTPF \emph{up} mode contribution $G_{U}^{\text{up}}\left(x,x'\right)$
in terms of Eddington modes is achieved by substituting
Eq.~\eqref{eq:gUP-fUP} in Eq.~\eqref{eq:G_U^up}: 
\begin{align}
 & G_{U}^{\text{up}}\left(x,x'\right)=\label{eq:Gup1}\\
 & \frac{\hbar}{4\pi^{2}}\sum_{\hat{J}}\int_{0}^{\infty}\frac{\text{d}\hat{\omega}}{\hat{\omega}}\sum_{J}\int_{-\infty}^{\infty}\text{d}\omega_{+}\sqrt{\left|\omega_{+}\right|}\alpha_{\hat{\omega}\omega_{+}}C_{\hat{J}J}^{\omega_{+}}\sum_{J'}\int_{-\infty}^{\infty}\text{d}\omega'_{+}\sqrt{\left|\omega'_{+}\right|}\, \alpha_{\hat{\omega}\omega'_{+}}^{*}C_{\hat{J}J'}^{\omega'_{+}*}\left\{ f_{\omega_{+}J}^{\text{up}\left(+\right)}\left(x\right),f_{\omega'_{+}J'}^{\text{up}\left(+\right)*}\left(x'\right)\right\} \,.\nonumber 
\end{align}

We conveniently rearrange Eq.~\eqref{eq:Gup1} as follows \footnote{\label{fn:re-arrangement}This re-arrangement involves interchanges
of the summation over $\hat{J}$ and the integration over $\hat{\omega}$
with all subsequent operations (summations and integrations). We do
not attempt to rigorously justify this manipulation (or similar ones
that appear later on). Nevertheless, after implementing this re-arrangement,
we do recover the correct, well known, result quoted in Eq.~\eqref{eq:TPFoutside}
above. This may be considered  a  justification for the
manipulations entailed.}:
\begin{align}
 & G_{U}^{\text{up}}\left(x,x'\right)=\label{eq:Gup2}\\
 & \frac{\hbar}{4\pi^{2}}\sum_{J}\int_{-\infty}^{\infty}\text{d}\omega_{+}\sqrt{\left|\omega_{+}\right|}\sum_{J'}\int_{-\infty}^{\infty}\text{d}\omega'_{+}\sqrt{\left|\omega'_{+}\right|}\left\{ f_{\omega_{+}J}^{\text{up}\left(+\right)}\left(x\right),f_{\omega'_{+}J'}^{\text{up}\left(+\right)*}\left(x'\right)\right\} \sum_{\hat{J}}C_{\hat{J}J}^{\omega_{+}}C_{\hat{J}J'}^{\omega'_{+}*}\int_{0}^{\infty}\frac{\text{d}\hat{\omega}}{\hat{\omega}}\alpha_{\hat{\omega}\omega_{+}}\alpha_{\hat{\omega}\omega'_{+}}^{*}\,.\nonumber 
\end{align}

We can now perform the $\hat{\omega}$-integral appearing in the above
equation, using Eq.~\eqref{eq:alpha^past_res}: \footnote{To obtain this integral, one may rewrite it as $\frac{1}{\kappa_{+}}e^{\left(\omega_{+}+\omega'_{+}\right)\pi/2\kappa_{+}}\Gamma\left(-i\omega_{+}/\kappa_{+}\right)\Gamma\left(i\omega'_{+}/\kappa_{+}\right)\int_{-\infty}^{\infty}\text{d}s\,e^{is\left(\omega_{+}-\omega'_{+}\right)}$,
where $s\equiv\frac{1}{\kappa_{+}}\ln\left(\hat{\omega}/\kappa_{+}\right)$,
and use the relation $\left|\Gamma\left(i\omega_{+}/\kappa_{+}\right)\right|^{2}=\pi\left(\kappa_{+}/\omega_{+}\right)/\sinh\left(\pi\omega_{+}/\kappa_{+}\right)$.} 
\begin{equation}
\int_{0}^{\infty}\frac{\text{d}\hat{\omega}}{\hat{\omega}}\alpha_{\hat{\omega}\omega_{+}}\alpha_{\hat{\omega}\omega'_{+}}^{*}=\frac{4\pi^{2}}{\omega_{+}}\frac{1}{1-e^{-2\pi\omega_{+}/\kappa_{+}}}\delta\left(\omega_{+}-\omega'_{+}\right)\,.\label{eq:int_alpha_alpha*}
\end{equation}
With this identity, Eq.~\eqref{eq:Gup2} reduces, after performing
the trivial integration of the $\delta$-function over $\omega'_{+}$,
to 
\[
G_{U}^{\text{up}}\left(x,x'\right)=\hbar\sum_{J}\int_{-\infty}^{\infty}\text{d}\omega_{+}\frac{\left|\omega_{+}\right|}{\omega_{+}}\frac{1}{1-e^{-2\pi\omega_{+}/\kappa_{+}}}\sum_{J'}\left\{ f_{\omega_{+}J}^{\text{up}\left(+\right)}\left(x\right),f_{\omega{}_{+}J'}^{\text{up}\left(+\right)*}\left(x'\right)\right\} \sum_{\hat{J}}C_{\hat{J}J}^{\omega_{+}}C_{\hat{J}J'}^{\omega{}_{+}*}\,.
\]
Next we use Eq.~\eqref{eq:sum_C*C} to obtain 
\begin{equation}
G_{U}^{\text{up}}\left(x,x'\right)=\hbar\sum_{J}\int_{-\infty}^{\infty}\text{d}\omega_{+}\text{\,sign}\left(\omega_{+}\right)\frac{1}{1-e^{-2\pi\omega_{+}/\kappa_{+}}}\left\{ f_{\omega_{+}J}^{\text{up}\left(+\right)}\left(x\right),f_{\omega_{+}J}^{\text{up}\left(+\right)*}\left(x'\right)\right\} \,.\label{eq:Gup3}
\end{equation}

Finally, we would like to ``fold'' the $\omega_{+}$-integral in
this equation so that only modes with $\omega_{+}>0$ show up. To
this end, we note that since the mapping $\left(\omega,m\right)\mapsto\left(-\omega,-m\right)$
is equivalent to $\left(\omega_{+},m\right)\mapsto\left(-\omega_{+},-m\right)$,
we may rewrite Eq.~\eqref{eq:ExtEddModesInvariance} (with $\Lambda$ taken as "up") as 
\begin{equation}
f_{\left(-\omega_{+}\right)l\left(-m\right)}^{\text{up}\left(+\right)}=(-1)^{m}f_{\omega_{+}lm}^{\text{up}\left(+\right)*}\,.\label{eq:f^up_wp_invariance}
\end{equation}
We now recall that $\sum_{J}\equiv\sum_{l=0}^{\infty}\sum_{m=-l}^{l}$ and concentrate on the summation over $m$.
For each given $l$ we have 
\[
\sum_{m=-l}^{l}\left\{ f_{\left(-\omega_{+}\right)lm}^{\text{up}\left(+\right)}\left(x\right),f_{\left(-\omega_{+}\right)lm}^{\text{up}\left(+\right)*}\left(x'\right)\right\} =\sum_{m=-l}^{l}\left\{ f_{\omega_{+}l\left(-m\right)}^{\text{up}\left(+\right)}\left(x\right),f_{\omega_{+}l\left(-m\right)}^{\text{up}\left(+\right)*}\left(x'\right)\right\} =\sum_{m=-l}^{l}\left\{ f_{\omega_{+}lm}^{\text{up}\left(+\right)}\left(x\right),f_{\omega_{+}lm}^{\text{up}\left(+\right)*}\left(x'\right)\right\} \,,
\]
where the first equality follows from Eq.~\eqref{eq:f^up_wp_invariance}, and the last equality simply involves a renaming of the summation
index $m\mapsto-m$. Using the group index $J$, this may be expressed
as

\[
\sum_{J}\left\{ f_{\left(-\omega_{+}\right)J}^{\text{up}\left(+\right)}\left(x\right),f_{\left(-\omega_{+}\right)J}^{\text{up}\left(+\right)*}\left(x'\right)\right\} =\sum_{J}\left\{ f_{\omega_{+}J}^{\text{up}\left(+\right)}\left(x\right),f_{\omega_{+}J}^{\text{up}\left(+\right)*}\left(x'\right)\right\} \,.
\]
Eq.~\eqref{eq:Gup3} can thus be rewritten as 
\[
G_{U}^{\text{up}}\left(x,x'\right)=\hbar\sum_{J}\int_{0}^{\infty}\text{d}\omega_{+}\left[\frac{1}{1-e^{-2\pi\omega_{+}/\kappa_{+}}}-\frac{1}{1-e^{2\pi\omega_{+}/\kappa_{+}}}\right]\left\{ f_{\omega_{+}J}^{\text{up}\left(+\right)}\left(x\right),f_{\omega{}_{+}J}^{\text{up}\left(+\right)*}\left(x'\right)\right\} \,.
\]
As one can easily see, the term in square brackets is $\coth\left(\pi\omega_{+}/\kappa_{+}\right)$.
Our final result for $G_{U}^{\text{up}}$ is, therefore, 
\begin{equation}
G_{U}^{\text{up}}\left(x,x'\right)=\hbar\sum_{J}\int_{0}^{\infty}\text{d}\omega_{+}\coth\left(\frac{\pi\omega_{+}}{\kappa_{+}}\right)\left\{ f_{\omega_{+}J}^{\text{up}\left(+\right)}\left(x\right),f_{\omega{}_{+}J}^{\text{up}\left(+\right)*}\left(x'\right)\right\} \,.\label{eq:G_up_ext}
\end{equation}

As prescribed in Eq.~\eqref{eq:G_U}, we may now put together the
\emph{up} and \emph{in} mode contributions, as given in Eqs.~\eqref{eq:G_in_ext}
and \eqref{eq:G_up_ext}, to yield $G_{U}^{\left(1\right)}\left(x,x'\right)$:
\begin{equation}
G_{U}^{(1)}(x,x')=\hbar\sum_{l,m}\left[\int_{0}^{\infty}\text{d}\omega\left\{ f_{\omega lm}^{\text{in}}\left(x\right),f_{\omega lm}^{\text{in}*}\left(x'\right)\right\} +\int_{0}^{\infty}\text{d}\omega_{+}\coth\left(\frac{\pi\omega_{+}}{\kappa_{+}}\right)\left\{ f_{\omega_{+}lm}^{\text{up}\left(+\right)}\left(x\right),f_{\omega_{+}lm}^{\text{up}\left(+\right)*}\left(x'\right)\right\} \right]\,.\label{eq:result_outside_wp}
\end{equation}

Finally, we may retrieve the standard $\omega$-indexed notation,
replacing $f_{\omega_{+}lm}^{\text{up}\left(+\right)}$ by $f_{\omega lm}^{\text{up}}$
(since, as mentioned above, they represent the same object). Then,
the mode-sum expression of the Unruh-state HTPF outside a Kerr BH,
in terms of Eddington modes, is as previously quoted (in
Eq.~\eqref{eq:TPFoutside}): 

\begin{equation}
G_{U}^{(1)}(x,x')=\hbar\sum_{l,m}\left[\int_{0}^{\infty}\text{d}\omega\left\{ f_{\omega lm}^{\text{in}}\left(x\right),f_{\omega lm}^{\text{in}*}\left(x'\right)\right\} +\int_{0}^{\infty}\text{d}\omega_{+}\coth\left(\frac{\pi\omega_{+}}{\kappa_{+}}\right)\left\{ f_{\omega lm}^{\text{up}}\left(x\right),f_{\omega lm}^{\text{up}*}\left(x'\right)\right\} \right]\,.\label{eq:our_final_result_outside}
\end{equation}

It may be shown (however, outside the scope of this paper) that at small $\omega_{+}$, the \emph{up} radial
function $\psi_{\omega lm}^{\text{up}}$ behaves (to leading order) as $\omega_{+}$,
which ensures regularity of the $\int_{0}^{\infty}\text{d}\omega_{+}$
integral at small $\omega_{+}$. (Regularity of the $\int_{0}^{\infty}\text{d}\omega$
integral at small $\omega$ is similarly ensured.)

\subsection{Invariance to the choice of the angular functions $\hat{Z}_{\hat{l}\hat{m}}^{\hat{\omega}}\left(\theta,\varphi_{+}\right)$
\label{subsec:Invariance-to-Zwlm-choice}}

Finally, we comment on the invariance of our final result \eqref{eq:our_final_result_outside}
in the exterior of a Kerr BH with respect to the choice of the angular
functions $\hat{Z}_{\hat{l}\hat{m}}^{\hat{\omega}}\left(\theta,\varphi_{+}\right)$,
used for prescribing the \emph{up} Unruh-modes initial data at $H_{\text{past}}\cup H_{L}$.
As discussed in Subsec.~\ref{subsec:The-up-Unruh}, $\hat{Z}_{\hat{l}\hat{m}}^{\hat{\omega}}\left(\theta,\varphi_{+}\right)$
can be any set of angular functions which is orthonormal and complete
on the 2-sphere (and in particular it may depend on the mode's Kruskal
frequency $\hat{\omega}$). However, for the sake of concreteness
(as well as simplicity and brevity), in the analysis above we made
the specific choice $\hat{Z}_{\hat{l}\hat{m}}^{\hat{\omega}}\left(\theta,\varphi_{+}\right)=Y_{\hat{l}\hat{m}}\left(\theta,\varphi_{+}\right)$.
Here we shall briefly consider how the analysis would proceed, and
ultimately what would be the final resultant mode structure of the
HTPF, if one chose to work with generic angular functions $\hat{Z}_{\hat{l}\hat{m}}^{\hat{\omega}}$
rather than the specific functions $Y_{\hat{l}\hat{m}}$.

Let us examine the consequence of replacing $Y_{\hat{l}\hat{m}}$
everywhere by $\hat{Z}_{\hat{l}\hat{m}}^{\hat{\omega}}$ (and likewise
$Y_{\hat{J}}$ by $\hat{Z}_{\hat{J}}^{\hat{\omega}}$), starting at
Eq.~\eqref{eq:gup_Hpast}. The $C$-coefficients, relating the spheroidal
harmonics with the \emph{up} Unruh-modes initial angular functions
(which now depend on $\hat{\omega}$), accordingly acquire an extra
index $\hat{\omega}$. That is, Eq.~\eqref{eq:C_def} is replaced
by 
\begin{equation}
C_{\hat{J}J}^{\hat{\omega}\omega_{+}}\equiv\int_{0}^{2\pi}\text{d}\varphi_{+}\int_{0}^{\pi}\text{d}\theta\sin\theta\,\,\left[Z_{J}^{\omega(\omega_{+},m)}\left(\theta,\varphi_{+}\right)\right]^{*}\hat{Z}_{\hat{J}}^{\hat{\omega}}\left(\theta,\varphi_{+}\right)\,,\label{eq:C_def-generic}
\end{equation}
and every instance of $C_{\hat{J}J}^{\omega_{+}}$ is replaced by
$C_{\hat{J}J}^{\hat{\omega}\omega_{+}}$. Under these replacements,
all equations up to Eq.~\eqref{eq:Gup1} (inclusive) hold in their
new analogous form. For later use, we quote in particular the new
form of Eq.~\eqref{eq:sum_C*C}, simply adding the index $\hat{\omega}$
to both coefficients: 
\begin{equation}
\sum_{\hat{J}}C_{\hat{J}J}^{\hat{\omega}\omega_{+}}C_{\hat{J}J'}^{\hat{\omega}\omega{}_{+}*}=\delta_{JJ'}\,.\label{eq:sum_C*C1}
\end{equation}

Things become slightly more delicate when arriving at (the $Y_{\hat{J}}\mapsto\hat{Z}_{\hat{J}}^{\hat{\omega}}$
counterpart of) Eq.~\eqref{eq:Gup2}: Here, the term $\sum_{\hat{J}}C_{\hat{J}J}^{\omega_{+}}C_{\hat{J}J'}^{\omega'_{+}*}$
should of course be replaced by $\sum_{\hat{J}}C_{\hat{J}J}^{\hat{\omega}\omega_{+}}C_{\hat{J}J'}^{\hat{\omega}\omega'_{+}*}$.
Naively one might be concerned that since this term exhibits explicit
dependence on $\hat{\omega}$, it would be necessary to keep it inside
the $\hat{\omega}$ integral (i.e. the integral at the very end of
the RHS of Eq.~\eqref{eq:Gup2}, which is then evaluated in Eq.~\eqref{eq:int_alpha_alpha*}).
It is therefore crucial to note that (as we shall shortly justify)
\begin{equation}
\sum_{\hat{J}}C_{\hat{J}J}^{\hat{\omega}\omega_{+}}C_{\hat{J}J'}^{\hat{\omega}\omega'_{+}*}=\tilde{C}_{JJ'}^{\omega_{+}\omega'_{+}*}\,,\label{eq:sum_C*C2}
\end{equation}
where $\tilde{C}_{JJ'}^{\omega_{+}\omega'_{+}}$ is defined by 
\begin{equation}
\tilde{C}_{JJ'}^{\omega_{+}\omega'_{+}}\equiv\int_{0}^{2\pi}\text{d}\varphi_{+}\int_{0}^{\pi}\text{d}\theta\sin\theta\,\,\left[Z_{J'}^{\omega\left(\omega'_{+},m'\right)}\left(\theta,\varphi_{+}\right)\right]^{*}Z_{J}^{\omega\left(\omega_{+},m\right)}\left(\theta,\varphi_{+}\right),\label{eq:Ctilde_def}
\end{equation}
which by construction is independent of $\hat{\omega}$. \footnote{As one can easily see by performing the integration over $\varphi_{+}$,
$\tilde{C}_{JJ'}^{\omega_{+}\omega'_{+}}$ vanishes for any $m'\neq m$;
but this specific property is not needed here.}

Note that taking $\omega'_{+}=\omega_{+}$ in the RHS of Eq.~\eqref{eq:Ctilde_def}
reduces to the LHS of the orthonormality relation \eqref{eq:ang_orth_phi+}.
Hence
\begin{equation}
\tilde{C}_{JJ'}^{\omega_{+}\omega{}_{+}}=\delta_{JJ'},\label{eq:Ctilde_wpwp}
\end{equation}
and Eq.~\eqref{eq:sum_C*C2} thus reduces appropriately to Eq.~\eqref{eq:sum_C*C1}.

In order to establish Eq.~\eqref{eq:sum_C*C2} we first note (using
an argument similar to the one employed above for the justification
of Eq.~\eqref{eq:C_sum_J}) that $\tilde{C}_{JJ'}^{\omega_{+}\omega'_{+}}$
are actually the coefficients relating the two sets of spheroidal
harmonics $Z_{J}^{\omega\left(\omega_{+},m\right)}$ and $Z_{J'}^{\omega\left(\omega'_{+},m'\right)}$,
via \footnote{Explicitly, the derivation is as follows: From the (fixed-$\omega_{+}$)
completeness of the spheroidal harmonics, one can write (in analogy
to Eq.~\eqref{eq:Y_decom_Z}) 
\[
Z_{J}^{\omega\left(\omega_{+},m\right)}\left(\theta,\varphi_{+}\right)=\sum_{J'}P_{JJ'}^{\omega_{+}\omega'_{+}}Z_{J'}^{\omega(\omega'_{+},m')}\left(\theta,\varphi_{+}\right),
\]
where $P_{JJ'}^{\omega_{+}\omega'_{+}}$ are the coefficients of the
decomposition. Substituting this decomposition into Eq.~\eqref{eq:Ctilde_def},
and using the orthonormality of the spheroidal harmonics, one obtains
$P_{JJ'}^{\omega_{+}\omega'_{+}}=\tilde{C}_{JJ'}^{\omega_{+}\omega'_{+}}$.} 
\begin{equation}
Z_{J}^{\omega\left(\omega_{+},m\right)}\left(\theta,\varphi_{+}\right)=\sum_{J'}\tilde{C}_{JJ'}^{\omega_{+}\omega'_{+}}Z_{J'}^{\omega\left(\omega'_{+},m'\right)}\left(\theta,\varphi_{+}\right)\,.\label{eq:Ctilde_sum_J'}
\end{equation}
Then, Eq.~\eqref{eq:sum_C*C2} naturally follows from the completeness
and orthonormality of each of the three involved families of angular
functions, namely $Z_{J}^{\omega\left(\omega_{+},m\right)}$, $Z_{J'}^{\omega\left(\omega'_{+},m'\right)}$
and $\hat{Z}_{\hat{J}}^{\hat{\omega}}$, by a slight generalization
of the argument described right after Eq.~\eqref{eq:sum_C*C} \footnote{More precisely, one would need to combine the $Y_{\hat{J}}\mapsto\hat{Z}_{\hat{J}}^{\hat{\omega}}$
counterparts of Eqs.~\eqref{eq:C_sum_J} and \eqref{eq:Z_decomposition},
taking $J\mapsto J'$ and $\omega_{+}\mapsto\omega'_{+}$ in the former,
to yield the (slightly generalized) counterpart of Eq.~\eqref{eq:Z'_Z_decomposition}:
\[
Z_{J}^{\omega(\omega_{+},m)}\left(\theta,\varphi_{+}\right)=\sum_{J'}\left[\sum_{\hat{J}}C_{\hat{J}J}^{\hat{\omega}\omega_{+}*}C_{\hat{J}J'}^{\hat{\omega}\omega'_{+}}\right]Z_{J'}^{\omega(\omega'_{+},m')}\left(\theta,\varphi_{+}\right)\,.
\]
Recalling the coefficients of the decomposition in Eq.~\eqref{eq:Ctilde_sum_J'}
are unique, one obtains 
\[
\sum_{\hat{J}}C_{\hat{J}J}^{\hat{\omega}\omega_{+}*}C_{\hat{J}J'}^{\hat{\omega}\omega'_{+}}=\tilde{C}_{JJ'}^{\omega_{+}\omega'_{+}}\,.
\]
The desired result, Eq.~\eqref{eq:sum_C*C2}, is then achieved by
complex conjugation.}.

Now, owing to Eq.~\eqref{eq:sum_C*C2}, we are allowed to place the
term $\sum_{\hat{J}}C_{\hat{J}J}^{\hat{\omega}\omega_{+}}C_{\hat{J}J'}^{\hat{\omega}\omega'_{+}*}$
out of the $\hat{\omega}$ integral, just as in Eq.~\eqref{eq:Gup2}.
From this point on the analysis proceeds in a completely analogous
manner (recalling Eq.~\eqref{eq:sum_C*C1}) to the previous subsection,
and the final result \eqref{eq:our_final_result_outside} is again
obtained, this time using the generic angular functions $\hat{Z}_{\hat{l}\hat{m}}^{\hat{\omega}}$
rather than the spherical harmonics $Y_{\hat{l}\hat{m}}$.

To avoid confusion, we also emphasize that in this final expression
for the mode structure of the Unruh-state HTPF, the modes that appear
are the Eddington modes $f_{\omega lm}^{\text{in}}$ and
$f_{\omega lm}^{\text{up}}$, which are of course separable in terms
of \emph{spheroidal harmonics} (this is regardless of the nature of
the angular functions $\hat{Z}_{\hat{l}\hat{m}}^{\hat{\omega}}\left(\theta,\varphi_{+}\right)$
that were chosen earlier in the process).


\section{Constructing the Unruh state HTPF in the interior of a Kerr BH\label{sec:TPF_interior}}

In this section we shall finally construct the mode-sum expression
for the Unruh-state HTPF \textit{inside} a Kerr BH in terms of Eddington
modes. We shall follow here an analogous procedure to the one carried
out in Ref.~\cite{Group:2018} in the RN case, while noting that
the presence of rotation induces some essential differences. Basically
it is the procedure demonstrated already in Sec.~\ref{sec:TPF_exterior}
for the exterior of a Kerr BH, although there are some notable technical
differences. In the first subsection we shall carry out the actual
derivation of the expression for the HTPF and in the second one we
shall prove that the expression is invariant with respect to the choice
of the initial angular functions.

\subsection{Mode decomposition of the interior Unruh HTPF}

In the analysis to follow, just like in the exterior counterpart presented
in Sec. \ref{sec:TPF_exterior}, we shall use the $\omega_{+}$-indexed
versions (that is, objects carrying an index $\omega_{+}$ rather
than $\omega$) of the Eddington modes $f_{\omega lm}^{R}$,
$f_{\omega lm}^{L}$ and $f_{\omega lm}^{\text{up}}$. The $\w_+$ version of $f_{\omega lm}^{\text{up}}$
has been introduced in Eq.~\eqref{eq:f^up_wp}, and
for the interior
modes we define in 
a similar
manner:
\[
f_{\omega_{+}lm}^{\Lambda\left(+\right)}\equiv f_{\omega\left(\omega_{+},m\right)lm}^{\Lambda},
\]
where $\Lambda$ is either ``\emph{R}" or ``\emph{L}", and $\omega\left(\omega_{+},m\right)\equiv\omega_{+}+m\Omega_{+}$.
In addition, we shall use the notation 
\[
\rho_{\omega_{+}lm}^{\text{up}\left(+\right)}\equiv\rho_{\omega\left(\omega_{+},m\right)lm}^{\text{up}}
\]
as the $\omega_{+}$-indexed version of $\rho_{\omega lm}^{\text{up}}$.
All relations and equations containing $f_{\omega lm}^{R}$, $f_{\omega lm}^{L}$,
$f_{\omega lm}^{\text{up}}$ and $\rho_{\omega lm}^{\text{up}}$ are
to be carried from previous sections to this section along with a
simple replacement of the $\omega$-indexed objects with their $\omega_{+}$-indexed
counterparts, as defined above (see also footnote \eqref{fn:wp_index_explained}).

We begin with Eqs.~\eqref{eq:G_U}-\eqref{eq:G_U^up}, which are valid in the interior as well as the exterior of the BH, aiming for
a mode-sum decomposition of both the \emph{in} and \emph{up} contributions
in terms of the interior Eddington modes. Starting with
$G_{U}^{\text{in}}\left(x,x'\right)$, we may readily use the relation
between the \emph{in} Unruh modes and the \emph{right} Eddington
modes, as given in Eq.~\eqref{eq:gIN-fR}, which holds throughout
the BH interior. Then, Eq.~\eqref{eq:G_U^in} may be written as 
\begin{equation}
G_{U}^{\text{in}}\left(x,x'\right)=\hbar\sum_{l,m}\int_{0}^{\infty}\text{d}\omega\frac{\left|\omega_{+}\right|}{\omega}\left|\tau_{\omega lm}^{\text{in}}\right|^{2}\left\{ f_{\omega_{+}lm}^{R\left(+\right)}\left(x\right),f_{\omega_{+}lm}^{R\left(+\right)*}\left(x'\right)\right\} \,.\label{eq:G_U^in_int}
\end{equation}

The contribution from the \emph{up} Unruh modes is, as expected (see
discussion in Subsec.~\ref{subsec:Unruh-modes}), less straightforward
to decompose in terms of Eddington modes. We find it convenient
to start by writing $G_{U}^{\text{up}}\left(x,x'\right)$ in Eq.~\eqref{eq:G_U^up}
in terms of the two $g_{\hat{\omega}\hat{l}\hat{m}}^{\text{up}}$
components: $g_{\hat{\omega}\hat{l}\hat{m}}^{\text{past}}$ and $g_{\hat{\omega}\hat{l}\hat{m}}^{L}$
(see Eqs.~\eqref{eq:gL_asym}-\eqref{eq:gPast_asym}). Recalling
that 
\[
g_{\hat{\omega}\hat{l}\hat{m}}^{\text{up}}\left(x\right)=g_{\hat{\omega}\hat{l}\hat{m}}^{\text{past}}\left(x\right)+g_{\hat{\omega}\hat{l}\hat{m}}^{L}\left(x\right)\,,
\]
which is valid throughout the united domain, and substituting this
relation into Eq.~\eqref{eq:G_U^up}, we readily obtain

\begin{align}
G_{U}^{\text{up}}\left(x,x'\right) & =\hbar\sum_{\hat{J}}\int_{0}^{\infty}\text{d}\hat{\omega}\left[\left\{ g_{\hat{\omega}\hat{J}}^{\text{past}}\left(x\right),g_{\hat{\omega}\hat{J}}^{\text{past}*}\left(x'\right)\right\} +\left\{ g_{\hat{\omega}\hat{J}}^{L}\left(x\right),g_{\hat{\omega}\hat{J}}^{L*}\left(x'\right)\right\} \right.\label{eq:G_up_past+L}\\
 & \left.+\left\{ g_{\hat{\omega}\hat{J}}^{\text{past}}\left(x\right),g_{\hat{\omega}\hat{J}}^{L*}\left(x'\right)\right\} +\left\{ g_{\hat{\omega}\hat{J}}^{L}\left(x\right),g_{\hat{\omega}\hat{J}}^{\text{past}*}\left(x'\right)\right\} \right]\,,\nonumber 
\end{align}
where the integration is over the Kruskal frequency $\hat{\omega}$
and we use the notation previously introduced, $\hat{J}=\left(\hat{l},\hat{m}\right)$.

Aiming for a decomposition of both $g_{\hat{\omega}\hat{J}}^{\text{past}}$
and $g_{\hat{\omega}\hat{J}}^{L}$ in terms of Eddington
modes, we shall follow the same reasoning as in Sec.~\ref{sec:TPF_exterior},
where we defined coefficients relating the frequential and angular
factors of the Unruh and Eddington modes under consideration,
constrained to the relevant asymptotic null surfaces. The angular
coefficients $C_{\hat{J}J}^{\omega_{+}}$, defined in Eq.~\eqref{eq:C_def},
will be utilized in the BH interior exactly as they were in the BH
exterior. However, as we shall see, adjusting the various frequential
factors will require defining an additional set of Fourier coefficients,
along with the ones already defined in the BH exterior. For future
use, we rename the $\alpha_{\hat{\omega}\omega_{+}}$ coefficients,
as defined in Eq.~\eqref{eq:alpha_def}, by $\alpha_{\hat{\omega}\omega_{+}}^{\text{past}}$
(adding a superscript ``past''). That is, 
\begin{equation}
\alpha_{\hat{\omega}\omega_{+}}^{\text{past}}\equiv\int_{-\infty}^{\infty}\text{d}u_{\text{ext}}\,e^{-i\hat{\omega}U\left(u_{\text{ext}}\right)}e^{i\omega_{+}u_{\text{ext}}}\,,\label{eq:alpha^past_def}
\end{equation}
and it is explicitly given by (see Eq. \eqref{eq:alpha^past_res})
\begin{equation}
\alpha_{\hat{\omega}\omega_{+}}^{\text{past}}=\frac{1}{\kappa_{+}}\left(\frac{\hat{\omega}}{\kappa_{+}}\right)^{i\omega_{+}/\kappa_{+}}e^{\pi\omega_{+}/2\kappa_{+}}\Gamma\left(-i\frac{\omega_{+}}{\kappa_{+}}\right)\,.\label{eq:alpha^past_value}
\end{equation}

We shall begin with $g_{\hat{\omega}\hat{J}}^{\text{past}}$, emerging
from $H_{\text{past}}$ and vanishing on $H_{L}$ and PNI, hence identical
to $g_{\hat{\omega}\hat{J}}^{\text{up}}$ when restricted to the BH
exterior (compare Eqs.~\eqref{eq:gUP_asym} and \eqref{eq:gPast_asym}).
This allows us to relate to the analysis carried out in Sec.~\ref{sec:TPF_exterior},
and replace $g_{\hat{\omega}\hat{J}}^{\text{up}}$ by $g_{\hat{\omega}\hat{J}}^{\text{past}}$
in the LHS of Eq.~\eqref{eq:gUP-fUP}. Explicitly, this relates $g_{\hat{\omega}\hat{J}}^{\text{past}}$
and $f_{\omega_{+}J}^{\text{up}\left(+\right)}$ throughout the BH
exterior as follows:
\[
\sqrt{\hat{\omega}}\,g_{\hat{\omega}\hat{J}}^{\text{past}}=\frac{1}{2\pi}\sum_{J}\int_{-\infty}^{\infty}\text{d}\omega_{+}\sqrt{\left|\omega_{+}\right|}\, \alpha_{\hat{\omega}\omega_{+}}^{\text{past}}C_{\hat{J}J}^{\omega_{+}}f_{\omega_{+}J}^{\text{up}\left(+\right)}\,\,,\,\,\,\,\,\,\,\,\,\,r\geq r_{+}\,.
\]
Aiming at the BH interior, we carry the above relation over to the
common boundary of the BH exterior and interior -- namely the hypersurface
$H_{R}$ (the EH). This yields 
\begin{equation}
\sqrt{\hat{\omega}}\left.g_{\hat{\omega}\hat{J}}^{\text{past}}\right|_{H_{R}}=\frac{1}{2\pi}\sum_{J}\int_{-\infty}^{\infty}\text{d}\omega_{+}\sqrt{\left|\omega_{+}\right|}\, \alpha_{\hat{\omega}\omega_{+}}^{\text{past}}C_{\hat{J}J}^{\omega_{+}}\left.f_{\omega_{+}J}^{\text{up}\left(+\right)}\right|_{H_{R}}\,.\label{eq:g^past-f^up-at_HR}
\end{equation}
Now, we wish to re-express this in terms of the \emph{interior} Eddington
modes instead of the exterior \emph{up} modes. By comparing Eq.~\eqref{eq:fUP_future_asym}
with Eq.~\eqref{eq:fR_asym}, we register their relation on $H_{R}$:
\begin{equation}
\left.f_{\omega_{+}J}^{\text{up}\left(+\right)}\right|_{H_{R}}=\rho_{\omega_{+}J}^{\text{up}\left(+\right)}\left.f_{\omega_{+}J}^{R\left(+\right)}\right|_{H_{R}}\,.\label{eq:f_up-f_R-at_HR}
\end{equation}
Substituting this into Eq.~\eqref{eq:g^past-f^up-at_HR} we find
\begin{equation}
\sqrt{\hat{\omega}}\left.g_{\hat{\omega}\hat{J}}^{\text{past}}\right|_{H_{R}}=\frac{1}{2\pi}\sum_{J}\int_{-\infty}^{\infty}\text{d}\omega_{+}\sqrt{\left|\omega_{+}\right|}\, \alpha_{\hat{\omega}\omega_{+}}^{\text{past}}C_{\hat{J}J}^{\omega_{+}}\rho_{\omega_{+}J}^{\text{up}\left(+\right)}\left.f_{\omega_{+}J}^{R\left(+\right)}\right|_{H_{R}}\,.\label{eq:gPast-fR-HR}
\end{equation}
In addition, both $g_{\hat{\omega}\hat{J}}^{\text{past}}$ and $f_{\omega_{+}J}^{R\left(+\right)}$
vanish on $H_{L}$ (see Eqs.~\eqref{eq:gPast_asym} and \eqref{eq:fR_asym}),
implying that this relation actually holds throughout the BH interior:
\begin{equation}
\sqrt{\hat{\omega}}g_{\hat{\omega}\hat{J}}^{\text{past}}(x)=\frac{1}{2\pi}\sum_{J}\int_{-\infty}^{\infty}\text{d}\omega_{+}\sqrt{\left|\omega_{+}\right|}\, \alpha_{\hat{\omega}\omega_{+}}^{\text{past}}C_{\hat{J}J}^{\omega_{+}}\rho_{\omega_{+}J}^{\text{up}\left(+\right)}f_{\omega_{+}J}^{R\left(+\right)}(x)\,\,,\,\,\,\,\,\,\,\,\,\,r_{-}\leq r\leq r_{+}\,.\label{eq:gPast-fR}
\end{equation}

We now proceed similarly with $g_{\hat{\omega}\hat{J}}^{L}$, whose
form on $H_{L}$ is (see Eq.~\eqref{eq:gL_asym})

\begin{align*}
\left.g_{\hat{\omega}\hat{J}}^{L}\left(x\right)\right|_{H_{L}} & =\frac{Y_{\hat{J}}\left(\theta,\varphi_{+}\right)}{\sqrt{4\pi\hat{\omega}\left(r_{+}^{2}+a^{2}\right)}}e^{-i\hat{\omega}U\left(u_{\text{int}}\right)}\,,
\end{align*}
aiming to relate it to the \emph{left} Eddington mode,
whose form on $H_{L}$ is (see Eq.~\eqref{eq:fL_asymZ})

\[
\left.f_{\omega_{+}J}^{L\left(+\right)}\left(x\right)\right|_{H_{L}}=\frac{Z_{J}^{\omega}\left(\theta,\varphi_{+}\right)}{\sqrt{4\pi\left|\omega_{+}\right|\left(r_{+}^{2}+a^{2}\right)}}e^{i\omega_{+}u_{\text{int}}}\,.
\]

This resembles the case of decomposing $g_{\hat{\omega}\hat{J}}^{\text{up}}$
in terms of $f_{\omega_{+}J}^{\text{up}\left(+\right)}$ on $H_{\text{past}}$,
as carried out in Sec.~\ref{sec:TPF_exterior}, with a modification
in the frequential factors: here we have $e^{-i\hat{\omega}U\left(u_{\text{int}}\right)}$
and $e^{i\omega_{+}u_{\text{int}}}$ as the Unruh and Eddington
frequential factors, respectively, instead of $e^{-i\hat{\omega}U\left(u_{\text{ext}}\right)}$
and $e^{-i\omega_{+}u_{\text{ext}}}$, respectively.

We thus define new Fourier coefficients relating $e^{-i\hat{\omega}U\left(u_{\text{int}}\right)}$
and $e^{i\omega_{+}u_{\text{int}}}$, which we shall denote by $\alpha_{\hat{\omega}\omega_{+}}^{L}$,
given by the inverse Fourier transform: 
\begin{equation}
\alpha_{\hat{\omega}\omega_{+}}^{L}\equiv\int_{-\infty}^{\infty}\text{d}u_{\text{int}}e^{-i\hat{\omega}U\left(u_{\text{int}}\right)}e^{-i\omega_{+}u_{\text{int}}}\,.\label{eq:alpha^L}
\end{equation}
This integral may be found by inspecting Eq.~\eqref{eq:alpha^L}
alongside Eq.~\eqref{eq:alpha^past_def}, changing the integration variable from $u_{\text{int}}$ to $-u_{\text{ext}}$ (which also implies $U(u_{\text{int}})\mapsto -U(u_{\text{ext}}$), see Eqs.~\eqref{eq:extKrusCoor}-\eqref{eq:intKrusCoor}), which results in $\alpha_{\hat{\omega}\omega_{+}}^{L}=\alpha_{\hat{\omega}\left(-\omega_{+}\right)}^{\text{past}*}$.
Applying this relation to Eq.~\eqref{eq:alpha^past_value} then yields
 \footnote{This expression is also given in Eq.~(3.5) in Ref.~\cite{Group:2018},
however, there are some notational differences in this case that need
to be bridged: Basically, the notation in Ref.~\cite{Group:2018}
is related to ours as prescribed in footnote \ref{fn:alpha^past_notation},
namely $\tilde{\omega}\mapsto\omega_{+},\omega\mapsto\hat{\omega}$.
However, in the RHS of Eq.~\eqref{eq:alpha^L_res} our $\omega_{+}$
should actually be mapped to $-\tilde{\omega}$ (this extra change
of sign has to do with the issue discussed earlier in footnote \ref{fn:omega_unlike_RN}).\label{fn:alpha^L_notation}}:
\begin{equation}
\alpha_{\hat{\omega}\omega_{+}}^{L}=\frac{1}{\kappa_{+}}\left(\frac{\hat{\omega}}{\kappa_{+}}\right)^{i\omega_{+}/\kappa_{+}}e^{-\pi\omega_{+}/2\kappa_{+}}\Gamma\left(-i\frac{\omega_{+}}{\kappa_{+}}\right)\,.\label{eq:alpha^L_res}
\end{equation} 

Comparing Eq.~\eqref{eq:alpha^L_res} with Eq.~\eqref{eq:alpha^past_value},
we find $\alpha_{\hat{\omega}\omega_{+}}^{L}$ in terms of $\alpha_{\hat{\omega}\omega_{+}}^{\text{past}}$:

\begin{equation}
\alpha_{\hat{\omega}\omega_{+}}^{L}=\alpha_{\hat{\omega}\omega_{+}}^{\text{past}}e^{-\pi\omega_{+}/\kappa_{+}}\label{eq:alpha^L_ito_alpha^past}
\end{equation}
 \footnote{One might be concerned about the notable difference between our Eq.~\eqref{eq:alpha^L_ito_alpha^past}, describing the relation between
$\alpha_{\hat{\omega}\omega_{+}}^{L}$ and $\alpha_{\hat{\omega}\omega_{+}}^{\text{past}}$,
and the rightmost side of Eq.~(3.5) in Ref.~\cite{Group:2018}
(which states that $\alpha_{\omega\tilde{\omega}}^{L}=\alpha_{\omega\tilde{\omega}}^{\text{past}*}$).
To reconcile this difference, note that when transforming our $\alpha_{\hat{\omega}\omega_{+}}^{\text{past}}$
and $\alpha_{\hat{\omega}\omega_{+}}^{L}$ to the notation of Ref.~\cite{Group:2018}, the former becomes $\alpha_{\omega\tilde{\omega}}^{\text{past}}$,
but $\alpha_{\hat{\omega}\omega_{+}}^{L}$ is translated to $\alpha_{\omega\left(-\tilde{\omega}\right)}^{L}$
(see footnotes \ref{fn:alpha^past_notation} and \ref{fn:alpha^L_notation},
as well as \ref{fn:omega_unlike_RN}). Indeed, considering the relation
between $\alpha_{\omega\left(-\tilde{\omega}\right)}^{L}$ and $\alpha_{\omega\tilde{\omega}}^{\text{past}}$
therein would yield the analog of Eq.~\eqref{eq:alpha^L_ito_alpha^past}
(namely, $\alpha_{\omega\left(-\tilde{\omega}\right)}^{L}=\alpha_{\omega\tilde{\omega}}^{\text{past}}e^{-\pi\tilde{\omega}/\kappa_{+}}$).}.

The two $H_{L}$-projected functions $\left.g_{\hat{\omega}\hat{J}}^{L}\left(x\right)\right|_{H_{L}}$
and $\left.f_{\omega_{+}J}^{L\left(+\right)}\left(x\right)\right|_{H_{L}}$
may now be related using the conversion coefficients $\alpha_{\hat{\omega}\omega_{+}}^{L}$
and $C_{\hat{J}J}^{\omega_{+}}$ (in full analogy with Eq.~\eqref{eq:gUP-fUP},
replacing $\alpha_{\hat{\omega}\omega_{+}}=\alpha_{\hat{\omega}\omega_{+}}^{\text{past}}$
by $\alpha_{\hat{\omega}\omega_{+}}^{L}$): 
\begin{equation}
\sqrt{\hat{\omega}}\left.g_{\hat{\omega}\hat{J}}^{L}\right|_{H_{L}}=\frac{1}{2\pi}\sum_{J}\int_{-\infty}^{\infty}\text{d}\omega_{+}\sqrt{\left|\omega_{+}\right|}\alpha_{\hat{\omega}\omega_{+}}^{L}C_{\hat{J}J}^{\omega_{+}}\left.f_{\omega_{+}J}^{L\left(+\right)}\right|_{H_{L}}\,.\label{eq:gL-fL-HL}
\end{equation}
Since both the \emph{left} Unruh and Eddington modes vanish
on $H_{R}$, this relation between $g_{\hat{\omega}\hat{J}}^{L}$
and $f_{\omega_{+}J}^{L\left(+\right)}$ actually holds throughout
the BH interior: 
\begin{equation}
\sqrt{\hat{\omega}}g_{\hat{\omega}\hat{J}}^{L}\left(x\right)=\frac{1}{2\pi}\sum_{J}\int_{-\infty}^{\infty}\text{d}\omega_{+}\sqrt{\left|\omega_{+}\right|}\alpha_{\hat{\omega}\omega_{+}}^{L}C_{\hat{J}J}^{\omega_{+}}f_{\omega_{+}J}^{L\left(+\right)}(x)\,\,,\,\,\,\,\,\,\,\,r_{-}\leq r\leq r_{+}\,.\label{eq:gL-fL}
\end{equation}

Now, substituting Eqs.~\eqref{eq:gPast-fR} and \eqref{eq:gL-fL}
into Eq.~\eqref{eq:G_up_past+L}, $G_{U}^{\text{up}}\left(x,x'\right)$
may be written as 
\begin{equation}
G_{U}^{\text{up}}\left(x,x'\right)=\hbar\left(I_{RR}+I_{LL}+I_{RL}+I_{LR}\right)\,,\label{eq:Gup_4I}
\end{equation}
where

\begin{align}
 & I_{RR}\equiv\label{eq:I_RR}\\
 & \sum_{\hat{J}}\int_{0}^{\infty}\frac{\text{d}\hat{\omega}}{4\pi^{2}\hat{\omega}}\sum_{J}\int_{-\infty}^{\infty}\text{d}\omega_{+}\sqrt{\left|\omega_{+}\right|}\alpha_{\hat{\omega}\omega_{+}}^{\text{past}}C_{\hat{J}J}^{\omega_{+}}\rho_{\omega_{+}J}^{\text{up}\left(+\right)}\sum_{J'}\int_{-\infty}^{\infty}\text{d}\omega'_{+}\sqrt{\left|\omega'_{+}\right|}\alpha_{\hat{\omega}\omega'_{+}}^{\text{past}*}C_{\hat{J}J'}^{\omega'_{+}*}\rho_{\omega'_{+}J'}^{\text{up}\left(+\right)*}\left\{ f_{\omega_{+}J}^{R\left(+\right)}\left(x\right),f_{\omega'_{+}J'}^{R\left(+\right)*}\left(x'\right)\right\} \,,\nonumber 
\end{align}
\begin{align}
I_{LL} & \equiv\sum_{\hat{J}}\int_{0}^{\infty}\frac{\text{d}\hat{\omega}}{4\pi^{2}\hat{\omega}}\sum_{J}\int_{-\infty}^{\infty}\text{d}\omega_{+}\sqrt{\left|\omega_{+}\right|}\alpha_{\hat{\omega}\omega_{+}}^{L}C_{\hat{J}J}^{\omega_{+}}\sum_{J'}\int_{-\infty}^{\infty}\text{d}\omega'_{+}\sqrt{\left|\omega'_{+}\right|}\alpha_{\hat{\omega}\omega'_{+}}^{L*}C_{\hat{J}J'}^{\omega'_{+}*}\left\{ f_{\omega_{+}J}^{L\left(+\right)}\left(x\right),f_{\omega'_{+}J'}^{L\left(+\right)*}\left(x'\right)\right\} \,,\label{eq:I_LL}
\end{align}

\begin{align}
I_{RL} & \equiv\sum_{\hat{J}}\int_{0}^{\infty}\frac{\text{d}\hat{\omega}}{4\pi^{2}\hat{\omega}}\sum_{J}\int_{-\infty}^{\infty}\text{d}\omega_{+}\sqrt{\left|\omega_{+}\right|}\alpha_{\hat{\omega}\omega_{+}}^{\text{past}}C_{\hat{J}J}^{\omega_{+}}\rho_{\omega_{+}J}^{\text{up}\left(+\right)}\sum_{J'}\int_{-\infty}^{\infty}\text{d}\omega'_{+}\sqrt{\left|\omega'_{+}\right|}\alpha_{\hat{\omega}\omega'_{+}}^{L*}C_{\hat{J}J'}^{\omega'_{+}*}\left\{ f_{\omega_{+}J}^{R\left(+\right)}\left(x\right),f_{\omega'_{+}J'}^{L\left(+\right)*}\left(x'\right)\right\} \,,\label{eq:I_RL}
\end{align}

\begin{align}
I_{LR} & \equiv\sum_{\hat{J}}\int_{0}^{\infty}\frac{\text{d}\hat{\omega}}{4\pi^{2}\hat{\omega}}\sum_{J}\int_{-\infty}^{\infty}\text{d}\omega_{+}\sqrt{\left|\omega_{+}\right|}\alpha_{\hat{\omega}\omega_{+}}^{L}C_{\hat{J}J}^{\omega_{+}}\sum_{J'}\int_{-\infty}^{\infty}\text{d}\omega'_{+}\sqrt{\left|\omega'_{+}\right|}\alpha_{\hat{\omega}\omega'_{+}}^{\text{past}*}C_{\hat{J}J'}^{\omega'_{+}*}\rho_{\omega'_{+}J'}^{\text{up}\left(+\right)*}\left\{ f_{\omega_{+}J}^{L\left(+\right)}\left(x\right),f_{\omega'_{+}J'}^{R\left(+\right)*}\left(x'\right)\right\} \,.\label{eq:I_LR}
\end{align}
We rearrange Eqs.~\eqref{eq:I_RR}-\eqref{eq:I_LR} into a form similar
to that of Eq.~\eqref{eq:Gup2}:{\small{}{} 
\begin{align}
I_{RR} & =\frac{1}{4\pi^{2}}\sum_{J}\int_{-\infty}^{\infty}\text{d}\omega_{+}\sqrt{\left|\omega_{+}\right|}\sum_{J'}\int_{-\infty}^{\infty}\text{d}\omega'_{+}\sqrt{\left|\omega'_{+}\right|}\rho_{\omega_{+}J}^{\text{up}\left(+\right)}\rho_{\omega'_{+}J'}^{\text{up}\left(+\right)*}\cdot\nonumber \\
 & \left\{ f_{\omega_{+}J}^{R\left(+\right)}\left(x\right),f_{\omega'_{+}J'}^{R\left(+\right)*}\left(x'\right)\right\} \sum_{\hat{J}}C_{\hat{J}J}^{\omega_{+}}C_{\hat{J}J'}^{\omega'_{+}*}\int_{0}^{\infty}\frac{\text{d}\hat{\omega}}{\hat{\omega}}\alpha_{\hat{\omega}\omega_{+}}^{\text{past}}\alpha_{\hat{\omega}\omega'_{+}}^{\text{past}*}\,,\label{eq:I_RR1}
\end{align}
\begin{align}
I_{LL} & =\frac{1}{4\pi^{2}}\sum_{J}\int_{-\infty}^{\infty}\text{d}\omega_{+}\sqrt{\left|\omega_{+}\right|}\sum_{J'}\int_{-\infty}^{\infty}\text{d}\omega'_{+}\sqrt{\left|\omega'_{+}\right|}\cdot\nonumber \\
 & \left\{ f_{\omega_{+}J}^{L\left(+\right)}\left(x\right),f_{\omega'_{+}J'}^{L\left(+\right)*}\left(x'\right)\right\} \sum_{\hat{J}}C_{\hat{J}J}^{\omega_{+}}C_{\hat{J}J'}^{\omega'_{+}*}\int_{0}^{\infty}\frac{\text{d}\hat{\omega}}{\hat{\omega}}\alpha_{\hat{\omega}\omega_{+}}^{L}\alpha_{\hat{\omega}\omega'_{+}}^{L*}\,,\label{eq:I_LL1}
\end{align}
\begin{align}
I_{RL} & =\frac{1}{4\pi^{2}}\sum_{J}\int_{-\infty}^{\infty}\text{d}\omega_{+}\sqrt{\left|\omega_{+}\right|}\sum_{J'}\int_{-\infty}^{\infty}\text{d}\omega'_{+}\sqrt{\left|\omega'_{+}\right|}\rho_{\omega_{+}J}^{\text{up}\left(+\right)}\cdot\nonumber \\
 & \left\{ f_{\omega_{+}J}^{R\left(+\right)}\left(x\right),f_{\omega'_{+}J'}^{L\left(+\right)*}\left(x'\right)\right\} \sum_{\hat{J}}C_{\hat{J}J}^{\omega_{+}}C_{\hat{J}J'}^{\omega'_{+}*}\int_{0}^{\infty}\frac{\text{d}\hat{\omega}}{\hat{\omega}}\alpha_{\hat{\omega}\omega_{+}}^{\text{past}}\alpha_{\hat{\omega}\omega'_{+}}^{L*}\,,\label{eq:I_RL1}
\end{align}
\begin{align}
I_{LR} & =\frac{1}{4\pi^{2}}\sum_{J}\int_{-\infty}^{\infty}\text{d}\omega_{+}\sqrt{\left|\omega_{+}\right|}\sum_{J'}\int_{-\infty}^{\infty}\text{d}\omega'_{+}\sqrt{\left|\omega'_{+}\right|}\rho_{\omega'_{+}J'}^{\text{up}\left(+\right)*}\cdot\nonumber \\
 & \left\{ f_{\omega_{+}J}^{L\left(+\right)}\left(x\right),f_{\omega'_{+}J'}^{R\left(+\right)*}\left(x'\right)\right\} \sum_{\hat{J}}C_{\hat{J}J}^{\omega_{+}}C_{\hat{J}J'}^{\omega'_{+}*}\int_{0}^{\infty}\frac{\text{d}\hat{\omega}}{\hat{\omega}}\alpha_{\hat{\omega}\omega_{+}}^{L}\alpha_{\hat{\omega}\omega'_{+}}^{\text{past}*}\,.\label{eq:I_LR1}
\end{align}
}At this point it becomes clear (after renaming the indices $\omega_{+}\leftrightarrow\omega'_{+}$
and $J\leftrightarrow J'$ in the last equation) that $I_{LR}=I_{RL}^{*}$.

In order to proceed as we did in the BH exterior, we need to perform
the integrals on the rightmost side of each equality, of the form
$\int_{0}^{\infty}\frac{\text{d}\hat{\omega}}{\hat{\omega}}\alpha_{\hat{\omega}\omega_{+}}^{\Lambda_{1}}\alpha_{\hat{\omega}\omega'_{+}}^{\Lambda_{2}*}$
with $\Lambda_{1,2}$ either ``past" or ``\emph{L}". The result
of the integral $\int_{0}^{\infty}\frac{\text{d}\hat{\omega}}{\hat{\omega}}\alpha_{\hat{\omega}\omega_{+}}^{\text{past}}\alpha_{\hat{\omega}\omega'_{+}}^{\text{past}*}$
has already been given in Eq.~\eqref{eq:int_alpha_alpha*}. Next,
we use the relation between $\alpha_{\hat{\omega}\omega_{+}}^{L}$
and $\alpha_{\hat{\omega}\omega_{+}}^{\text{past}}$ given in Eq.
\eqref{eq:alpha^L_ito_alpha^past}, along with the previously mentioned
integral (Eq.~\eqref{eq:int_alpha_alpha*}), to find 
\begin{equation}
\int_{0}^{\infty}\frac{\text{d}\hat{\omega}}{\hat{\omega}}\alpha_{\hat{\omega}\omega_{+}}^{L}\alpha_{\hat{\omega}\omega'_{+}}^{L*}=e^{-\pi\left(\omega_{+}+\omega'_{+}\right)/\kappa_{+}}\int_{0}^{\infty}\frac{\text{d}\hat{\omega}}{\hat{\omega}}\alpha_{\hat{\omega}\omega_{+}}^{\text{past}}\alpha_{\hat{\omega}\omega'_{+}}^{\text{past}*}=\frac{4\pi^{2}}{\omega_{+}}\frac{1}{e^{2\pi\omega_{+}/\kappa_{+}}-1}\delta\left(\omega_{+}-\omega'_{+}\right)\label{eq:int_alphaL_alphaL*}
\end{equation}
and 
\begin{equation}
\int_{0}^{\infty}\frac{\text{d}\hat{\omega}}{\hat{\omega}}\alpha_{\hat{\omega}\omega_{+}}^{\text{past}}\alpha_{\hat{\omega}\omega'_{+}}^{L*}=e^{-\pi\omega'_{+}/\kappa_{+}}\int_{0}^{\infty}\frac{\text{d}\hat{\omega}}{\hat{\omega}}\alpha_{\hat{\omega}\omega_{+}}^{\text{past}}\alpha_{\hat{\omega}\omega'_{+}}^{\text{past}*}=\frac{4\pi^{2}}{\omega_{+}}\frac{1}{e^{\pi\omega_{+}/\kappa_{+}}-e^{-\pi\omega_{+}/\kappa_{+}}}\delta\left(\omega_{+}-\omega'_{+}\right)\,.\label{eq:int_alphaP_alphaL*}
\end{equation}
Performing a computation very similar to the one carried out in Eq.~\eqref{eq:Gup3}
in the BH exterior, substituting the integrals in Eqs.~\eqref{eq:int_alpha_alpha*},
\eqref{eq:int_alphaL_alphaL*} and \eqref{eq:int_alphaP_alphaL*}
into Eqs.~\eqref{eq:I_RR1}, \eqref{eq:I_LL1} and \eqref{eq:I_RL1}
respectively, and making use of Eq.~\eqref{eq:sum_C*C}, we obtain,
after translating the exponential factors into corresponding hyperbolic-geometric
functions:

{\small{}{} 
\begin{align}
I_{RR} & =\frac{1}{2}\sum_{J}\int_{-\infty}^{\infty}\text{d}\omega_{+}\frac{\left|\omega_{+}\right|}{\omega_{+}}\left[\coth\left(\frac{\pi\omega_{+}}{\kappa_{+}}\right)+1\right]\left[\left|\rho_{\omega_{+}J}^{\text{up}\left(+\right)}\right|^{2}\left\{ f_{\omega_{+}J}^{R\left(+\right)}\left(x\right),f_{\omega{}_{+}J}^{R\left(+\right)*}\left(x'\right)\right\} \right]\,,\label{eq:I_RR2}
\end{align}
\begin{align}
I_{LL} & =\frac{1}{2}\sum_{J}\int_{-\infty}^{\infty}\text{d}\omega_{+}\frac{\left|\omega_{+}\right|}{\omega_{+}}\left[\coth\left(\frac{\pi\omega_{+}}{\kappa_{+}}\right)-1\right]\left[\left\{ f_{\omega_{+}J}^{L\left(+\right)}\left(x\right),f_{\omega{}_{+}J}^{L\left(+\right)*}\left(x'\right)\right\} \right]\,,\label{eq:I_LL2}
\end{align}
and 
\begin{equation}
I_{RL}=I_{LR}^{*}=\frac{1}{2}\sum_{J}\int_{-\infty}^{\infty}\text{d}\omega_{+}\frac{\left|\omega_{+}\right|}{\omega_{+}}\left[\text{cosech}\left(\frac{\pi\omega_{+}}{\kappa_{+}}\right)\right]\left[\rho_{\omega_{+}J}^{\text{up}\left(+\right)}\left\{ f_{\omega_{+}J}^{R\left(+\right)}\left(x\right),f_{\omega{}_{+}J}^{L\left(+\right)*}\left(x'\right)\right\} \right]\,,\label{eq:I_RL2}
\end{equation}
}where $\text{cosech}\equiv1/\sinh$.

Next, we would like to fold these three integrals through $\omega_{+}=0$,
just as we did in Sec.~\ref{sec:TPF_exterior} for the BH exterior.
To this end, we first note that in all three equations \eqref{eq:I_RR2}-\eqref{eq:I_RL2},
the RHS is of the general form 
\begin{equation}
I\equiv\frac{1}{2}\sum_{J}\int_{-\infty}^{\infty}\text{d}\omega_{+}\frac{\left|\omega_{+}\right|}{\omega_{+}}H\left(\omega_{+}\right)F_{\omega_{+}J}\left(x,x'\right)=\frac{1}{2}\sum_{l,m}\int_{-\infty}^{\infty}\text{d}\omega_{+}\,\text{sign}\left(\omega_{+}\right)H\left(\omega_{+}\right)F_{\omega_{+}lm}\left(x,x'\right)\,,\label{eq:general_structure}
\end{equation}
where $H\left(\omega_{+}\right)$ and $F_{\omega_{+}lm}\left(x,x'\right)$
respectively stand for the first and second terms in square brackets
in each of these three equations. Furthermore, we once again note
that since $\left(\omega,m\right)\mapsto\left(-\omega,-m\right)$
is identical to $\left(\omega_{+},m\right)\mapsto\left(-\omega_{+},-m\right)$,
we may rewrite Eq.~\eqref{eq:IntEddModesInvariance} as 
\begin{equation}
f_{\left(-\omega_{+}\right)l\left(-m\right)}^{\Lambda\left(+\right)}=\left(-1\right){}^{m}f_{\omega_{+}lm}^{\Lambda\left(+\right)*}\label{eq:int_f_wp_invariance}
\end{equation}
(with $\Lambda$ either ``\emph{R}" or ``\emph{L}") and the invariance relation
of $\rho^{\text{up}}$ included in Eq.~\eqref{eq:rho_inv} as
\begin{equation}
\rho_{\left(-\omega_{+}\right)l\left(-m\right)}^{\text{up}\left(+\right)}=\rho_{\omega_{+}lm}^{\text{up}\left(+\right)*}\,.\label{eq:rho^up_wp_invariance}
\end{equation}
Eqs. \eqref{eq:int_f_wp_invariance} and \eqref{eq:rho^up_wp_invariance}
may now be used to show that the function $F_{\omega_{+}lm}\left(x,x'\right)$
in all three cases \eqref{eq:I_RR2}-\eqref{eq:I_RL2} satisfies 
\[
F_{\left(-\omega_{+}\right)lm}\left(x,x'\right)=F_{\omega_{+}l\left(-m\right)}^{*}\left(x,x'\right)\,.
\]
Then, summing over $l$ and $m$ (recalling $\sum_{l,m}=\sum_{l=0}^{\infty}\sum_{m=-l}^{l}$),
we obtain 
\[
\sum_{l,m}F_{\left(-\omega_{+}\right)lm}\left(x,x'\right)=\sum_{l,m}F_{\omega_{+}lm}^{*}\left(x,x'\right)=\sum_{J}F_{\omega_{+}J}^{*}\left(x,x'\right)\,.
\]
Therefore, by folding the $\omega_{+}$-integral in Eq.~\eqref{eq:general_structure},
we obtain the following explicit form: 
\begin{equation}
I=\frac{1}{2}\sum_{J}\int_{0}^{\infty}\text{d}\omega_{+}\left[H\left(\omega_{+}\right)F_{\omega_{+}J}\left(x,x'\right)-H\left(-\omega_{+}\right)F_{\omega_{+}J}^{*}\left(x,x'\right)\right]\,.\label{eq:general_folding}
\end{equation}

Note that the function $H\left(\omega_{+}\right)$ potentially has
both a part that is an even function of $\omega_{+}$and a part that
is an odd function. Then, from Eq.~\eqref{eq:general_folding} it
is clear that the even part of $H\left(\omega_{+}\right)$ leaves
out the imaginary part of $F_{\omega_{+}J}\left(x,x'\right)$, while
the odd part of $H\left(\omega_{+}\right)$ leaves out the real part
of $F_{\omega_{+}J}\left(x,x'\right)$. Applying this general folding
structure to Eqs.~\eqref{eq:I_RR2}-\eqref{eq:I_RL2} and noticing
that $F_{\omega_{+}J}\left(x,x'\right)$ is actually real for $I_{RR}$
and $I_{LL}$ (Eqs.~\eqref{eq:I_RR2} and \eqref{eq:I_LL2}, while
for Eq.~\eqref{eq:I_RL2} this is not the case), we obtain the three
desired folded integrals:

\begin{align}
I_{RR} & =\sum_{J}\int_{0}^{\infty}\text{d}\omega_{+}\,\coth\left(\frac{\pi\omega_{+}}{\kappa_{+}}\right)\left|\rho_{\omega_{+}J}^{\text{up}\left(+\right)}\right|^{2}\left\{ f_{\omega_{+}J}^{R\left(+\right)}\left(x\right),f_{\omega{}_{+}J}^{R\left(+\right)*}\left(x'\right)\right\} \,,\label{eq:I_RR_fin}
\end{align}
\begin{equation}
I_{LL}=\sum_{J}\int_{0}^{\infty}\text{d}\omega_{+}\,\coth\left(\frac{\pi\omega_{+}}{\kappa_{+}}\right)\left\{ f_{\omega_{+}J}^{L\left(+\right)}\left(x\right),f_{\omega_{+}J}^{L\left(+\right)*}\left(x'\right)\right\} \,,\label{eq:I_LL_fin}
\end{equation}
and 
\begin{equation}
I_{RL}=I_{LR}=\sum_{J}\int_{0}^{\infty}\text{d}\omega_{+}\,\text{cosech}\left(\frac{\pi\omega_{+}}{\kappa_{+}}\right)\Re\left(\rho_{\omega_{+}J}^{\text{up}\left(+\right)}\left\{ f_{\omega_{+}J}^{R\left(+\right)}\left(x\right),f_{\omega_{+}J}^{L\left(+\right)*}\left(x'\right)\right\} \right)\,.\label{eq:I_RL_fin}
\end{equation}
Notably, all four individual contributions to $G_{U}^{\text{up}}\left(x,x'\right)$,
namely $I_{RR}$, $I_{LL}$, and $I_{RL}=I_{LR}$, are real. {\small{}{}}Combining
now Eqs.~\eqref{eq:I_RR_fin}-\eqref{eq:I_RL_fin}, we obtain the
\emph{up} contribution: 
\begin{align}
 & G_{U}^{\text{up}}\left(x,x'\right)=\nonumber \\
 & \hbar\sum_{l,m}\int_{0}^{\infty}\text{d}\omega_{+}\left[\coth\left(\frac{\pi\omega_{+}}{\kappa_{+}}\right)\left(\left\{ f_{\omega_{+}lm}^{L\left(+\right)}\left(x\right),f_{\omega_{+}lm}^{L\left(+\right)*}\left(x'\right)\right\} +\left|\rho_{\omega_{+}lm}^{\text{up}\left(+\right)}\right|^{2}\left\{ f_{\omega_{+}lm}^{R\left(+\right)}\left(x\right),f_{\omega_{+}lm}^{R\left(+\right)*}\left(x'\right)\right\} \right)\right.\nonumber \\
 & \left.+2\, \text{cosech}\left(\frac{\pi\omega_{+}}{\kappa_{+}}\right)\Re\left(\rho_{\omega_{+}lm}^{\text{up}\left(+\right)}\left\{ f_{\omega_{+}lm}^{R\left(+\right)}\left(x\right),f_{\omega_{+}lm}^{L\left(+\right)*}\left(x'\right)\right\} \right)\right]\,.\label{eq:G_U^up_int}
\end{align}

Finally, combining the \emph{in} contribution (Eq.~\eqref{eq:G_U^in_int})
with the \emph{up} contribution (Eq.~\eqref{eq:G_U^up_int}), we
obtain the full HTPF in the BH interior: 

\begin{align}
 & G_{U}^{(1)}\left(x,x'\right)=\nonumber \\
 & \hbar\sum_{l,m}\int_{0}^{\infty}\text{d}\omega_{+}\left[\coth\left(\frac{\pi\omega_{+}}{\kappa_{+}}\right)\left(\left\{ f_{\omega_{+}lm}^{L\left(+\right)}\left(x\right),f_{\omega_{+}lm}^{L\left(+\right)*}\left(x'\right)\right\} +\left|\rho_{\omega_{+}lm}^{\text{up}\left(+\right)}\right|^{2}\left\{ f_{\omega_{+}lm}^{R\left(+\right)}\left(x\right),f_{\omega_{+}lm}^{R\left(+\right)*}\left(x'\right)\right\} \right)\right.\nonumber \\
 & \left.+2\, \text{cosech}\left(\frac{\pi\omega_{+}}{\kappa_{+}}\right)\Re\left(\rho_{\omega_{+}lm}^{\text{up}\left(+\right)}\left\{ f_{\omega_{+}lm}^{R\left(+\right)}\left(x\right),f_{\omega_{+}lm}^{L\left(+\right)*}\left(x'\right)\right\} \right)\right]+\hbar\sum_{l,m}\int_{0}^{\infty}\text{d}\omega\frac{\left|\omega_{+}\right|}{\omega}\left|\tau_{\omega lm}^{\text{in}}\right|^{2}\left\{ f_{\omega_{+}lm}^{R\left(+\right)}\left(x\right),f_{\omega_{+}lm}^{R\left(+\right)*}\left(x'\right)\right\} \,,\label{eq:G_U_int_wp}
\end{align}
and, translating back to the standard $\omega$-indexing notation,
we reach our final result:
\begin{align}
 & G_{U}^{(1)}\left(x,x'\right)=\nonumber \\
 & \hbar\sum_{l,m}\int_{0}^{\infty}\text{d}\omega_{+}\left[\coth\left(\frac{\pi\omega_{+}}{\kappa_{+}}\right)\left(\left\{ f_{\omega lm}^{L}\left(x\right),f_{\omega lm}^{L*}\left(x'\right)\right\} +\left|\rho_{\omega lm}^{\text{up}}\right|^{2}\left\{ f_{\omega lm}^{R}\left(x\right),f_{\omega lm}^{R*}\left(x'\right)\right\} \right)\right.\nonumber \\
 & \left.+2\, \text{cosech}\left(\frac{\pi\omega_{+}}{\kappa_{+}}\right)\Re\left(\rho_{\omega lm}^{\text{up}}\left\{ f_{\omega lm}^{R}\left(x\right),f_{\omega lm}^{L*}\left(x'\right)\right\} \right)\right]+\hbar\sum_{l,m}\int_{0}^{\infty}\text{d}\omega\frac{\left|\omega_{+}\right|}{\omega}\left|\tau_{\omega lm}^{\text{in}}\right|^{2}\left\{ f_{\omega lm}^{R}\left(x\right),f_{\omega lm}^{R*}\left(x'\right)\right\} \,.\label{eq:G_U_int}
\end{align}

One may be concerned about the apparent singularity of the mode-sum at $\omega_{+}\to0$  (and, similarly, at $\omega\to0$). In Appendix~\ref{App:small freq}, we address this issue
and show that the integrands in Eq.~\eqref{eq:G_U_int} are in fact entirely
regular at both limits.

\subsection{Invariance to the choice of the angular functions $\hat{Z}_{\hat{l}\hat{m}}^{\hat{\omega}}\left(\theta,\varphi_{+}\right)$ }

\label{subsec:Invariance-to-Zwlm-choice-int}

In Subsec.~\ref{subsec:Invariance-to-Zwlm-choice} we established that
in the construction of the HTPF outside the BH, the final result remains
unchanged if in the definition of the \emph{up} Unruh modes (described
in Subsec.~\ref{subsec:The-up-Unruh}) one replaces the spherical harmonics
$Y_{\hat{l}\hat{m}}\left(\theta,\varphi_{+}\right)$ by any other
complete orthonormal set of angular functions $\hat{Z}_{\hat{l}\hat{m}}^{\hat{\omega}}\left(\theta,\varphi_{+}\right)$
(fulfilling Eq.~\eqref{eq:Zhat_ort}). As one can easily verify,
all the arguments and considerations made there are equally valid
for the interior.

More specifically, in order to carry out this generalization in the
construction of the interior HTPF, throughout the analysis in the
present section one simply has to replace everywhere $Y_{\hat{l}\hat{m}}$
by $\hat{Z}_{\hat{l}\hat{m}}^{\hat{\omega}}$ and $C_{\hat{J}J}^{\omega_{+}}$
by $C_{\hat{J}J}^{\hat{\omega}\omega_{+}}$ (and similarly replace
$Y_{\hat{J}}$ by $\hat{Z}_{\hat{J}}^{\hat{\omega}}$). Then, recalling
Eq.~\eqref{eq:sum_C*C2}, in (the $Y_{\hat{J}}\mapsto\hat{Z}_{\hat{J}}^{\hat{\omega}}$
counterparts of) each of the four equations \eqref{eq:I_RR1}-\eqref{eq:I_LR1},
the factor $\sum_{\hat{J}}C_{\hat{J}J}^{\hat{\omega}\omega_{+}}C_{\hat{J}J'}^{\hat{\omega}\omega'_{+}*}$
may be justifiably kept out of the $\hat{\omega}$ integral. From
that point on, making use of Eq.~\eqref{eq:sum_C*C1}, the analysis
proceeds with no further modifications.

We conclude that the result in Eq.~\eqref{eq:G_U_int} for the mode-sum
expression of the Unruh HTPF inside the BH is invariant with respect
to the choice of the initial angular functions $\hat{Z}_{\hat{l}\hat{m}}^{\hat{\omega}}\left(\theta,\varphi_{+}\right)$
in the \emph{up} Unruh modes, as anticipated.

\subsection{Alternative forms of the final result}

We propose here alternative forms of the final result for the HTPF
given in Eq.~\eqref{eq:G_U_int}, which may prove to be useful in
future applications. In particular, we shall provide an expression
in which the integral over $\omega_{+}$ (arising from the \emph{up} contribution) is replaced by an integral over $\omega$. To this
end, we shall proceed as follows:

We begin by introducing the functions $\tilde{f}_{\omega lm}^{R}$
and $\tilde{f}_{\omega lm}^{L}$ related to the standard interior
Eddington modes $f_{\omega lm}^{R}$ and $f_{\omega lm}^{L}$
(see Eq.~\eqref{eq:EddModesRL}) by eliminating the normalization
factor, that is: 
\begin{align}
&\tilde{f}_{\omega lm}^{R}\equiv\sqrt{4\pi\left|\omega_{+}\right|}f_{\omega lm}^{R}=\frac{1}{\sqrt{r^{2}+a^{2}}}Z_{lm}^{\omega}\left(\theta,\varphi\right)e^{-i\omega t}\psi_{\omega lm}^{\text{int}}\, ,\nonumber \\
&\tilde{f}_{\omega lm}^{L}\equiv\sqrt{4\pi\left|\omega_{+}\right|}f_{\omega lm}^{L}=\frac{1}{\sqrt{r^{2}+a^{2}}}Z_{lm}^{\omega}\left(\theta,\varphi\right)e^{-i\omega t}\psi_{\omega lm}^{\text{int}*}\,.\label{eq:fLtilde}
\end{align}

Rewriting Eq.~\eqref{eq:G_U_int} in terms of $\tilde{f}_{\omega lm}^{R}$
and $\tilde{f}_{\omega lm}^{L}$, we obtain 
\begin{align}
 & G_{U}^{(1)}\left(x,x'\right)=\nonumber \\
 & \hbar\sum_{l,m}\int_{0}^{\infty}\frac{\text{d}\omega_{+}}{4\pi\omega_{+}}\left[\coth\left(\frac{\pi\omega_{+}}{\kappa_{+}}\right)\left(\left\{ \tilde{f}_{\omega lm}^{L}\left(x\right),\tilde{f}_{\omega lm}^{L*}\left(x'\right)\right\} +\left|\rho_{\omega lm}^{\text{up}}\right|^{2}\left\{ \tilde{f}_{\omega lm}^{R}\left(x\right),\tilde{f}_{\omega lm}^{R*}\left(x'\right)\right\} \right)\right.\nonumber \\
 & \left.+2\, \text{cosech}\left(\frac{\pi\omega_{+}}{\kappa_{+}}\right)\Re\left(\rho_{\omega lm}^{\text{up}}\left\{ \tilde{f}_{\omega lm}^{R}\left(x\right),\tilde{f}_{\omega lm}^{L*}\left(x'\right)\right\} \right)\right]+\hbar\sum_{l,m}\int_{0}^{\infty}\frac{\text{d}\omega}{4\pi\omega}\left|\tau_{\omega lm}^{\text{in}}\right|^{2}\left\{ \tilde{f}_{\omega lm}^{R}\left(x\right),\tilde{f}_{\omega lm}^{R*}\left(x'\right)\right\} \,,\label{eq:G_U_ftilde}
\end{align}
where we note that the first integral is over \emph{positive} $\omega_{+}$
only, crucially allowing us to replace $\left|\omega_{+}\right|$
by $\omega_{+}$ there.

Recall that there are two parts making up $G_{U}^{\left(1\right)}\left(x,x'\right)$:
the $\int_{0}^{\infty}\text{d}\omega_{+}$ integral arising from the
\emph{up} modes contribution $G_{U}^{\text{up}}\left(x,x'\right)$,
and the $\int_{0}^{\infty}\text{d}\omega$ integral corresponding
to $G_{U}^{\text{in}}\left(x,x'\right)$. We now concentrate on the
former.

Inspecting the form of the integrand of $G_{U}^{\text{up}}\left(x,x'\right)$
as written in the first part of Eq.~\eqref{eq:G_U_ftilde} (namely,
the integrand of $\int_{0}^{\infty}\text{d}\omega_{+}$ there), we
find that it is invariant under the simultaneous sign changes $m\mapsto-m$
and $\omega\mapsto-\omega$. To see this, apply the symmetries given
in Eqs. \eqref{eq:IntEddModesInvariance} and \eqref{eq:rho_inv}
along with the odd nature of the $\coth$ and $\text{cosech}$ functions
(recalling that $\left(\omega,m\right)\mapsto\left(-\omega,-m\right)$
also implies $\omega_{+}\mapsto-\omega_{+}$).

We shall now establish the following statement: Given a function $E_{m}\left(\omega\right)$
with the property 
\[
E_{\left(-m\right)}\left(-\omega\right)=E_{m}\left(\omega\right)\,,
\]
we may formally write
\begin{equation}
\sum_{m=-l}^{l}\int_{0}^{\infty}\text{d}\omega_{+}E_{m}\left(\omega\right)=\sum_{m=-l}^{l}\int_{0}^{\infty}\text{d}\omega E_{m}\left(\omega\right)\,.\label{eq:w+_to_w}
\end{equation}
That is, one may replace $\sum_{l,m}\int_{0}^{\infty}\text{d}\omega_{+}$
(in the case of a symmetric integrand as described) by $\sum_{l,m}\int_{0}^{\infty}\text{d}\omega$.

To see this, we may denote 
\[
I_{m}\equiv\int_{0}^{\infty}\text{d}\omega_{+}E_{m}\left(\omega\right)\,,
\]
and
then express $I_{m}$ and $I_{-m}$ as follows: 
\[
I_{\pm m}=\int_{\pm m\Omega_{+}}^{0}\text{d}\omega\,E_{\pm m}\left(\omega\right)+\int_{0}^{\infty}\text{d}\omega\,E_{\pm m}\left(\omega\right)\,.
\]

We now concentrate on the finite-domain integration term of $I_{-m}$:
\[
\int_{-m\Omega_{+}}^{0}\text{d}\omega\,E_{-m}\left(\omega\right)=-\int_{m\Omega_{+}}^{0}\text{d}\omega\,E_{-m}\left(-\omega\right)=-\int_{m\Omega_{+}}^{0}\text{d}\omega\,E_{m}\left(\omega\right)
\]
where we have changed variables from $\omega$ to $-\omega$ and then
used the symmetry of $E_{m}\left(\omega\right)$. We readily see that
this term exactly cancels the finite-domain integration term in the
corresponding expression for $I_{m}$. Summing over $m$ in pairs
of $\pm m$ (and noting that the $m=0$ term does not contain such
a finite-domain integration piece), we are thus left with the desired
relation \eqref{eq:w+_to_w}.

Following the discussion above, this property may now be applied to
the integral $\int_{0}^{\infty}\text{d}\omega_{+}$ in Eq.~\eqref{eq:G_U_ftilde},
replacing it by an integral $\int_{0}^{\infty}\text{d}\omega$. Then,
the entire $G_{U}^{\left(1\right)}\left(x,x'\right)$ may be written
in terms of an integral over $\omega$:  
\begin{align}
 & G_{U}^{(1)}\left(x,x'\right)=\nonumber \\
 & \hbar\sum_{l,m}\int_{0}^{\infty}\frac{\text{d}\omega}{4\pi\omega_{+}}\left[\coth\left(\frac{\pi\omega_{+}}{\kappa_{+}}\right)\left(\left\{ \tilde{f}_{\omega lm}^{L}\left(x\right),\tilde{f}_{\omega lm}^{L*}\left(x'\right)\right\} +\left|\rho_{\omega lm}^{\text{up}}\right|^{2}\left\{ \tilde{f}_{\omega lm}^{R}\left(x\right),\tilde{f}_{\omega lm}^{R*}\left(x'\right)\right\} \right)\right.\nonumber \\
 & \left.+2\, \text{cosech}\left(\frac{\pi\omega_{+}}{\kappa_{+}}\right)\Re\left(\rho_{\omega lm}^{\text{up}}\left\{ \tilde{f}_{\omega lm}^{R}\left(x\right),\tilde{f}_{\omega lm}^{L*}\left(x'\right)\right\} \right)+\frac{\omega_{+}}{\omega}\left|\tau_{\omega lm}^{\text{in}}\right|^{2}\left\{ \tilde{f}_{\omega lm}^{R}\left(x\right),\tilde{f}_{\omega lm}^{R*}\left(x'\right)\right\} \right]\,.\label{eq:G_U_ftilde-1}
\end{align}
The integrand in this equation is regular at both $\omega\to0$ and
$\omega_{+}\to0$, as mentioned after Eq.~\eqref{eq:G_U_int}. 

One can also modify the form of Eq.~\eqref{eq:G_U_ftilde-1} (or likewise
Eq.~\eqref{eq:G_U_int}) by various applications of the Wronskian
relations, given in Eq.~\eqref{eq:rho_tau_Wron}. In particular,
one may replace $\frac{\omega_{+}}{\omega}\left|\tau_{\omega lm}^{\text{in}}\right|^{2}$
by $1-\left|\rho_{\omega lm}^{\text{up}}\right|^{2}$, thereby entirely
eliminating $\tau$ from the final expression.

In Appendix~\ref{app:fluxes} we  harness Eq.~\eqref{eq:G_U_ftilde-1}
in order to construct the bare mode-sum expressions for the Unruh
fluxes $\langle T_{uu}\rangle^{U}$ and $\langle T_{vv}\rangle^{U}$ (where $T_{\mu\nu}$ is the  stress-energy tensor)
for a minimally-coupled massless scalar quantum field, starting at
a general $r$ value in the BH interior and then taking it to the
horizons $r\to r_{\pm}$. The resulting expressions will serve as
a basis for future research.

\section{{Acknowledgments.}}

M.C.\ acknowledges partial financial support by CNPq (Brazil), process
number 314824/2020-0, and by the Scientific Council of the Paris Observatory during a visit.
 A.O. and
N.Z. were supported by the Israel Science Foundation under Grant No.
600/18. N.Z. also acknowledges support by the Israeli Planning and Budgeting Committee.

\appendix

\section{The HTPF integrands at small frequencies}\label{App:small freq}

Equation~\eqref{eq:G_U_int} expresses the HTPF as a sum of an \emph{up} part,
 involving an integral over positive $\omega_{+}$,  and an \emph{in} part,  which involves an integral over positive $\omega$.  The
expressions appearing in Eq.~\eqref{eq:G_U_int} for these two integrals may
raise concerns about possible divergences at two specific frequencies:
The \emph{up} piece includes terms proportional to $\coth\left(\pi\omega_{+}/\kappa_{+}\right)$
or $\text{cosech}\left(\pi\omega_{+}/\kappa_{+}\right)$, both diverging
at $\omega_{+}\to0$; and the \emph{in} piece includes a $1/\omega$
factor, which diverges at $\omega\to0$. In addition, both the \emph{up} and \emph{in} pieces include products of $f_{\omega lm}^{R}$ and/or
$f_{\omega lm}^{L}$ functions, each entailing a factor of $1/\sqrt{\left|\omega_{+}\right|}$
in its definition (see Eq.~\eqref{eq:EddModesRL}), which also contribute to a potential
divergence at $\omega_{+}\to0$. Our goal in this Appendix is to analyze
these potential divergences and to show that no divergence actually
occurs in neither $\omega_{+}\to0$ nor the $\omega\to0$ limit. We
shall show this separately for the \emph{up} and \emph{in} pieces. 

\subsection{The \emph{up} integrand }

We begin with the \emph{up} part of the HTPF, given by
\begin{align}
G_{U}^{\text{up}}\left(x,x'\right) & =\hbar\sum_{l,m}\int_{0}^{\infty}\text{d}\omega_{+}G_{\omega lm}^{\text{up}}\left(x,x'\right)\label{eq:Gup}
\end{align}
with the individual mode contribution
\begin{align}
G_{\omega lm}^{\text{up}} & \left(x,x'\right)=\coth\left(\frac{\pi\omega_{+}}{\kappa_{+}}\right)\left(\left\{ f_{\omega lm}^{L}\left(x\right),f_{\omega lm}^{L*}\left(x'\right)\right\} +\left\{ f_{\omega lm}^{R}\left(x\right),f_{\omega lm}^{R*}\left(x'\right)\right\} \left|\rho_{\omega lm}^{\text{up}}\right|^{2}\right)\nonumber \\
 & +2\, \text{cosech}\left(\frac{\pi\omega_{+}}{\kappa_{+}}\right)\Re\left(\rho_{\omega lm}^{\text{up}}\left\{ f_{\omega lm}^{R}\left(x\right),f_{\omega lm}^{L*}\left(x'\right)\right\} \right)\,,\label{eq:GwlmUp}
\end{align}
where $x=\left(t,r,\theta,\varphi\right)$ and $x'=\left(t',r',\theta',\varphi'\right)$.
We denote $\delta t\equiv t-t'$ and $\delta\varphi\equiv\varphi-\varphi'$.

Eq.~\eqref{eq:GwlmUp} is composed of the  inner Eddington
modes $f_{\omega lm}^{R}\left(x\right)$ and $f_{\omega lm}^{L}\left(x\right)$,
given in Eq.~\eqref{eq:EddModesRL}. These mode functions involve in their definition
a factor $\left|\omega_{+}\right|^{-1/2}$. In order to explicitly
reveal this divergent factor, we shall here rewrite $f_{\omega lm}^{R}\left(x\right)$
and $f_{\omega lm}^{L}\left(x\right)$ in the form:
\begin{equation}
f_{\omega lm}^{\Lambda}\left(x\right)=\frac{S_{lm}^{\omega}\left(\theta\right)}{\sqrt{8\pi^{2}\left|\omega_{+}\right|\left(r^{2}+a^{2}\right)}}\overline{f}_{\omega lm}^{\Lambda}\left(x\right)\label{eq:f}
\end{equation}
(for $\Lambda$ either ``\emph{R}" or ``\emph{L}"), where we denote
\begin{equation}
\overline{f}_{\omega lm}^{\Lambda}\left(x\right)\equiv e^{im\varphi}e^{-i\omega t}\psi_{\omega lm}^{\Lambda}\left(r\right)\,.\label{eq:fbar}
\end{equation}
As before, $\psi_{\omega lm}^{\Lambda}$ is the radial function and
we have $\psi_{\omega lm}^{R}=\psi_{\omega lm}^{L*}\equiv\psi_{\omega lm}^{\text{int}}$.
At this stage it becomes clear that there is a potential divergence
in $G_{\omega lm}^{\text{up}}$ that goes like $1/\omega_{+}^{2}$,
due to the $\left|\omega_{+}\right|$ factor appearing in the radical
in the denominator of Eq.~(\ref{eq:f}), combined with the $\coth\left(\pi\omega_{+}/\kappa_{+}\right)$
or $\sinh^{-1}\left(\pi\omega_{+}/\kappa_{+}\right)$ factors in Eq.
(\ref{eq:GwlmUp}). 

In order to facilitate the analysis of this potential divergence
at $\omega_{+}\to0$, we next write $\{f_{\omega lm}^{\Lambda_{1}}\left(x\right),f_{\omega lm}^{\Lambda_{2}*}\left(x'\right)\}$
(with $\Lambda_{1},\Lambda_{2}$ either ``\emph{R}" or ``\emph{L}") as

\begin{equation}
\left\{ f_{\omega lm}^{\Lambda_{1}}\left(x\right),f_{\omega lm}^{\Lambda_{2}*}\left(x'\right)\right\} =\frac{S_{lm}^{\omega}\left(\theta\right)S_{lm}^{\omega}\left(\theta'\right)}{8\pi^{2}\left|\omega_{+}\right|\sqrt{\left(r^{2}+a^{2}\right)\left(r'^{2}+a^{2}\right)}}\left\{ \overline{f}_{\omega lm}^{\Lambda_{1}}\left(x\right),\overline{f}_{\omega lm}^{\Lambda_{2}*}\left(x'\right)\right\} \,.\label{eq:=00007Bf,f*=00007D}
\end{equation}
Then $\{\overline{f}^{\Lambda_{1}}(x),\overline{f}^{\Lambda_{2}*}(x')\}$
is given by:
\begin{equation}
\left\{ \overline{f}_{\omega lm}^{\Lambda_{1}}\left(x\right),\overline{f}_{\omega lm}^{\Lambda_{2}*}\left(x'\right)\right\} =e^{-iK}\psi_{\omega lm}^{\Lambda_{1}}\left(r\right)\psi_{\omega lm}^{\Lambda_{2}*}\left(r'\right)+e^{iK}\psi_{\omega lm}^{\Lambda_{2}*}\left(r\right)\psi_{\omega lm}^{\Lambda_{1}}\left(r'\right)\label{eq:=00007Bfbar,fbar*=00007D}
\end{equation}
where $K\equiv\omega\delta t-m\delta\varphi$.

In what follows, we suppress the superscript ``int'' for brevity
(that is, $\psi_{\omega lm}^{\text{int}}$ is to be denoted by $\psi_{\omega lm}$).
Explicitly, the three relevant cases for $\{\overline{f}_{\omega lm}^{\Lambda_{1}}(x),\overline{f}_{\omega lm}^{\Lambda_{2}*}(x')\}$
are: 

\begin{align}
\left\{ \overline{f}_{\omega lm}^{L}\left(x\right),\overline{f}_{\omega lm}^{L*}\left(x'\right)\right\}  & =2\, \Re\left[e^{-iK}\psi_{\omega lm}^{*}\left(r\right)\psi_{\omega lm}\left(r'\right)\right]\,,\label{eq:=00007BfbarL,fbarL*=00007D}\\
\left\{ \overline{f}_{\omega lm}^{R}\left(x\right),\overline{f}_{\omega lm}^{R*}\left(x'\right)\right\}  & =2\, \Re\left[e^{-iK}\psi_{\omega lm}\left(r\right)\psi_{\omega lm}^{*}\left(r'\right)\right]\,,\label{eq:=00007BfbarR,fbarR*=00007D}\\
\left\{ \overline{f}_{\omega lm}^{R}\left(x\right),\overline{f}_{\omega lm}^{L*}\left(x'\right)\right\}  & =2\cos\left(K\right)\psi_{\omega lm}\left(r\right)\psi_{\omega lm}\left(r'\right)\,.\label{eq:=00007BfbarR,fbarL*=00007D}
\end{align}

Recalling the small-$\omega_{+}$ expansions $\coth\left(\pi\omega_{+}/\kappa_{+}\right)=\kappa_{+}/\pi\omega_{+}+O\left(\omega_{+}\right)$
and $\text{cosech}\left(\pi\omega_{+}/\kappa_{+}\right)=\kappa_{+}/\pi\omega_{+}+O\left(\omega_{+}\right)$,
along with Eqs.~(\ref{eq:=00007BfbarL,fbarL*=00007D})-(\ref{eq:=00007BfbarR,fbarL*=00007D})
for the various $\{\overline{f}_{\omega lm}^{\Lambda_{1}}(x),\overline{f}_{\omega lm}^{\Lambda_{2}*}(x')\}$
factors, we obtain:
\begin{equation}
G_{\omega lm}^{\text{up}}(x,x')=\frac{S_{lm}^{\omega}\left(\theta'\right)S_{lm}^{\omega}\left(\theta\right)}{8\pi^{2}\sqrt{\left(r^{2}+a^{2}\right)\left(r'^{2}+a^{2}\right)}}\frac{\kappa_{+}}{\pi}\left[G_{\omega lm}^{\text{up}\left(A\right)}(r,r')+G_{\omega lm}^{\text{up}\left(B\right)}(r,r')+G_{\omega lm}^{\text{up}\left(C\right)}(r,r')\right]+O\left(\omega_{+}^{0}\right)\label{eq:Gup_sum}
\end{equation}
where
\begin{align}
G_{\omega lm}^{\text{up}\left(A\right)} & (r,r')\equiv\frac{2}{\omega_{+}^{2}}\Re\left[e^{iK}\psi_{\omega lm}\left(r\right)\psi_{\omega lm}^{*}\left(r'\right)\right]\,,\label{eq:GupA}\\
G_{\omega lm}^{\text{up}\left(B\right)} & (r,r')\equiv\frac{2}{\omega_{+}^{2}}\left|\rho_{\omega lm}^{\text{up}}\right|^{2}\Re\left[e^{-iK}\psi_{\omega lm}\left(r\right)\psi_{\omega lm}^{*}\left(r'\right)\right]\,,\label{eq:GupB}\\
G_{\omega lm}^{\text{up}\left(C\right)} & (r,r')\equiv\frac{4}{\omega_{+}^{2}}\cos\left(K\right)\Re\left[\rho_{\omega lm}^{\text{up}}\psi_{\omega lm}\left(r\right)\psi_{\omega lm}\left(r'\right)\right]\,.\label{eq:GupC}
\end{align}

We next resort to the small-$\omega_{+}$ expansion of $\psi_{\omega lm}(r)$
and $\rho_{\omega lm}^{\text{up}}$. For $\psi_{\omega lm}(r)$, the
analysis in Subsec.~\ref{subsec:psi_expansion}  below (see in particular
Eqs.~(\ref{eq:Ansatz}) and (\ref{eq:psi0Real})) implies that 
\begin{equation}
\psi_{\omega lm}\left(r\right)=\psi_{lm}^{\left(0\right)}\left(r\right)+\psi_{lm}^{\left(1\right)}\left(r\right)\omega_{+}+o\left(\omega_{+}\right)\,\label{eq:psi_expansion}
\end{equation}
where $\psi_{lm}^{\left(0\right)}\left(r\right)$ is a \emph{real}
function, and $o\left(\omega_{+}\right)$ denotes terms whose decay
rate at $\omega_{+}\to0$ is faster than $\omega_{+}$.   For
$\rho_{\omega lm}^{\text{up}}$, it can be shown \footnote{We analytically derived this small-$\omega_{+}$ expansion of $\rho_{\omega lm}^{\text{up}}$,
both for $m\neq0$ and $m=0$ (in which case $\omega_{+}\to0$ means
$\omega\to0$). We also verified this small-$\omega_{+}$ expansion
numerically. We do not provide the analytical derivation here, as
this issue (being solely related to wave scattering \emph{outside}
the BH) is beyond the scope of the present paper. \label{fn:small_w}} that
\begin{equation}
\rho_{\omega lm}^{\text{up}}=-1+\rho_{lm}^{\left(1\right)}\omega_{+}+o\left(\omega_{+}\right)\,.\label{eq:rho_expansion}
\end{equation}
 Naturally, the expansion coefficients $\psi_{lm}^{\left(0\right)}\left(r\right),$
$\psi_{lm}^{\left(1\right)}\left(r\right)$ and $\rho_{lm}^{\left(1\right)}$
are independent of $\omega$ (or $\omega_{+}$), as their  lower
indices (being solely $lm$) also indicate.

To facilitate the analysis below, we now rewrite Eqs.~(\ref{eq:psi_expansion})
and (\ref{eq:rho_expansion}) by absorbing their $o\left(\omega_{+}\right)$
parts into the first-order coefficients $\psi_{lm}^{\left(1\right)}\left(r\right)$
and $\rho_{lm}^{\left(1\right)}$. As a result, these first-order
coefficients now become $\omega$-dependent, and correspondingly
we denote them by $\psi_{\omega lm}^{\left(1\right)}\left(r\right)$
and $\rho_{\omega lm}^{\left(1\right)}$. (Nevertheless, this dependence
on $\omega$ will not cause any complication: The only relevant fact
is that both $\psi_{\omega lm}^{\left(1\right)}\left(r\right)$ and
$\rho_{\omega lm}^{\left(1\right)}$ remain finite as $\omega_{+}\to0$.)
Thus, we rewrite Eqs.~(\ref{eq:psi_expansion}) and (\ref{eq:rho_expansion})
as follows: 

\begin{equation}
\psi_{\omega lm}\left(r\right)=\psi_{lm}^{\left(0\right)}\left(r\right)+\psi_{\omega lm}^{\left(1\right)}\left(r\right)\omega_{+}\,,\label{eq:psi_expansion-1}
\end{equation}

\begin{equation}
\rho_{\omega lm}^{\text{up}}=-1+\rho_{\omega lm}^{\left(1\right)}\omega_{+}\,.\label{eq:rho_expansion-1}
\end{equation}
 The last equation also implies

\begin{equation}
\left|\rho_{\omega lm}^{\text{up}}\right|^{2}=1-2\omega_{+}\Re\left(\rho_{\omega lm}^{\left(1\right)}\right)+O\left(\omega_{+}^{2}\right)\,.\label{eq:abs(rho)^2_expansion}
\end{equation}

We now plug the expansions (\ref{eq:psi_expansion-1}-\ref{eq:abs(rho)^2_expansion})
of $\psi_{\omega lm}$, $\rho_{\omega lm}^{\text{up}}$ and $\left|\rho_{\omega lm}^{\text{up}}\right|^{2}$
  into Eqs.~(\ref{eq:GupA})-(\ref{eq:GupC}).  The three quantities
$G_{\omega lm}^{\text{up}\left(A\right)}$, $G_{\omega lm}^{\text{up}\left(B\right)}$
and $G_{\omega lm}^{\text{up}\left(C\right)}$ then split accordingly
into terms multiplying $\omega_{+}^{-2}$ and $\omega_{+}^{-1}$
(plus  an $O\left(\omega_{+}^{0}\right)$ term), and they all take
the form 

\begin{equation}
G_{\omega lm}^{\text{up}\left(X\right)}(r,r')=\frac{1}{\omega_{+}^{2}}G_{\omega lm}^{\text{up}\left(X_{-2}\right)}(r,r')+\frac{1}{\omega_{+}}G_{\omega lm}^{\text{up}\left(X_{-1}\right)}(r,r')+O\left(\omega_{+}^{0}\right)\,,\label{eq:GupABC_sum}
\end{equation}
where $X$ stands here for either $A$, $B$ or $C$. The computation
of the $2\times3$ coefficients $G_{\omega lm}^{\text{up}\left(X_{-2}\right)}$
and $G_{\omega lm}^{\text{up}\left(X_{-1}\right)}$ is straightforward
(and uses the fact that $\psi_{lm}^{\left(0\right)}\left(r\right)$
is real).   For $X=A$ the two coefficients are
\begin{equation}
G_{\omega lm}^{\text{up}\left(A_{-2}\right)}(r,r')=2\cos\left(K\right)\psi_{lm}^{\left(0\right)}\left(r\right)\psi_{lm}^{\left(0\right)}\left(r'\right)\label{eq:GupA2}
\end{equation}
and
\begin{equation}
G_{\omega lm}^{\text{up}\left(A_{-1}\right)}(r,r')=2\, \psi_{lm}^{\left(0\right)}\left(r\right)\Re\left[e^{-iK}\psi_{\omega lm}^{\left(1\right)}\left(r'\right)\right]+2\psi_{lm}^{\left(0\right)}\left(r'\right)\Re\left[e^{iK}\psi_{\omega lm}^{\left(1\right)}\left(r\right)\right]\,,\label{eq:GupA1}
\end{equation}
 for $X=B$ the coefficients are
\begin{equation}
G_{\omega lm}^{\text{up}\left(B_{-2}\right)}(r,r')=2\cos\left(K\right)\psi_{lm}^{\left(0\right)}\left(r\right)\psi_{lm}^{\left(0\right)}\left(r'\right)\label{eq:GupB2}
\end{equation}
and
\begin{align}
G_{\omega lm}^{\text{up}\left(B_{-1}\right)}(r,r')= & 2\, \psi_{lm}^{\left(0\right)}\left(r\right)\Re\left[e^{iK}\psi_{\omega lm}^{\left(1\right)}\left(r'\right)\right]+2\psi_{lm}^{\left(0\right)}\left(r'\right)\Re\left[e^{-iK}\psi_{\omega lm}^{\left(1\right)}\left(r\right)\right]\label{eq:GupB1}\\
 & -4\cos\left(K\right)\Re\left[\rho_{\omega lm}^{\left(1\right)}\right]\psi_{lm}^{\left(0\right)}\left(r\right)\psi_{lm}^{\left(0\right)}\left(r'\right)\,,\nonumber 
\end{align}
and for  $X=C$:
\begin{equation}
G_{\omega lm}^{\text{up}\left(C_{-2}\right)}(r,r')=-4\cos\left(K\right)\psi_{lm}^{\left(0\right)}\left(r\right)\psi_{lm}^{\left(0\right)}\left(r'\right)\,\label{eq:GupC2}
\end{equation}
and 
\begin{align}
G_{\omega lm}^{\text{up}\left(C_{-1}\right)}(r,r')= & -4\cos\left(K\right)\left(\psi_{lm}^{\left(0\right)}\left(r\right)\Re\left[\psi_{\omega lm}^{\left(1\right)}\left(r'\right)\right]+\psi_{lm}^{\left(0\right)}\left(r'\right)\Re\left[\psi_{\omega lm}^{\left(1\right)}\left(r\right)\right]\right)\label{eq:GupC1}\\
 & +4\cos\left(K\right)\Re\left[\rho_{\omega lm}^{\left(1\right)}\right]\psi_{lm}^{\left(0\right)}\left(r\right)\psi_{lm}^{\left(0\right)}\left(r'\right)\,.\nonumber 
\end{align}

Concentrating first on the three $\propto1/\omega_{+}^{2}$ coefficients,
namely $G_{\omega lm}^{\text{up}\left(X_{-2}\right)}$,  notice
that
\begin{equation}
G_{\omega lm}^{\text{up}\left(A_{-2}\right)}+G_{\omega lm}^{\text{up}\left(B_{-2}\right)}=-G_{\omega lm}^{\text{up}\left(C_{-2}\right)}\label{eq:A2+B2=00003D-C2}
\end{equation}
so they cancel out in $G_{\omega lm}^{\text{up}}$.  Turning next
to the three $\propto1/\omega_{+}$ coefficients $G_{\omega lm}^{\text{up}\left(X_{-1}\right)}$,
we see the same structure here again: 
\begin{equation}
G_{\omega lm}^{\text{up}\left(A_{-1}\right)}+G_{\omega lm}^{\text{up}\left(B_{-1}\right)}=-G_{\omega lm}^{\text{up}\left(C_{-1}\right)}\label{eq:A1+B1=00003D-C1}
\end{equation}
so that part is cancelled out as well. Substituting this back into
 Eq.~(\ref{eq:Gup_sum}), we are left with
\begin{equation}
G_{\omega lm}^{\text{up}}\left(x,x'\right)=O\left(\omega_{+}^{0}\right)\label{eq:Gup_final}
\end{equation}
at $\omega_{+}\to0$; that is,  the potential divergence of the individual
\emph{up} mode contributions at $\omega_{+}=0$ is gone. 

Finally we consider the behavior of $G_{\omega lm}^{\text{up}}\left(x,x'\right)$
at $\omega\to0$. Both $\psi_{\omega lm}$ and $\rho_{\omega lm}^{\text{up}}$
are  regular at that limit \footnote{For $\psi_{\omega lm}$  in the case $m\neq0$, this regularity naturally
follows from the definition of $\psi_{\omega lm}$  based on boundary
conditions specified at the EH  in terms of $\omega_{+}$ rather
than $\omega$. For $\rho_{\omega lm}^{\text{up}}$  in the case
$m\neq0$, we analytically computed $\rho_{\omega lm}^{\text{up}}$
at $\omega\to0$ and found it to be finite and well-defined (and it
satisfies $\left|\rho_{(\omega=0)lm}^{\text{up}}\right|=1$), but
again, this analysis of $\rho_{\omega lm}^{\text{up}}$  is beyond the scope of
this paper.  We also numerically verified smoothness of $\rho_{\omega lm}^{\text{up}}$
at the limit $\omega\to0$.  In the other case $m=0$, the limit
$\omega\to0$ coincides with the limit  $\omega_{+}\to0$, for which
regularity of $\psi_{\omega lm}$ and $\rho_{\omega lm}^{\text{up}}$
has already been established above  (see Eqs.~(\ref{eq:psi_expansion})
and (\ref{eq:rho_expansion}), and also footnote \ref{fn:small_w}).
}. For $m\neq0$ (in which case $\omega\to0$ implies that $\omega_{+}$
stays remote from zero), no divergence can occur in $G_{\omega lm}^{\text{up}}$
at $\omega\to0$. In the special case $m=0$, taking the limit $\omega\to0$
also implies $\omega_{+}\to0$ (which in turn implies there are potentially-divergent
terms in the above expression for $G_{\omega lm}^{\text{up}}$);
nevertheless, it was already shown above that the overall expression
for $G_{\omega lm}^{\text{up}}$ is actually regular at $\omega_{+}\to0$. 

We therefore conclude that $G_{\omega lm}^{\text{up}}\left(x,x'\right)$
is regular at both limits $\omega_{+}\to0$ and $\omega\to0$.

\subsection{The \emph{in} integrand }

We now consider the \emph{in} mode contribution, which is much simpler,
and   is given by
\begin{equation}
G_{U}^{\text{in}}\left(x,x'\right)=\hbar\sum_{l,m}\int_{0}^{\infty}\text{d}\omega\,G_{\omega lm}^{\text{in}}\left(x,x'\right)\,,\label{eq:Gin}
\end{equation}
with the integrand
\begin{equation}
G_{\omega lm}^{\text{in}}\left(x,x'\right)=\frac{\left|\omega_{+}\right|}{\omega}\left|\tau_{\omega lm}^{\text{in}}\right|^{2}\left\{ f_{\omega lm}^{R}\left(x\right),f_{\omega lm}^{R*}\left(x'\right)\right\} \,.\label{eq:GwlmIn}
\end{equation}
The Wronskian relations in Eq.~\eqref{eq:rho_tau_Wron} yield 

\[
\frac{\omega_{+}}{\omega}\left|\tau_{\omega lm}^{\text{in}}\right|^{2}=1-\left|\rho_{\omega lm}^{\text{in}}\right|^{2}\,,
\]
and therefore (recalling $\left|\rho_{\omega lm}^{\text{in}}\right|=\left|\rho_{\omega lm}^{\text{up}}\right|$)
also 
\[
\frac{\left|\omega_{+}\right|}{\omega}\left|\tau_{\omega lm}^{\text{in}}\right|^{2}=\text{sign}\left(\omega_{+}\right)\left(1-\left|\rho_{\omega lm}^{\text{up}}\right|^{2}\right)\,.
\]
 Plugging this relation along with Eq.~(\ref{eq:=00007Bf,f*=00007D})
into Eq.~(\ref{eq:GwlmIn}), we obtain:
\begin{equation}
G_{\omega lm}^{\text{in}}\left(x,x'\right)=\frac{1}{\omega_{+}}\left(1-\left|\rho_{\omega lm}^{\text{up}}\right|^{2}\right)\frac{S_{lm}^{\omega}\left(\theta'\right)S_{lm}^{\omega}\left(\theta\right)}{8\pi^{2}\sqrt{\left(r^{2}+a^{2}\right)\left(r'^{2}+a^{2}\right)}}\left\{ \overline{f}_{\omega lm}^{R}\left(x\right),\overline{f}_{\omega lm}^{R*}\left(x'\right)\right\} \,,\label{eq:Gin_fbar}
\end{equation}
which highlights the potential divergence at $\omega_{+}\to0$. However,
Eq.~(\ref{eq:abs(rho)^2_expansion}) reads
\begin{equation}
1-\left|\rho_{\omega lm}^{\text{up}}\right|^{2}=2\, \omega_{+}\Re\left(\rho_{1}\right)+O\left(\omega_{+}^{2}\right)\,,\label{eq:1-rho^2}
\end{equation}
and the $\omega_{+}$ factor at the right-hand side cancels out
the $1/\omega_{+}$ factor in Eq.~(\ref{eq:Gin_fbar}).  Also, $\overline{f}_{\omega lm}^{R}$
is regular at $\omega_{+}\to0$, as  directly follows from Eqs.~(\ref{eq:fbar})
and (\ref{eq:psi_expansion}).  We are therefore left with
\begin{equation}
G_{\omega lm}^{\text{in}}\left(x,x'\right)=O\left(\omega_{+}^{0}\right)\label{eq:Gin_final}
\end{equation}
at the limit $\omega_{+}\to0$.

The form of Eq.~(\ref{eq:Gin_fbar}) also guarantees that no irregularity
occurs at $\omega\to0$ either. \footnote{As before, we use the fact that $\rho_{\omega lm}^{\text{up}}$ and
$\psi_{\omega lm}$ (and hence also $\overline{f}_{\omega lm}^{R}$)
are  regular at $\omega\to0$. We also recall that in the special
case $m=0$, for which the $1/\omega_{+}$ factor in Eq.~(\ref{eq:Gin_fbar})
diverges as $\omega\to0$ (because now this limit also implies $\omega_{+}\to0$),
this potential divergence is already handled in the above analysis,
which showed that $G_{\omega lm}^{\text{in}}$ is actually regular
at $\omega_{+}\to0$.}

\subsection{$\psi_{\omega lm}^{\text{int}}$ at small $\omega_{+}$\label{subsec:psi_expansion}}

The function $\psi_{\omega lm}(r)$ (which, recall, denotes $\psi_{\omega lm}^{\text{int}}\left(r\right)$
in this Appendix) satisfies the radial equation

\begin{equation}
\psi_{\omega lm,r_{*}r_{*}}=-V_{\omega lm}\left(r\right)\psi_{\omega lm}\label{eq:ODE}
\end{equation}
with an effective potential $V_{\omega lm}\left(r\right)$ given explicitly
in Eqs.~(2.19)-(2.20). The initial condition for this ODE is specified
at the EH (corresponding to $r_{*}\to-\infty$) by 
\begin{equation}
\psi_{\omega lm}\simeq e^{-i\omega_{+}r_{*}}\equiv\psi_{\omega lm}^{(\text{init})}\qquad(r_{*}\to-\infty)\,.\label{eq:initial}
\end{equation}
(Recall, the symbol ``$\simeq$'' denotes equality at the relevant
asymptotic boundary, namely $r_{*}\to-\infty$ in the present case.) 

Our goal is to analyze the behavior of $\psi_{\omega lm}^{\text{}}(r)$
at small $\omega_{+}$. To this end, we first write the asymptotic
behavior of the initial condition (\ref{eq:initial}) at small $\omega_{+}$:
\begin{equation}
\psi_{\omega lm}^{(\text{init})}=1-i\omega_{+}r_{*}-\omega_{+}^{2}\frac{r_{*}^{2}}{2}+...\label{eq:initial-2}
\end{equation}
The effective potential $V_{\omega lm}\left(r\right)$, too, can be
decomposed in a power series in $\omega_{+}$ around $\omega_{+}=0$
(see below). We are therefore motivated to adopt the following Ansatz
for the form of $\psi_{\omega lm}^{\text{}}(r)$ at small $\omega_{+}$,
as a power series in $\omega_{+}$:
\begin{equation}
\psi_{\omega lm}(r)=\psi_{lm}^{\left(0\right)}(r)+\psi_{lm}^{\left(1\right)}(r)\,\omega_{+}+\psi_{lm}^{\left(2\right)}(r)\,\omega_{+}^{2}+...\label{eq:Ansatz}
\end{equation}
Each coefficient $\psi_{lm}^{\left(n\right)}(r)$ in this expansion
should satisfy its own ODE (with its own initial condition at the
EH), as we shall now describe.

To find the specific ODE that each term $\psi_{lm}^{\left(n\right)}(r)$
satisfies, we have to expand the potential $V_{\omega lm}(r)$ in
powers of $\omega_{+}$. Since $V_{\omega lm}$ depends on the angular
eigenvalue $\lambda_{\omega lm}$, we first expand this eigenvalue: \footnote{For $m\neq0$ it is trivial, as the formulation of the angular eigenvalue
problem is insensitive to the $\omega_{+}\to0$ limit. For $m=0$,
it has been shown \cite{StarobinskiChurilov:1974} that such a power-series
expansion exists.}
\begin{equation}
\lambda_{\omega lm}=\lambda_{lm}^{\left(0\right)}+\lambda_{lm}^{\left(1\right)}a\omega_{+}+\lambda_{lm}^{\left(2\right)}\left(a\omega_{+}\right)^{2}+...\label{eq:lambda_expansion}
\end{equation}
Then we can expand the effective potential in the same manner:
\begin{equation}
V_{\omega lm}(r)=V_{lm}^{\left(0\right)}(r)+V_{lm}^{\left(1\right)}(r)\,\omega_{+}+V_{lm}^{\left(2\right)}(r)\,\omega_{+}^{2}+...\label{eq:V_expansion}
\end{equation}
The leading-order coefficient is given by
\begin{equation}
V_{lm}^{\left(0\right)}(r)=\left(m\Omega_{+}\right)^{2}\left(\frac{r^{2}-r_{+}^{2}}{r^{2}+a^{2}}\right)^{2}-G^{2}-\frac{\Delta}{r^{2}+a^{2}}\frac{\text{d}G}{\text{d}r}-\frac{\lambda_{lm}^{\left(0\right)}\Delta}{\left(r^{2}+a^{2}\right)^{2}}\,\label{eq:V0}
\end{equation}
 where, recall, $\Omega_{+}=a/\left(r_{+}^{2}+a^{2}\right)$ and
$G=r\Delta/\left(r^{2}+a^{2}\right)^{2}$. The first-order coefficient
is
\begin{equation}
V_{lm}^{\left(1\right)}(r)=2m\Omega_{+}\frac{r^{2}-r_{+}^{2}}{r^{2}+a^{2}}-\frac{a\lambda_{lm}^{\left(1\right)}\Delta}{\left(r^{2}+a^{2}\right)^{2}}\,\label{eq:V1}
\end{equation}
and the second-order coefficient:
\begin{equation}
V_{lm}^{\left(2\right)}(r)=1-\frac{a^{2}\lambda_{lm}^{\left(2\right)}\Delta}{\left(r^{2}+a^{2}\right)^{2}}\,.\label{eq:V2}
\end{equation}
(In fact, all higher-order coefficients are of the same simple form:
$V_{lm}^{\left(n>2\right)}=-a^{n}\lambda_{lm}^{\left(n\right)}\Delta/\left(r^{2}+a^{2}\right)^{2}$).

It is important to recall that the potential $V_{\omega lm}(r)$ is
\emph{real} -- and so are all its expansion coefficients $V_{lm}^{\left(n\right)}$.
In addition, note that both coefficients $V_{lm}^{\left(0\right)}$
and $V_{lm}^{\left(1\right)}$ vanish at the EH like $\propto\Delta\propto r-r_{+}$
 -- hence, they both decay exponentially with $r_{*}$ at the EH
limit $r_{*}\to-\infty$. In fact, at the EH we have $V_{\omega lm}=\omega_{+}^{2}$.
\footnote{Correspondingly $V_{lm}^{\left(2\right)}(r)=1$ at the EH, as one
can also see from Eq.~(\ref{eq:V2}). } 

Inserting the Ansatz (\ref{eq:Ansatz}) for $\psi_{\omega lm}$ into
the radial equation (\ref{eq:ODE}) with the expanded form (\ref{eq:V_expansion})
of $V_{\omega lm}$, and grouping powers of $\omega_{+}$, we obtain
the following hierarchy of ODEs:
\begin{align}
\psi_{lm,r_{*}r_{*}}^{\left(0\right)} & +V_{lm}^{\left(0\right)}\psi_{lm}^{\left(0\right)}=0\,,\nonumber \\
\psi_{lm,r_{*}r_{*}}^{\left(1\right)} & +V_{lm}^{\left(0\right)}\psi_{lm}^{\left(1\right)}=-V_{lm}^{\left(1\right)}\psi_{lm}^{\left(0\right)}\,,\label{eq:ODEs}\\
\psi_{lm,r_{*}r_{*}}^{\left(2\right)} & +V_{lm}^{\left(0\right)}\psi_{lm}^{\left(2\right)}=-V_{lm}^{\left(1\right)}\psi_{lm}^{\left(1\right)}-V_{lm}^{\left(2\right)}\psi_{lm}^{\left(0\right)}\,,\nonumber \\
{\color{blue}} & \dots\nonumber 
\end{align}
Note that $\psi_{lm}^{\left(0\right)}(r)$ satisfies a homogeneous
ODE, but all other functions $\psi_{lm}^{\left(n\right)}(r)$ satisfy
inhomogeneous ones (having $V_{lm}^{\left(0\right)}$ as their potential,
and a source term involving other coefficients in the expansion of
$V_{\omega lm}$).

The initial conditions for these ODEs are to be specified at the EH
limit, just like those of the original function  $\psi_{\omega lm}\left(r\right)$.
They are naturally obtained by the Taylor expansion (\ref{eq:initial-2})
of the original initial data $\psi_{\omega lm}^{(\text{init})}$:
\begin{equation}
\psi_{lm}^{\left(0\right)}\simeq1\qquad(r_{*}\to-\infty)\,,\label{eq:initial_0}
\end{equation}
\begin{equation}
\psi_{lm}^{\left(1\right)}\simeq-ir_{*}\qquad(r_{*}\to-\infty)\,,\label{eq:initial_1}
\end{equation}
\begin{equation}
\psi_{lm}^{\left(2\right)}\simeq-\frac{r_{*}^{2}}{2}\qquad(r_{*}\to-\infty)\,,\label{eq:initial_2}
\end{equation}
etc. 

Of particular importance to our analysis is the leading-order function
$\psi_{lm}^{\left(0\right)}(r)$. It satisfies a real ODE (as $V_{lm}^{\left(0\right)}$
is real), with real initial conditions. It therefore follows that
$\psi_{lm}^{\left(0\right)}(r)$ is a \emph{real} function: 
\begin{equation}
\psi_{lm}^{\left(0\right)}(r)\in\mathbb{R}\,.\label{eq:psi0Real}
\end{equation}

Although not necessary for the regularity analysis carried out in
this Appendix, it may be interesting to consider the properties of
$\psi_{lm}^{\left(1\right)}$ as well. It still satisfies a real ODE
(because both its potential $V_{lm}^{\left(0\right)}$ and its source
term $-V_{lm}^{\left(1\right)}\psi_{lm}^{\left(0\right)}$ are real);
However, its initial condition at the EH limit ($\simeq-ir_{*}$)
is imaginary. Therefore, $\psi_{lm}^{\left(1\right)}(r)$ is not real.
It is not purely imaginary either, because it  is fed by a real source
term $-V_{lm}^{\left(1\right)}\psi_{lm}^{\left(0\right)}$. All higher-order
terms $\psi_{lm}^{\left(n>1\right)}$ are expected to be complex too.

\section{The Unruh-state bare flux expressions inside the BH}\label{app:fluxes}

This appendix is dedicated to developing the  mode-sum expressions
for the Unruh-state RSET components $T_{uu}$ and $T_{vv}$ (where
hereafter $u=u_{\text{int}}$ and $v$ are the interior Eddington
coordinates given in Eq.~\eqref{eq:intEddCoor}). We first construct this mode-sum
expression at a general point $r_{-}<r<r_{+}$ in the BH interior,
then we concentrate on the horizon limits $r\to r_{-}$ and $r\to r_{+}$.

We focus here on $T_{uu}$ and $T_{vv}$ because these two components
are especially meaningful for the semiclassical study of backreaction
on BH interiors (in particular, at the IH vicinity). At the horizons,
these components play the role of energy fluxes \footnote{Note that the Eddington coordinates $u$ and $v$ are spacelike at
$r_{-}<r<r_{+}$ but they become asymptotically null at $r\to r_{-}$
and $r\to r_{+}$. (To be more precise, we can look at the corresponding
Kruskal coordinates, which are found to be spacelike between the horizons
and null at the horizons. These properties are then carried over to
the corresponding Eddington coordinates.)}, and we shall thus refer to them as the \emph{flux components} or,
in short, the \emph{fluxes}. In addition, these components reveal
notable simplicity at the IH limit, as we shall briefly note later
on.

\subsection{Bare fluxes at general $r$}

 Before we begin with the construction,  we note that the components
of a tensor such as $T_{\alpha\beta}$ clearly depend on the underlying
coordinate system. Here we shall particularly be interested in three
coordinate systems, which only differ from each other by the choice
of the azimuthal coordinate. We collectively denote these three coordinate
systems as $\left(u,v,\theta,\tilde{\varphi}\right)$, where $\tilde{\varphi}$
stands for either $\varphi$ , $\varphi_{+}$ or $\varphi_{-}$. Recall
that $\varphi$ is the original Boyer-Lindquist azimuthal coordinate,
while $\varphi_{+}$ and $\varphi_{-}$ are the two modified azimuthal
coordinates constructed to be regular respectively at the EH and IH,
and they are given by $\varphi\equiv\varphi_{\pm}-\Omega_{\pm}t$
(see Subsec.~\ref{subsec:metric_and_coordinates}). Thus, we may generally define $\tilde{\varphi}$
as  
\begin{equation}
\tilde{\varphi}\equiv\varphi-\tilde{\Omega}t\label{eq:phiTilde}
\end{equation}
where the constant $\tilde{\Omega}$ is either zero, $\Omega_{+}$
or $\Omega_{-}$, for $\varphi$, $\varphi_{+}$ and $\varphi_{-}$
respectively. \footnote{We point out that, specifically, the choice of the azimuthal coordinate
$\tilde{\varphi}$ does affect the values of the flux components $T_{uu}$
and $T_{vv}$. }

We shall restrict our attention here to a minimally-coupled massless
scalar field (i.e. $m=\xi=0$ in Eq.~\eqref{eq:generalKG}). Then, at the classical
level, the stress-energy tensor $T_{\alpha\beta}$ of this field may
be expressed as 
\begin{equation}
T_{\alpha\beta}=\overline{T}_{\alpha\beta}-(1/2)g_{\alpha\beta}\overline{T}_{\,\,\,\,\mu}^{\mu}\,,\label{eq:new1}
\end{equation}
where $\overline{T}_{\alpha\beta}$ (the \emph{trace-reversed} stress-energy
tensor) is given in terms of the first-order scalar field derivatives
by
\begin{equation}
\overline{T}_{\alpha\beta}=\Phi_{,\alpha}\Phi_{,\beta}\,.\label{eq:new2}
\end{equation}
For the analysis below, it will be useful to re-express $\overline{T}_{\alpha\beta}$
as a second-order differential operator acting on a certain quantity
bi-linear in $\Phi$ (this form will later allow us to conveniently
express the quantum expectation value of $T_{\alpha\beta}$ in terms
of a differential operator acting on the quantity $G_{U}^{\left(1\right)}\left(x,x'\right)$
that is already available to us). To this end, we re-express $\overline{T}_{\alpha\beta}$
(still at the classical level) as 
\begin{equation}
\overline{T}_{\alpha\beta}\left(x\right)=\lim_{x'\to x}\left[\Phi\left(x\right)\Phi\left(x'\right)\right]_{,\alpha\beta'}\,.\label{eq:new3}
\end{equation}
The symbol $_{,\alpha\beta'}$ denotes differentiation with respect
to $x^{\alpha}$ and $x'^{\beta}$, where $\partial/\partial x^{\alpha}$
acts on functions of the spacetime point $x$ while $\partial/\partial x'^{\beta}$
acts on functions of the spacetime point $x'$. We then further rewrite
it in the form 
\begin{equation}
\overline{T}_{\alpha\beta}\left(x\right)=\frac{1}{2}\lim_{x'\to x}\left[\Phi\left(x\right)\Phi\left(x'\right)+\Phi\left(x'\right)\Phi\left(x\right)\right]_{,\alpha\beta'}\label{eq:new4}
\end{equation}
(which, although trivial, sets the stage for the quantum treatment
that will now follow).

Transitioning from the classical- to the quantum-field context, we
want to compute the expectation value of $\overline{T}_{\alpha\beta}\left(x\right)$,
for our minimally-coupled quantum field $\hat{\Phi}$, in the Unruh
state \footnote{Note that in the quantum context, both $T_{\alpha\beta}$ and $\overline{T}_{\alpha\beta}$
 are treated as quantum operators. To avoid notational complications
(especially for $\overline{T}_{\alpha\beta}$), we do not add any
special symbol (e.g. an over-hat) to make this quantum nature explicitly
visible. Nevertheless, in the equations below, the expectation-value
symbol $\left\langle ...\right\rangle ^{U}$ will always reveal the
quantum-operator nature of $T_{\alpha\beta}$ and $\overline{T}_{\alpha\beta}$.}. Applying the $\left\langle ...\right\rangle _{U}$ expectation value
operation to the two sides of Eq.~(\ref{eq:new4}) (which are viewed
now as quantum operators), and recalling that $G_{U}^{\left(1\right)}\left(x,x'\right)=\left\langle \hat\Phi\left(x\right)\hat\Phi\left(x'\right)+\hat\Phi\left(x'\right)\hat\Phi\left(x\right)\right\rangle _{U}$
(see Eq.~\eqref{eq:G_U}) we obtain the following formal expression for $\left\langle \overline{T}_{\alpha\beta}\right\rangle _{\text{bare}}^{U}$:
\begin{equation}
\left\langle \overline{T}_{\alpha\beta}\right\rangle _{\text{bare}}^{U}\left(x\right)=\frac{1}{2}\lim_{x'\to x}\left[G_{U}^{\left(1\right)}\left(x,x'\right)_{,\alpha\beta'}\right]\,.\label{eq:  point_split}
\end{equation}
 This is complemented by the quantum version of Eq.~(\ref{eq:new1}),
namely
\begin{equation}
\left\langle T_{\alpha\beta}\right\rangle _{\text{bare}}^{U}=\left\langle \overline{T}_{\alpha\beta}\right\rangle _{\text{bare}}^{U}-(1/2)g_{\alpha\beta}\left\langle \overline{T}_{\,\,\,\,\mu}^{\mu}\right\rangle _{\text{bare}}^{U}\,.\label{eq:new5}
\end{equation}

To avoid confusion we emphasize again that the split appearing in
the right-hand side of Eqs.~(\ref{eq:new3}), (\ref{eq:new4}) and
(\ref{eq:  point_split}) is \emph{not} aimed for regularization (recall
we  deal in Eqs.~(\ref{eq:new3}),(\ref{eq:new4}) with the \emph{classical}
expressions, and in Eq.~(\ref{eq:  point_split}) with the \emph{bare}
quantum expression): The only purpose of this split is to allow differentiation
with respect to $x$ and $x'$ separately -- in order to eventually
express the RSET in terms of  the already-known function $G_{U}^{\left(1\right)}\left(x,x'\right)$. 

To proceed, we write the Unruh-state HTPF $G_{U}^{\left(1\right)}\left(x,x'\right)$
(for points inside the BH)
given in Eq.~\eqref{eq:G_U_ftilde-1} as the mode-sum
\begin{equation}
G_{U}^{\left(1\right)}\left(x,x'\right)=\hbar\sum_{l,m}\int_{0}^{\infty}G_{\omega lm}\left(x,x'\right)\text{d}\omega\,,\label{eq:Gmodesum}
\end{equation}
where 
\begin{align}
 & G_{\omega lm}\left(x,x'\right)=\nonumber \\
 & \frac{1}{4\pi\omega_{+}}\left[\coth\left(\frac{\pi\omega_{+}}{\kappa_{+}}\right)\left(\left\{ \tilde{f}_{\omega lm}^{L}\left(x\right),\tilde{f}_{\omega lm}^{L*}\left(x'\right)\right\} +\left|\rho_{\omega lm}^{\text{up}}\right|^{2}\left\{ \tilde{f}_{\omega lm}^{R}\left(x\right),\tilde{f}_{\omega lm}^{R*}\left(x'\right)\right\} \right)\right.\label{eq:Ghat}\\
 & \left.+2\, \text{cosech}\left(\frac{\pi\omega_{+}}{\kappa_{+}}\right)\Re\left(\rho_{\omega lm}^{\text{up}}\left\{ \tilde{f}_{\omega lm}^{R}\left(x\right),\tilde{f}_{\omega lm}^{L*}\left(x'\right)\right\} \right)+\frac{\omega_{+}}{\omega}\left|\tau_{\omega lm}^{\text{in}}\right|^{2}\left\{ \tilde{f}_{\omega lm}^{R}\left(x\right),\tilde{f}_{\omega lm}^{R*}\left(x'\right)\right\} \right]\,.\nonumber 
\end{align}
Correspondingly we may then rewrite Eq.~(\ref{eq:  point_split})
as the mode sum
\begin{equation}
\left\langle \overline{T}_{\alpha\beta}\right\rangle _{\text{bare}}^{U}\left(x\right)=\sum_{l,m}\int_{0}^{\infty}\overline{T}_{\alpha\beta\left(\omega lm\right)}\text{d}\omega\,,\label{eq:  point_split-2}
\end{equation}
where the integrand is defined as
\begin{equation}
\overline{T}_{\alpha\beta\left(\omega lm\right)}\equiv\frac{\hbar}{2}\lim_{x'\to x}\left[G_{\omega lm}\left(x,x'\right)_{,\alpha\beta'}\right]\,.\label{eq:Twlm}
\end{equation}

From this point on we shall concentrate on the two flux components,
taking $\alpha\beta$ to be  $yy$, with $y$ hereafter denoting
either $u$ or $v$. The computation of $\overline{T}_{yy\left(\omega lm\right)}$
then involves two simple stages: (i) differentiating $G_{\omega lm}\left(x,x'\right)$
with respect to $y$ and $y'$, and then (ii) taking the coincidence
limit $y\to y'$. The RHS of Eq.~(\ref{eq:Ghat}) consists of several
terms of the form 
\[
\left\{ F_{1}\left(x\right),F_{2}\left(x'\right)\right\} =F_{1}\left(x\right)F_{2}\left(x'\right)+F_{1}\left(x'\right)F_{2}\left(x\right)\,.
\]
 Applying these stages (i) and (ii) to the term $F_{1}\left(x\right)F_{2}\left(x'\right)$
simply yields $F_{1,y}F_{2,y}$ (evaluated at the point $x$), and
applying it to the other term $F_{1}\left(x'\right)F_{2}\left(x\right)$
yields exactly the same result; that is, 
\[
\lim_{y'\to y}\left\{ F_{1}\left(x\right),F_{2}\left(x'\right)\right\} _{,yy'}=2F_{1,y}F_{2,y}
\]
(evaluated at the point $x$ as mentioned above). Implementing this
in Eq.~(\ref{eq:Twlm}), we obtain

\begin{align}
 & \overline{T}_{yy\left(\omega lm\right)}=\nonumber \\
 & \frac{\hbar}{4\pi\omega_{+}}\left[\coth\left(\frac{\pi\omega_{+}}{\kappa_{+}}\right)\left(\tilde{f}_{\omega lm,y}^{L}\tilde{f}_{\omega lm,y}^{L*}+\left|\rho_{\omega lm}^{\text{up}}\right|^{2}\tilde{f}_{\omega lm,y}^{R}\tilde{f}_{\omega lm,y}^{R*}\right)\right.\label{eq:Twlm1}\\
 & \left.+2\, \text{cosech}\left(\frac{\pi\omega_{+}}{\kappa_{+}}\right)\Re\left(\rho_{\omega lm}^{\text{up}}\tilde{f}_{\omega lm,y}^{R}\tilde{f}_{\omega lm,y}^{L*}\right)+\frac{\omega_{+}}{\omega}\left|\tau_{\omega lm}^{\text{in}}\right|^{2}\tilde{f}_{\omega lm,y}^{R}\tilde{f}_{\omega lm,y}^{R*}\right]\,.\nonumber 
\end{align}

The functions $\tilde{f}_{\omega lm}^{\Lambda}$ (with $\Lambda$
denoting either ``\emph{R}" or ``\emph{L}") were defined in Eq.~\eqref{eq:fLtilde}. Recalling
that 
\[
Z_{lm}^{\omega}\left(\theta,\varphi\right)=\frac{1}{\sqrt{2\pi}}S_{lm}^{\omega}\left(\theta\right)e^{im\varphi}\,,
\]
we may rewrite these functions in the more explicit form:
\[
\tilde{f}_{\omega lm}^{R}=\frac{1}{\sqrt{2\pi\left(r^{2}+a^{2}\right)}}S_{lm}^{\omega}\left(\theta\right)e^{im\varphi}e^{-i\omega t}\psi_{\omega lm}^{\text{int}}\,,
\]
\[
\tilde{f}_{\omega lm}^{L}=\frac{1}{\sqrt{2\pi\left(r^{2}+a^{2}\right)}}S_{lm}^{\omega}\left(\theta\right)e^{im\varphi}e^{-i\omega t}\psi_{\omega lm}^{\text{int}*}\,.
\]
We need to differentiate these functions with respect to $y$ \emph{with
fixed $\tilde{\varphi}$} (rather than  fixed $\varphi$). To this
end, we define a general frequency-parameter $\tilde{\omega}$ of
the form:
\begin{equation}
\tilde{\omega}\equiv\omega-m\tilde{\Omega}\label{eq:wTilde}
\end{equation}
(the upper ''$\sim$'' symbol links $\tilde{\omega}$ to the choice
of the azimuthal coordinate $\tilde{\varphi}$). Noting that
\begin{equation}
e^{-i\omega t}e^{im\varphi}=e^{-i\tilde{\omega}t}e^{im\tilde{\varphi}}\,,\label{eq:tilde_exponent}
\end{equation}
we may now re-express $\tilde{f}_{\omega lm}^{\Lambda}$ as

\begin{equation}
\tilde{f}_{\omega lm}^{\Lambda}\left(x\right)=\frac{1}{\sqrt{8\pi}}S_{lm}^{\omega}\left(\theta\right)e^{im\tilde{\varphi}}\hat{f}_{\omega lm}^{\Lambda}\left(t,r\right)\label{eq:f-Lambda}
\end{equation}
where we define
\begin{equation}
\hat{f}_{\omega lm}^{R}\left(t,r\right)\equiv\frac{2}{\sqrt{r^{2}+a^{2}}}e^{-i\tilde{\omega}t}\psi_{\omega lm}^{\text{int}}\left(r\right)\,,\,\,\hat{f}_{\omega lm}^{L}\left(t,r\right)\equiv\frac{2}{\sqrt{r^{2}+a^{2}}}e^{-i\tilde{\omega}t}\psi_{\omega lm}^{\text{int}*}\left(r\right)\,.\label{eq:f^RL}
\end{equation}
(The variables $t$ and $r$ should be viewed here as functions of
the coordinates $u,v$.) Substituting Eq.~(\ref{eq:f-Lambda}) in
Eq.~(\ref{eq:Twlm1}) (recalling that $S_{lm}^{\omega}\left(\theta\right)$
is real), we may now entirely factor out the angular dependence:
\begin{equation}
\overline{T}_{yy\left(\omega lm\right)}=\hbar\frac{\left[S_{lm}^{\omega}\left(\theta\right)\right]^{2}}{32\pi^{2}\omega_{+}}\mathcal{\overline{T}}_{yy\left(\omega lm\right)}\label{eq:  point_split-4}
\end{equation}
where
\begin{align}
 & \mathcal{\overline{T}}_{yy\left(\omega lm\right)}\equiv\coth\left(\frac{\pi\omega_{+}}{\kappa_{+}}\right)\left(\hat{f}_{\omega lm,y}^{L}\hat{f}_{\omega lm,y}^{L*}+\left|\rho_{\omega lm}^{\text{up}}\right|^{2}\hat{f}_{\omega lm,y}^{R}\hat{f}_{\omega lm,y}^{R*}\right)\label{eq:Twlm2}\\
 & +2\, \text{cosech}\left(\frac{\pi\omega_{+}}{\kappa_{+}}\right)\Re\left(\rho_{\omega lm}^{\text{up}}\hat{f}_{\omega lm,y}^{R}\hat{f}_{\omega lm,y}^{L*}\right)+\frac{\omega_{+}}{\omega}\left|\tau_{\omega lm}^{\text{in}}\right|^{2}\hat{f}_{\omega lm,y}^{R}\hat{f}_{\omega lm,y}^{R*}\,.\nonumber 
\end{align}

To further process this expression, we next specify the four combinations
entailed in $\hat{f}_{\omega lm,y}^{\Lambda}$ (corresponding to $\Lambda=L,R$
and $y=u,v$). Using the relations
\begin{align*}
t & =\frac{v-u}{2},\,\,r_{*}=\frac{u+v}{2}\,,
\end{align*}
as well as $\text{d}r/\text{d}r_{*}=\Delta/\left(r^{2}+a^{2}\right),$we
find

\begin{equation}
\hat{f}_{\omega lm,u}^{R}=\frac{e^{-i\tilde{\omega}t}}{\sqrt{r^{2}+a^{2}}}\left[\left(i\tilde{\omega}\psi_{\omega lm}^{\text{int}}+\psi_{\omega lm,r_{*}}^{\text{int}}\right)-\mathcal{H}(r)\psi_{\omega lm}^{\text{int}}\Delta\right]\,,\label{eq:Single_Ru}
\end{equation}
\begin{equation}
\hat{f}_{\omega lm,v}^{R}=\frac{e^{-i\tilde{\omega}t}}{\sqrt{r^{2}+a^{2}}}\left[\left(-i\tilde{\omega}\psi_{\omega lm}^{\text{int}}+\psi_{\omega lm,r_{*}}^{\text{int}}\right)-\mathcal{H}(r)\psi_{\omega lm}^{\text{int}}\Delta\right]\,,\label{eq:Single_Rv}
\end{equation}
\begin{equation}
\hat{f}_{\omega lm,u}^{L}=\frac{e^{-i\tilde{\omega}t}}{\sqrt{r^{2}+a^{2}}}\left[\left(i\tilde{\omega}\psi_{\omega lm}^{\text{int}*}+\psi_{\omega lm,r_{*}}^{\text{int}*}\right)-\mathcal{H}(r)\psi_{\omega lm}^{\text{int}*}\Delta\right]\,,\label{eq:Single_Lu}
\end{equation}
\begin{equation}
\hat{f}_{\omega lm,v}^{L}=\frac{e^{-i\tilde{\omega}t}}{\sqrt{r^{2}+a^{2}}}\left[\left(-i\tilde{\omega}\psi_{\omega lm}^{\text{int}*}+\psi_{\omega lm,r_{*}}^{\text{int}*}\right)-\mathcal{H}(r)\psi_{\omega lm}^{\text{int*}}\Delta\right]\,,\label{eq:Single_Lv}
\end{equation}
where 
\[
\mathcal{H}(r)\equiv\frac{r}{\left(r^{2}+a^{2}\right)^{2}}\,.
\]
Notice that $\hat{f}_{\omega lm,v}^{\Lambda}$ and $\hat{f}_{\omega lm,u}^{\Lambda}$
are related by the transformation $\tilde{\omega}\mapsto-\tilde{\omega},t\mapsto-t$
 \footnote{In $\tilde{\omega}\mapsto-\tilde{\omega}$ we refer to any \emph{explicit}
occurrence of $\tilde{\omega}$ -- we do not touch the indices $\omega lm$
(as an illustration, see how the described transformation relates
Eqs.~(\ref{eq:Single_Ru}) and (\ref{eq:Single_Rv}), or (\ref{eq:Single_Lu})
and (\ref{eq:Single_Lv})).}. Also, $\hat{f}_{\omega lm,y}^{L}$ and $\hat{f}_{\omega lm,y}^{R}$
are related by the transformation $\tilde{\omega}\mapsto-\tilde{\omega}$
combined with overall complex conjugation. These relations will be
useful below. 

We now combine these derivatives to form the various bilinear combinations
of the form $\hat{f}_{\omega lm,y}^{\Lambda_{1}}\hat{f}_{\omega lm,y}^{\Lambda_{2}*}$
appearing in Eq.~(\ref{eq:Twlm2}), namely the three combinations
$\Lambda_{1}\Lambda_{2}=\left(LL,RR,RL\right)$.  One immediately
notices that the factors $e^{-i\tilde{\omega}t}$ (and indeed the
entire dependence on $t$) cancel out in all these combinations. It
is convenient to express each of these  contributions in the form
\begin{equation}
\hat{f}_{\omega lm,y}^{\Lambda_{1}}\hat{f}_{\omega lm,y}^{\Lambda_{2}*}=\frac{1}{r^{2}+a^{2}}\left[\mathcal{A}_{\omega lm(y)}^{\Lambda_{1}\Lambda_{2}}-2\mathcal{H}(r)\mathcal{B}_{\omega lm(y)}^{\Lambda_{1}\Lambda_{2}}\Delta+\mathcal{H}^{2}(r)\mathcal{C}_{\omega lm(y)}^{\Lambda_{1}\Lambda_{2}}\Delta^{2}\right]\,.\label{eq:Delta_LRY}
\end{equation}

By a direct substitution of Eqs.~(\ref{eq:Single_Ru})-(\ref{eq:Single_Lv}),
recalling the Wronskian relation 
\[
\psi_{\omega lm}^{\text{int}}\psi_{\omega lm,r_{*}}^{\text{int*}}-\psi_{\omega lm}^{\text{int*}}\psi_{\omega lm,r_{*}}^{\text{int}}=2i\omega_{+}\,,
\]
 we obtain the following expressions for the $\mathcal{A}$ coefficients:

\begin{equation}
\mathcal{A}_{\omega lm(v)}^{RR}=\mathcal{A}_{\omega lm(u)}^{LL}=\left|\psi_{\omega lm,r_{*}}^{\text{int}}\right|^{2}+\tilde{\omega}^{2}\left|\psi_{\omega lm}^{\text{int}}\right|^{2}+2\tilde{\omega}\omega_{+}\,,\label{eq:ARRvLLu}
\end{equation}

\begin{equation}
\mathcal{A}_{\omega lm(v)}^{LL}=\mathcal{A}_{\omega lm(u)}^{RR}=\left|\psi_{\omega lm,r_{*}}^{\text{int}}\right|^{2}+\tilde{\omega}^{2}\left|\psi_{\omega lm}^{\text{int}}\right|^{2}-2\tilde{\omega}\omega_{+}\,,\label{eq:ALLvRRu}
\end{equation}
\begin{equation}
\mathcal{A}_{\omega lm(u)}^{RL}=\mathcal{A}_{\omega lm(v)}^{RL}=\left(\psi_{\omega lm,r_{*}}^{\text{int}}\right)^{2}+\tilde{\omega}^{2}\left(\psi_{\omega lm}^{\text{int}}\right)^{2}\,.\label{eq:ARL}
\end{equation}
For the $\mathcal{B}$ and $\mathcal{C}$ coefficients we will omit
the $\left(y\right)$ subscript, as they attain the same value for
both $y=v$ and $y=u$. We obtain:
\begin{equation}
\mathcal{B}_{\omega lm}^{RR}=\mathcal{B}_{\omega lm}^{LL}=\Re\left(\psi_{\omega lm}^{\text{int}}\psi_{\omega lm,r_{*}}^{\text{int*}}\right)\,,\quad\mathcal{B}_{\omega lm}^{RL}=\psi_{\omega lm}^{\text{int}}\psi_{\omega lm,r_{*}}^{\text{int}}\,,\label{eq:Bterms}
\end{equation}
\begin{equation}
\mathcal{C}_{\omega lm}^{LL}=\mathcal{C}_{\omega lm}^{RR}=\left|\psi_{\omega lm}^{\text{int}}\right|^{2}\,,\quad\mathcal{C}_{\omega lm}^{RL}=\left(\psi_{\omega lm}^{\text{int}}\right)^{2}\,.\label{eq:Cterms}
\end{equation}
Note also that none of the $\mathcal{B},\mathcal{C}$ coefficients
depend explicitly on $\tilde{\omega}$ (but the $\mathcal{A}$ coefficients
do). 

We have mentioned above simple rules for the transformations $R\leftrightarrow L$
and $u\leftrightarrow v$ in the expressions for $\hat{f}_{\omega lm,y}^{\Lambda}$.
We can use them to derive corresponding rules for (overall) interchanges
$R\leftrightarrow L$ and/or $u\leftrightarrow v$ in $\hat{f}_{\omega lm,y}^{\Lambda_{1}}\hat{f}_{\omega lm,y}^{\Lambda_{2}*}$.
It follows that both transformations $RR\leftrightarrow LL$ and $u\leftrightarrow v$
amount to changing $\tilde{\omega}\mapsto-\tilde{\omega}$ \footnote{To this end, one should recall that (i) $\hat{f}_{\omega lm,y}^{\Lambda_{1}}\hat{f}_{\omega lm,y}^{\Lambda_{2}*}$
is independent of $t$, and (ii) all these $\mathcal{A},\mathcal{B},\mathcal{C}$
coefficients for $LL$ or $RR$ are real.} (and therefore the combined transformation $RR\leftrightarrow LL,u\leftrightarrow v$
leaves the expression unchanged). One can easily verify that the above
expressions (\ref{eq:ARRvLLu})-(\ref{eq:Cterms}) indeed satisfy
these simple exchange rules. 

The next stage would be to substitute Eq.~(\ref{eq:Delta_LRY}) in
Eq.~(\ref{eq:Twlm2}) for $\mathcal{\overline{T}}_{yy\left(\omega lm\right)}$.
It is again convenient to rewrite the latter (just as we did in the
former) explicitly in powers of $\Delta$, so we write: 
\begin{equation}
\mathcal{\overline{T}}_{yy\left(\omega lm\right)}=\frac{1}{r^{2}+a^{2}}\left[\mathcal{\overline{T}}_{yy\left(\omega lm\right)}^{\mathcal{A}}-2\mathcal{H}(r)\mathcal{\overline{T}}_{\left(\omega lm\right)}^{\mathcal{B}}\Delta+\mathcal{H}^{2}(r)\mathcal{\overline{T}}_{\left(\omega lm\right)}^{\mathcal{C}}\Delta^{2}\right]\,.\label{eq:Delta_calT}
\end{equation}
We find $\mathcal{\overline{T}}_{uu\left(\omega lm\right)}^{\mathcal{A}}$
to be given by:
\begin{align}
 & \mathcal{\overline{T}}_{uu\left(\omega lm\right)}^{\mathcal{A}}=\nonumber \\
 & \coth\left(\frac{\pi\omega_{+}}{\kappa_{+}}\right)\left[\left|\psi_{\omega lm,r_{*}}^{\text{int}}\right|^{2}+\tilde{\omega}^{2}\left|\psi_{\omega lm}^{\text{int}}\right|^{2}+2\tilde{\omega}\omega_{+}+\left|\rho_{\omega lm}^{\text{up}}\right|^{2}\left(\left|\psi_{\omega lm,r_{*}}^{\text{int}}\right|^{2}+\tilde{\omega}^{2}\left|\psi_{\omega lm}^{\text{int}}\right|^{2}-2\tilde{\omega}\omega_{+}\right)\right]\nonumber \\
 & +2\, \text{cosech}\left(\frac{\pi\omega_{+}}{\kappa_{+}}\right)\Re\left(\rho_{\omega lm}^{\text{up}}\left[\left(\psi_{\omega lm,r_{*}}^{\text{int}}\right)^{2}+\tilde{\omega}^{2}\left(\psi_{\omega lm}^{\text{int}}\right)^{2}\right]\right)+\frac{\omega_{+}}{\omega}\left|\tau_{\omega lm}^{\text{in}}\right|^{2}\left(\left|\psi_{\omega lm,r_{*}}^{\text{int}}\right|^{2}+\tilde{\omega}^{2}\left|\psi_{\omega lm}^{\text{int}}\right|^{2}-2\tilde{\omega}\omega_{+}\right)\,.\label{eq:TAuu}
\end{align}
The $vv$ counterpart, $\mathcal{\overline{T}}_{vv\left(\omega lm\right)}^{\mathcal{A}}$,
is obtained by taking $\tilde{\omega}\mapsto-\tilde{\omega}$ in the
above expression:
\begin{align}
 & \mathcal{\overline{T}}_{vv\left(\omega lm\right)}^{\mathcal{A}}=\nonumber \\
 & \coth\left(\frac{\pi\omega_{+}}{\kappa_{+}}\right)\left[\left|\psi_{\omega lm,r_{*}}^{\text{int}}\right|^{2}+\tilde{\omega}^{2}\left|\psi_{\omega lm}^{\text{int}}\right|^{2}-2\tilde{\omega}\omega_{+}+\left|\rho_{\omega lm}^{\text{up}}\right|^{2}\left(\left|\psi_{\omega lm,r_{*}}^{\text{int}}\right|^{2}+\tilde{\omega}^{2}\left|\psi_{\omega lm}^{\text{int}}\right|^{2}+2\tilde{\omega}\omega_{+}\right)\right]\nonumber \\
 & +2\, \text{cosech}\left(\frac{\pi\omega_{+}}{\kappa_{+}}\right)\Re\left(\rho_{\omega lm}^{\text{up}}\left[\left(\psi_{\omega lm,r_{*}}^{\text{int}}\right)^{2}+\tilde{\omega}^{2}\left(\psi_{\omega lm}^{\text{int}}\right)^{2}\right]\right)+\frac{\omega_{+}}{\omega}\left|\tau_{\omega lm}^{\text{in}}\right|^{2}\left(\left|\psi_{\omega lm,r_{*}}^{\text{int}}\right|^{2}+\tilde{\omega}^{2}\left|\psi_{\omega lm}^{\text{int}}\right|^{2}+2\tilde{\omega}\omega_{+}\right)\,.\label{eq:TAvv}
\end{align}
Finally, the $\mathcal{B}$ and $\mathcal{C}$ coefficients are given
by:
\begin{align}
 & \mathcal{\overline{T}}_{\left(\omega lm\right)}^{\mathcal{B}}=\coth\left(\frac{\pi\omega_{+}}{\kappa_{+}}\right)\Re\left(\psi_{\omega lm}^{\text{int}}\psi_{\omega lm,r_{*}}^{\text{int}*}\right)\left(1+\left|\rho_{\omega lm}^{\text{up}}\right|^{2}\right)\nonumber \\
 & +2\, \text{cosech}\left(\frac{\pi\omega_{+}}{\kappa_{+}}\right)\Re\left(\rho_{\omega lm}^{\text{up}}\psi_{\omega lm}^{\text{int}}\psi_{\omega lm,r_{*}}^{\text{int}}\right)+\frac{\omega_{+}}{\omega}\left|\tau_{\omega lm}^{\text{in}}\right|^{2}\Re\left(\psi_{\omega lm}^{\text{int}}\psi_{\omega lm,r_{*}}^{\text{int}*}\right)\,,\label{eq:TB}
\end{align}
\begin{align}
\mathcal{\overline{T}}_{\left(\omega lm\right)}^{\mathcal{C}} & =\coth\left(\frac{\pi\omega_{+}}{\kappa_{+}}\right)\left|\psi_{\omega lm}^{\text{int}}\right|^{2}\left(1+\left|\rho_{\omega lm}^{\text{up}}\right|^{2}\right)+2\, \text{cosech}\left(\frac{\pi\omega_{+}}{\kappa_{+}}\right)\Re\left(\rho_{\omega lm}^{\text{up}}\left(\psi_{\omega lm}^{\text{int}}\right)^{2}\right)+\frac{\omega_{+}}{\omega}\left|\tau_{\omega lm}^{\text{in}}\right|^{2}\left|\psi_{\omega lm}^{\text{int}}\right|^{2}\,.\label{eq:TC}
\end{align}
Note that the $\mathcal{B}$ and $\mathcal{C}$ contributions are
the same for $uu$ and $vv$ -- and the same  applies to $\overline{T}_{\left(\omega lm\right)}^{\mathcal{B},\mathcal{C}}$
defined below.

Finally, we substitute the expression (\ref{eq:Delta_calT}) for $\mathcal{\overline{T}}_{yy\left(\omega lm\right)}$
into Eq.~(\ref{eq:  point_split-4}) for $\overline{T}_{yy\left(\omega lm\right)}$.
Again, we rewrite the latter in powers of $\Delta$:
\begin{equation}
\overline{T}_{yy\left(\omega lm\right)}=\overline{T}_{yy\left(\omega lm\right)}^{\mathcal{A}}+\overline{T}_{\left(\omega lm\right)}^{\mathcal{B}}\Delta+\overline{T}_{\left(\omega lm\right)}^{\mathcal{C}}\Delta^{2}\,.\label{eq:That_sum}
\end{equation}

The coefficients $\overline{T}_{yy\left(\omega lm\right)}^{\mathcal{A}}$,
$\overline{T}_{\left(\omega lm\right)}^{\mathcal{B}}$ and $\overline{T}_{\left(\omega lm\right)}^{\mathcal{C}}$
are then given by:
\begin{align}
 & \overline{T}_{uu\left(\omega lm\right)}^{\mathcal{A}}=\hbar\frac{\left[S_{lm}^{\omega}\left(\theta\right)\right]^{2}}{32\pi^{2}\omega_{+}\left(r^{2}+a^{2}\right)}\label{eq:ThatAuu}\\
 & \left(\coth\left(\frac{\pi\omega_{+}}{\kappa_{+}}\right)\left[\left|\psi_{\omega lm,r_{*}}^{\text{int}}\right|^{2}+\tilde{\omega}^{2}\left|\psi_{\omega lm}^{\text{int}}\right|^{2}+2\tilde{\omega}\omega_{+}+\left|\rho_{\omega lm}^{\text{up}}\right|^{2}\left(\left|\psi_{\omega lm,r_{*}}^{\text{int}}\right|^{2}+\tilde{\omega}^{2}\left|\psi_{\omega lm}^{\text{int}}\right|^{2}-2\tilde{\omega}\omega_{+}\right)\right]\right.\nonumber \\
 & \left.+2\, \text{cosech}\left(\frac{\pi\omega_{+}}{\kappa_{+}}\right)\Re\left(\rho_{\omega lm}^{\text{up}}\left[\left(\psi_{\omega lm,r_{*}}^{\text{int}}\right)^{2}+\tilde{\omega}^{2}\left(\psi_{\omega lm}^{\text{int}}\right)^{2}\right]\right)+\left(1-\left|\rho_{\omega lm}^{\text{up}}\right|^{2}\right)\left(\left|\psi_{\omega lm,r_{*}}^{\text{int}}\right|^{2}+\tilde{\omega}^{2}\left|\psi_{\omega lm}^{\text{int}}\right|^{2}-2\tilde{\omega}\omega_{+}\right)\right)\,,\nonumber 
\end{align}
\begin{align}
 & \overline{T}_{vv\left(\omega lm\right)}^{\mathcal{A}}=\hbar\frac{\left[S_{lm}^{\omega}\left(\theta\right)\right]^{2}}{32\pi^{2}\omega_{+}\left(r^{2}+a^{2}\right)}\label{eq:ThatAvv}\\
 & \left(\coth\left(\frac{\pi\omega_{+}}{\kappa_{+}}\right)\left[\left|\psi_{\omega lm,r_{*}}^{\text{int}}\right|^{2}+\tilde{\omega}^{2}\left|\psi_{\omega lm}^{\text{int}}\right|^{2}-2\tilde{\omega}\omega_{+}+\left|\rho_{\omega lm}^{\text{up}}\right|^{2}\left(\left|\psi_{\omega lm,r_{*}}^{\text{int}}\right|^{2}+\tilde{\omega}^{2}\left|\psi_{\omega lm}^{\text{int}}\right|^{2}+2\tilde{\omega}\omega_{+}\right)\right]\right.\nonumber \\
 & \left.+2\, \text{cosech}\left(\frac{\pi\omega_{+}}{\kappa_{+}}\right)\Re\left(\rho_{\omega lm}^{\text{up}}\left[\left(\psi_{\omega lm,r_{*}}^{\text{int}}\right)^{2}+\tilde{\omega}^{2}\left(\psi_{\omega lm}^{\text{int}}\right)^{2}\right]\right)+\left(1-\left|\rho_{\omega lm}^{\text{up}}\right|^{2}\right)\left(\left|\psi_{\omega lm,r_{*}}^{\text{int}}\right|^{2}+\tilde{\omega}^{2}\left|\psi_{\omega lm}^{\text{int}}\right|^{2}+2\tilde{\omega}\omega_{+}\right)\right)\,,\nonumber 
\end{align}
\begin{align}
\overline{T}_{\left(\omega lm\right)}^{\mathcal{B}} & =-\hbar\frac{\left[S_{lm}^{\omega}\left(\theta\right)\right]^{2}r}{16\pi^{2}\omega_{+}\left(r^{2}+a^{2}\right)^{3}}\left(\coth\left(\frac{\pi\omega_{+}}{\kappa_{+}}\right)\Re\left(\psi_{\omega lm}^{\text{int}}\psi_{\omega lm,r_{*}}^{\text{int}*}\right)\left(1+\left|\rho_{\omega lm}^{\text{up}}\right|^{2}\right)\right.\label{eq:ThatB}\\
 & \left.+2\, \text{cosech}\left(\frac{\pi\omega_{+}}{\kappa_{+}}\right)\Re\left(\rho_{\omega lm}^{\text{up}}\psi_{\omega lm}^{\text{int}}\psi_{\omega lm,r_{*}}^{\text{int}}\right)+\left(1-\left|\rho_{\omega lm}^{\text{up}}\right|^{2}\right)\Re\left(\psi_{\omega lm}^{\text{int}}\psi_{\omega lm,r_{*}}^{\text{int}*}\right)\right)\,,\nonumber 
\end{align}
\begin{align}
 & \overline{T}_{\left(\omega lm\right)}^{\mathcal{C}}=\hbar\frac{\left[S_{lm}^{\omega}\left(\theta\right)\right]^{2}r^{2}}{32\pi^{2}\omega_{+}\left(r^{2}+a^{2}\right)^{5}}\label{eq:ThatC}\\
 & \left(\coth\left(\frac{\pi\omega_{+}}{\kappa_{+}}\right)\left|\psi_{\omega lm}^{\text{int}}\right|^{2}\left(1+\left|\rho_{\omega lm}^{\text{up}}\right|^{2}\right)+2\, \text{cosech}\left(\frac{\pi\omega_{+}}{\kappa_{+}}\right)\Re\left(\rho_{\omega lm}^{\text{up}}\left(\psi_{\omega lm}^{\text{int}}\right)^{2}\right)+\left(1-\left|\rho_{\omega lm}^{\text{up}}\right|^{2}\right)\left|\psi_{\omega lm}^{\text{int}}\right|^{2}\right)\,,\nonumber 
\end{align}
where we made the substitution $\left(\omega_{+}/\omega\right)\left|\tau_{\omega lm}^{\text{in}}\right|^{2}=1-\left|\rho_{\omega lm}^{\text{up}}\right|^{2}$
(see Wronskian relations, Eq.~\eqref{eq:rho_tau_Wron}) to eliminate $\tau$ from
the final results.

This provides the desired mode-sum expression for $\left\langle \overline{T}_{yy}\right\rangle _{\text{bare}}^{U}\left(x\right)$,
see Eq.~(\ref{eq:  point_split-2}). The translation from the trace-reversed
to the original bare RSET  then proceeds according to Eq.~(\ref{eq:new5})
(although, this last stage requires also the RSET trace mode sum,
which we have not addressed here).

We worked here in coordinates $\left(u,v,\theta,\tilde{\varphi}\right)$,
with an azimuthal coordinate $\tilde{\varphi}$ whose general form
is given in Eq.~(\ref{eq:phiTilde}). Evidently, dependence on the
choice of azimuthal coordinate $\tilde{\varphi}$ only appears (through
$\tilde{\omega}$) in $\overline{T}_{yy\left(\omega lm\right)}^{\mathcal{A}}$,
the part that does not vanish at the horizons (in particular, for
$\tilde{\varphi}=\varphi$, the parameter $\tilde{\omega}$ is replaced
by $\omega$; and  for $\tilde{\varphi}=\varphi_{\pm}$, the parameter
$\tilde{\omega}$ is replaced by $\omega_{\pm}$, respectively). 

\subsection{The difference $T_{uu}-T_{vv}$}

Note that, as mentioned, the terms $\overline{T}_{\left(\omega lm\right)}^{\mathcal{B}}$
and $\overline{T}_{\left(\omega lm\right)}^{\mathcal{C}}$ are shared
by the $uu$ and $vv$ components. In addition, the difference between
$\overline{T}_{uu\left(\omega lm\right)}^{\mathcal{A}}$ and $\overline{T}_{vv\left(\omega lm\right)}^{\mathcal{A}}$
is only in the sign of the three $\propto\tilde{\omega}$ terms. The
difference between $\left\langle \overline{T}_{uu}\right\rangle _{\text{bare}}^{U}$
and $\left\langle \overline{T}_{vv}\right\rangle _{\text{bare}}^{U}$
(which also equals the difference between $\left\langle T_{uu}\right\rangle _{\text{bare}}^{U}$
and $\left\langle T_{vv}\right\rangle _{\text{bare}}^{U}$, since
$g_{uu}=g_{vv}$ in coordinates $\left(u,v,\theta,\tilde{\varphi}\right)$),
therefore has a rather simple form:
\begin{align}
\left\langle T_{uu}\right\rangle _{\text{bare}}^{U}-\left\langle T_{vv}\right\rangle _{\text{bare}}^{U}=\hbar\sum_{l,m}\int_{0}^{\infty}\text{d}\omega\frac{\left[S_{lm}^{\omega}\left(\theta\right)\right]^{2}}{8\pi^{2}\left(r^{2}+a^{2}\right)}\,\tilde{\omega}\left[\coth\left(\frac{\pi\omega_{+}}{\kappa_{+}}\right)-1\right]\left(1-\left|\rho_{\omega lm}^{\text{up}}\right|^{2}\right)\,.\label{eq:Tuu-Tvv_bare}
\end{align}

Next we consider the renormalized version of Eq.~(\ref{eq:Tuu-Tvv_bare}).
Performing a coordinate transformation from $\left(u,v,\theta,\tilde{\varphi}\right)$
to $\left(t,r_{*},\theta,\tilde{\varphi}\right)$ and then to $\left(t,r_{*},\theta,\varphi\right)$,
yields
\begin{equation}
\left(T_{uu}-T_{vv}\right)_{\left(u,v,\theta,\tilde{\varphi}\right)}=\left(-T_{r_{*}t}\right)_{\left(t,r_{*},\theta,\tilde{\varphi}\right)}=-\left(\tilde{\Omega}T_{\varphi r_{*}}+T_{r_{*}t}\right)_{\left(t,r_{*},\theta,\varphi\right)}\,.\label{eq:Tuu-Tvv(t,r,th,phi)}
\end{equation}
From Eq.~(3.30)  in Ref. \cite{FrolovThorne:1989}, we see that
the counterterms $T_{r_{*}t}^{\text{div}}$ and $T_{\varphi r_{*}}^{\text{div}}$
vanish. Thus, the renormalized difference $\left\langle T_{uu}\right\rangle _{\text{ren}}^{U}-\left\langle T_{vv}\right\rangle _{\text{ren}}^{U}$
is equal to the bare difference $\left\langle T_{uu}\right\rangle _{\text{bare}}^{U}-\left\langle T_{vv}\right\rangle _{\text{bare}}^{U}$,
given in the RHS of Eq.~(\ref{eq:Tuu-Tvv_bare}) (in coordinates $\left(u,v,\theta,\tilde{\varphi}\right)$):
\begin{equation}
\left\langle T_{uu}\right\rangle _{\text{ren}}^{U}-\left\langle T_{vv}\right\rangle _{\text{ren}}^{U}=\hbar\sum_{l,m}\int_{0}^{\infty}\text{d}\omega\frac{\left[S_{lm}^{\omega}\left(\theta\right)\right]^{2}}{8\pi^{2}\left(r^{2}+a^{2}\right)}\,\tilde{\omega}\left[\coth\left(\frac{\pi\omega_{+}}{\kappa_{+}}\right)-1\right]\left(1-\left|\rho_{\omega lm}^{\text{up}}\right|^{2}\right)\,.\label{eq:Tuu-Tvv_ren}
\end{equation}
Evidently, $\left(r^{2}+a^{2}\right)\left(\left\langle T_{uu}\right\rangle _{\text{ren}}^{U}-\left\langle T_{vv}\right\rangle _{\text{ren}}^{U}\right)$
is independent of $r$ (reflecting energy-momentum conservation).

The mode-sum expression for the Hawking outflux (per solid angle)
may then be obtained from Eq.~(\ref{eq:Tuu-Tvv_ren}) by choosing
the Boyer-Lindquist azimuthal coordinate $\tilde{\varphi}=\varphi$
(that is, taking $\tilde{\omega}=\omega$) and multiplying by $\left(r^{2}+a^{2}\right)$.
This yields the expression
\begin{equation}
\left(r^{2}+a^{2}\right)\left(\left\langle T_{uu}\right\rangle _{\text{ren}}^{U}-\left\langle T_{vv}\right\rangle _{\text{ren}}^{U}\right)=\hbar\sum_{l,m}\int_{0}^{\infty}\text{d}\omega\frac{\left[S_{lm}^{\omega}\left(\theta\right)\right]^{2}}{8\pi^{2}}\,\omega\left[\coth\left(\frac{\pi\omega_{+}}{\kappa_{+}}\right)-1\right]\left(1-\left|\rho_{\omega lm}^{\text{up}}\right|^{2}\right)\,.\label{eq:Hawking}
\end{equation}
This is a well-known result (see e.g. Eq.~(5.5) in Ref. \cite{OttewillWinstanley:2000}
 \footnote{
\noindent 
Eq.~(5.5) in Ref. \cite{OttewillWinstanley:2000} gives
a quantity denoted by $K_{U-}\left(\theta\right)$, which coincides
with $-\left(r^{2}+a^{2}\right)\left(\left\langle T_{uu}\right\rangle _{\text{ren}}^{U}-\left\langle T_{vv}\right\rangle _{\text{ren}}^{U}\right)$.
For comparison with Eq.~(\ref{eq:Hawking}), note that in Ref. \cite{OttewillWinstanley:2000}
$\tilde{\omega}$ denotes $\omega_{+}$ and $B_{\omega lm}^{-}$ is
our $\tau_{\omega lm}^{\text{up}}$, and use the Wronskian relation
relating $\left|\tau_{\omega lm}^{\text{up}}\right|$ with $\left|\rho_{\omega lm}^{\text{up}}\right|$
(see Eq.~\eqref{eq:rho_tau_Wron}).
}). 

\subsection{Bare fluxes at the horizons}

We are particularly interested in the behavior of the fluxes at
the EH and IH of the BH. To this end, we take the limits $r\to r_{\pm}$
of the general-$r$ expressions for the fluxes $\left\langle \overline{T}_{yy}\right\rangle _{\text{bare}}^{U}$
(where $y$ is either $u$ or $v$). Since $\Delta$ vanishes at
$r=r_{\pm}$, $\overline{T}_{yy\left(\omega lm\right)}^{\mathcal{A}}$
is the only piece that contributes at the horizons (see Eq.~(\ref{eq:That_sum})). 

Hereafter, a superscript $\pm$ denotes $r\to r_{\pm}$ along with
the coordinate system in use -- $\left(u,v,\theta,\varphi_{\pm}\right)$
at $r_{\pm}$, respectively (recalling that the regular azimuthal
coordinate at $r_{\pm}$ is $\varphi_{\pm}$). Note that since the
coordinate system (at both horizons) is chosen such that, in particular,
$g_{uu}=g_{vv}=0$ there, each flux component coincides with its trace-reversed
counterpart at the horizons (see Eq.~(\ref{eq:new5})) \footnote{In the corresponding coordinate systems, $g_{vv}$ and $g_{uu}$ vanish
on approaching the horizons as $\delta r^{2}$ (where $\delta r\equiv r-r_{\pm}$
denotes the distance to the corresponding horizon). Thus, for the
$g_{yy}\left\langle \overline{T}_{\,\,\,\mu}^{\mu}\right\rangle _{\text{bare}}^{U}$
term to vanish there, we assume that the trace diverges at a sufficiently
slow rate as $\delta r\to0$. (This is, indeed, the case in the RN
counterpart -- see Eq.~(15) in Ref. \cite{GroupPhiRN:2019}, where
the trace divergence rate is weaker than $1/\delta r$.)}. 

We hereby introduce the summation/integration operator,
\begin{equation}
\hat{\sum}_{\pm}\left(\dots\right)\equiv\hbar\int_{0}^{\infty}\sum_{l=0}^{\infty}\sum_{m=-l}^{l}\frac{\left[S_{lm}^{\omega}\left(\theta\right)\right]^{2}}{8\pi^{2}\left(r_{\pm}^{2}+a^{2}\right)}\left(\dots\right)\text{d}\omega\,.\label{eq:Sigma}
\end{equation}

\subsection{The event horizon}

At the EH, as prescribed in Eq.~\eqref{eq:psi_nearEH_asym}, the radial function $\psi_{\omega lm}^{\text{int}}$
behaves as $e^{-i\omega_{+}r_{*}}$. Also, in Eqs.~(\ref{eq:ThatAvv})
and (\ref{eq:ThatAuu}) we now substitute $\tilde{\omega}=\omega_{+}$
(corresponding to the choice $\tilde{\Omega}=\Omega_{+}$ and hence
$\tilde{\varphi}=\varphi_{+}$). This leads to a remarkable simplification,
because now both combinations $\left(\psi_{\omega lm,r_{*}}^{\text{int}}\right)^{2}+\tilde{\omega}^{2}\left(\psi_{\omega lm}^{\text{int}}\right)^{2}$
and $\left|\psi_{\omega lm,r_{*}}^{\text{int}}\right|^{2}+\tilde{\omega}^{2}\left|\psi_{\omega lm}^{\text{int}}\right|^{2}-2\tilde{\omega}\omega_{+}$
vanish, and $\left|\psi_{\omega lm,r_{*}}^{\text{int}}\right|^{2}+\tilde{\omega}^{2}\left|\psi_{\omega lm}^{\text{int}}\right|^{2}+2\tilde{\omega}\omega_{+}$
simplifies to $4\omega_{+}^{2}$. Eq.~(\ref{eq:ThatAvv}) then reduces
to 
\begin{equation}
\left\langle T_{vv}^{+}\right\rangle _{\text{bare}}^{U}=\hat{\sum}_{+}\omega_{+}\left(\left|\rho_{\omega lm}^{\text{up}}\right|^{2}\left[\coth\left(\frac{\pi\omega_{+}}{\kappa_{+}}\right)-1\right]+1\right)\,,\label{eq:Tvv+}
\end{equation}
and Eq.~(\ref{eq:ThatAuu}) to 
\begin{equation}
\left\langle T_{uu}^{+}\right\rangle _{\text{bare}}^{U}=\hat{\sum}_{+}\omega_{+}\coth\left(\frac{\pi\omega_{+}}{\kappa_{+}}\right)\,.\label{eq:Tuu+}
\end{equation}

\subsection{The inner horizon}

We now turn to the IH. At $r\to r_{-}$, the radial function $\psi_{\omega lm}^{\text{int}}$
behaves asymptotically as given in Eq.~\eqref{eq:psi_nearIH_asym}, namely $\psi_{\omega lm}^{\text{int}}\simeq A_{\omega lm}e^{i\omega_{-}r_{*}}+B_{\omega lm}e^{-i\omega_{-}r_{*}}$.

We substitute $\tilde{\omega}=\omega_{-}$ in Eqs.~(\ref{eq:ThatAvv})
and (\ref{eq:ThatAuu}). Then, using the Wronskian relation in Eq.~\eqref{eq:A=000026B_Wron}, we find in the $r\to r_{-}$ limit:
\begin{align}
\omega_{-}^{2}\left|\psi_{\omega lm}^{\text{int}}\right|^{2}+2\omega_{+}\omega_{-}+\left|\psi_{\omega lm,r_{*}}^{\text{int}}\right|^{2} & =4\omega_{-}^{2}\left|B_{\omega lm}\right|^{2}\,,\label{eq:relationB^2}
\end{align}
\begin{align}
\omega_{-}^{2}\left|\psi_{\omega lm}^{\text{int}}\right|^{2}-2\omega_{+}\omega_{-}+\left|\psi_{\omega lm,r_{*}}^{\text{int}}\right|^{2} & =4\omega_{-}^{2}\left|A_{\omega lm}\right|^{2}\,,\label{eq:relationA^2}
\end{align}
and
\begin{equation}
\omega_{-}^{2}\left(\psi_{\omega lm}^{\text{int}}\right)^{2}+\left(\psi_{\omega lm,r_{*}}^{\text{int}}\right)^{2}=4\omega_{-}^{2}A_{\omega lm}B_{\omega lm}\,.\label{eq:relationAB}
\end{equation}

For $\left\langle T_{vv}^{-}\right\rangle _{\text{bare}}^{U}$, this
yields:
\begin{align}
\left\langle T_{vv}^{-}\right\rangle _{\text{bare}}^{U} & =\hat{\sum}_{-}\,\,\frac{\omega_{-}^{2}}{\omega_{+}}\left[\coth\left(\frac{\pi\omega_{+}}{\kappa_{+}}\right)\left(\left|A_{\omega lm}\right|^{2}+\left|\rho_{\omega lm}^{\text{up}}\right|^{2}\left|B_{\omega lm}\right|^{2}\right)\right.\nonumber \\
 & \left.+2\, \text{cosech}\left(\frac{\pi\omega_{+}}{\kappa_{+}}\right)\Re\left(\rho_{\omega lm}^{\text{up}}A_{\omega lm}B_{\omega lm}\right)+\left(1-\left|\rho_{\omega lm}^{\text{up}}\right|^{2}\right)\left|B_{\omega lm}\right|^{2}\right]\,.\label{eq:Tvv-}
\end{align}

Turning now to $\left\langle T_{uu}^{-}\right\rangle _{\text{bare}}^{U}$,
we note again that Eq.~(\ref{eq:TAuu}) differs from Eq.~(\ref{eq:TAvv})
by merely taking $\tilde{\omega}\mapsto-\tilde{\omega}$. This amounts
here to taking $\omega_{-}\mapsto-\omega_{-}$, which in turn interchanges
Eq.~(\ref{eq:relationB^2}) and Eq.~(\ref{eq:relationA^2}). Consequently,
$\left\langle T_{uu}^{-}\right\rangle _{\text{bare}}^{U}$ is obtained
by interchanging $A_{\omega lm}$ and $B_{\omega lm}$ in Eq.~(\ref{eq:Tvv-}):
\begin{align}
\left\langle T_{uu}^{-}\right\rangle _{\text{bare}}^{U} & =\hat{\sum}_{-}\,\,\frac{\omega_{-}^{2}}{\omega_{+}}\left[\coth\left(\frac{\pi\omega_{+}}{\kappa_{+}}\right)\left(\left|B_{\omega lm}\right|^{2}+\left|\rho_{\omega lm}^{\text{up}}\right|^{2}\left|A_{\omega lm}\right|^{2}\right)\right.\nonumber \\
 & \left.+2\, \text{cosech}\left(\frac{\pi\omega_{+}}{\kappa_{+}}\right)\Re\left(\rho_{\omega lm}^{\text{up}}A_{\omega lm}B_{\omega lm}\right)+\left(1-\left|\rho_{\omega lm}^{\text{up}}\right|^{2}\right)\left|A_{\omega lm}\right|^{2}\right]\,.\label{eq:Tuu-}
\end{align}

Note how, through relations (\ref{eq:relationB^2})-(\ref{eq:relationAB}),
all oscillatory factors (of the form $e^{\pm i\omega_{-}r_{*}}$,
as appear in the asymptotic behavior of $\psi_{\omega lm}^{\text{int}}$
at the IH) are canceled out -- and as a consequence, the individual
mode contribution to the flux components have a well-defined limiting
value at the IH, which depends only on the scattering parameters
$\rho_{\omega lm}^{\text{up}}$, $A_{\omega lm}$ and $B_{\omega lm}$.
In this respect, the flux components are simpler than other $T_{\alpha\beta}$
components at the IH limit.

Eq.~(\ref{eq:Tuu-Tvv_ren}) for the renormalized difference also applies
at the IH (in coordinates $\left(u,v,\theta,\varphi_{-}\right)$),
yielding:
\begin{equation}
\left\langle T_{uu}^{-}\right\rangle _{\text{ren}}^{U}-\left\langle T_{vv}^{-}\right\rangle _{\text{ren}}^{U}=\hat{\sum}_{-}\,\omega_{-}\left[\coth\left(\frac{\pi\omega_{+}}{\kappa_{+}}\right)-1\right]\left(1-\left|\rho_{\omega lm}^{\text{up}}\right|^{2}\right)\,.\label{eq:Tuu-Tvv_ren(IH)}
\end{equation}

\end{document}